\documentclass[12pt]{article}

\usepackage{enumerate}
\usepackage{amsmath}
\usepackage{amssymb}
\usepackage{graphicx}
\usepackage{xcolor}
\usepackage{booktabs}
\usepackage{adjustbox}
\usepackage{threeparttable}
\usepackage{threeparttablex}
\usepackage{array}
\usepackage{etoolbox}
\usepackage{natbib} 
\usepackage[linesnumbered,ruled,vlined]{algorithm2e}
\usepackage{amsthm}
\usepackage{url}

\usepackage[utf8]{inputenc}
\usepackage{textcomp}
\usepackage{dsfont}
\usepackage{array}
\usepackage{newtxtext}
\usepackage{newtxmath}
\usepackage{anyfontsize}
\usepackage{makecell}

\usepackage{multirow}
\usepackage{float}
\usepackage{tablefootnote}
\usepackage[skip=0.5\baselineskip]{caption}

\usepackage[colorlinks=true,linkcolor=black,citecolor=blue,urlcolor=blue]{hyperref} 
\usepackage{authblk} 
\newtheorem{assumption}{Assumption}

\newtheorem{theorem}{Theorem}
\newtheorem{example}{Example}
\newcommand{\indep}{\perp \!\!\! \perp}

\newenvironment{continuedexample}[1]{%
  \renewcommand\theexample{\ref*{#1} (continued)}%
  \begin{example}
}{%
  \end{example}
}

\newcommand{\nrct}{{n_{\mathcal{R}}}}
\newcommand{\nec}{{n_{\mathcal{E}}}}

\newcommand{\prct}{{\pi_{\mathcal{R}}}}

\newcommand{\ifrct}{{\mathbb{IF}_{\mathcal{R}}}}

\newcommand{\hthetaR}[1]{\hat{\theta}_{#1,\mathcal{R}}}

\newcommand{\taurd}{\tau_{\rm RD}}
\newcommand{\htaurdR}{\hat{\tau}_{{\rm RD},\mathcal{R}}}
\newcommand{\htaurd}{\hat{\tau}_{\rm RD}}

\newcommand{\taurr}{\tau_{\rm RR}}
\newcommand{\htaurrR}{\hat{\tau}_{{\rm RR},\mathcal{R}}}
\newcommand{\htaurr}{\hat{\tau}_{\rm RR}}

\newcommand{\tauor}{\tau_{\rm OR}}
\newcommand{\htauorR}{\hat{\tau}_{{\rm OR},\mathcal{R}}}
\newcommand{\htauor}{\hat{\tau}_{\rm OR}}

\newcommand\highlight[1]{\textcolor{black}{#1}}

\begin{document}

\def\spacingset#1{\renewcommand{\baselinestretch}%
{#1}\small\normalsize} \spacingset{1}

\title{\bf Robust Estimation and Inference in Hybrid Controlled Trials for Binary Outcomes: A Case Study on Non-Small Cell Lung Cancer}

\author[1]{Jiajun Liu}
\author[1,2]{Ke Zhu\footnote{Co-first author}}
\author[2]{Shu Yang}
\author[1]{Xiaofei Wang\footnote{Address for correspondence: Xiaofei Wang, Department of Biostatistics and Bioinformatics, Duke University, Durham, NC 27710, U.S.A. Email: xiaofei.wang@duke.edu}}

\affil[1]{\small Department of Biostatistics and Bioinformatics, Duke University, Durham, NC 27710, U.S.A.}

\affil[2]{\small Department of Statistics, North Carolina State University, Raleigh, NC 27695, U.S.A.}

\date{}
\maketitle

\begin{abstract}
Hybrid controlled trials (HCTs), which augment randomized controlled trials (RCTs) with external controls (ECs), are increasingly attracting attention to address limited power, slow accrual, and ethical concerns in clinical research. However, borrowing from ECs raises challenges in estimation and inference, especially for binary outcomes where hidden bias is difficult to detect and estimands such as risk difference, risk ratio, and odds ratio are of primary interest. We propose a framework that combines doubly robust estimators under covariate shift with conformal selective borrowing (CSB) to address outcome drift. CSB uses nearest-neighbor conformal scores and label-conditional extensions to construct conformal p-values measuring individual-level EC comparability, and uses an MSE minimization procedure to determine the selection threshold. To ensure finite-sample type I error control, we further use a Fisher randomization test for the sharp null with the CSB estimator as the test statistic. Extensive simulations demonstrate robust performance. We apply the method to CALGB 9633 and National Cancer Database data to evaluate adjuvant chemotherapy effects in Stage IB non-small-cell lung cancer and show that the proposed method mitigates hidden bias from full-borrowing approaches and improves efficiency over RCT-only analysis. We also provide practical guidance and the R package intFRT.

\end{abstract}

\bigskip
\noindent%
\vfill

\newpage
\spacingset{1.9} 

\section{Introduction}\label{sec:intro}

Randomized controlled trials (RCTs) are considered the gold standard for estimating treatment effects, as randomization eliminates both measured and unmeasured confounding. However, classical RCTs face several limitations in medical research: they may be underpowered to detect clinically meaningful effects in the target population due to limited sample sizes and slow accrual, particularly in rare diseases or urgent public health crises \citep{fda2019rare}; they may also raise ethical concerns when patients or physicians are unwilling to accept randomization due to lack of equipoise \citep{miller2011}; and they are often costly and time-consuming to conduct. In contrast, real-world data (RWD), such as data from historical RCTs or observational studies, often contain rich information that can supplement RCTs \citep{pocock1976,colnet2024causal}. This has led to growing interest in borrowing external controls (ECs) to enhance RCTs and support treatment effect evaluation, giving rise to innovative designs known as hybrid controlled trials (HCTs) \citep{ventz2022design,shan2022simulation,hampson2023innovative}. HCTs anchor on the internal validity of RCTs while improving efficiency by leveraging external information, accelerating timelines in urgent settings, and enabling a higher treatment allocation ratio, thereby improving patient outcomes and welfare. HCTs represent a promising approach to integrating RCTs and RWD in modern clinical research.

However, HCTs introduce new statistical challenges in both estimation and inference due to potential bias arising from ECs, which is primarily driven by five key concerns: selection bias, unmeasured confounding, lack of concurrency, data quality, and outcome validity \citep{fda2019rare}. From the statistical perspective, these concerns can be broadly classified into two types: the first is classified as baseline incomparability (also referred to as covariate shift or observed bias), while the remaining four are classified as outcome incomparability (also known as outcome drift or hidden bias).
To specifically address baseline incomparability, covariate balancing approaches from the causal inference literature such as matching, propensity score weighting, calibration weighting, and their augmented forms have been used \citep{chen2020propensity,li2023improving,Valancius2024}.
To address outcome incomparability, various Bayesian methods, including informative priors (e.g., power priors \citep{ibrahim2000power} and elastic priors \citep{jiang2023elastic}) and shrinkage approaches (e.g., Bayesian hierarchical models \citet{kaizer2018bayesian}), as well as frequentist methods, including prognostic adjustment \citep[e.g.][]{schuler2022increasing}, test-then-pool \citep[e.g.][]{viele2014use,yang2023elastic}, selective borrowing \citep[e.g.][]{Gao2024,Zhu2024}, bias-based integration \citep[e.g.][]{stuart2008matching,yang2024datafusion}, estimator averaging \citep[e.g.][]{chen2021minimax,rosenman2023combining}, and sensitivity analysis \citep[e.g.][]{gordon2025non}, have been proposed. For a broader review, we refer readers to \cite{zhu2026externally} and the references therein for a comprehensive review of these methods.

\highlight{Binary outcomes, which are commonly used endpoints in RCTs such as tumor response, hospitalization, and viral clearance in oncology and other clinical trials, remain underexplored in HCTs that jointly account for both covariate and outcome comparability of ECs. First, general estimands for binary outcomes, including the risk difference (RD), risk ratio (RR), and odds ratio (OR), are often of primary interest, yet semiparametric efficient estimators that accommodate covariate shift in ECs remain limited. Second, binary outcomes introduce unique challenges for identifying and adjusting for hidden bias, as they contain less information than continuous outcomes, making outcome incomparability more difficult to detect.}

\highlight{To address the first challenge, we derive efficient influence functions for general estimands, including the RD, RR, and OR, under covariate shift between the RCT and EC, and propose corresponding doubly robust estimators and inference procedures.}

\highlight{To address the second challenge, motivated by our real data application where only a subset of ECs remains comparable to randomized controls after covariate balancing, we propose selectively borrowing ECs based on their conditional outcome comparability.
We build on conformal inference, a flexible framework that provides finite-sample valid uncertainty quantification for individual predictions across diverse data types \citep{vovk2005algorithmic}, to perform unit-level exchangeability testing and guide borrowing decisions. In particular, we leverage nearest-neighbor conformal scores \citep{shafer2008tutorial}, which demonstrate strong selection performance for binary outcomes, and further improve performance through label-conditional conformal prediction \citep{vovk2012conditional}. These scores were selected from a broad class of candidate conformal scores through extensive numerical experiments and were found to outperform residual-based conformal scores in binary settings. Collectively, these contributions extend the conformal selective borrowing (CSB) framework \citep{Zhu2024} to broader applications, providing a more flexible and robust alternative to existing global or model-based borrowing methods in HCTs.}

The third fundamental challenge lies in type I error rate control and power gain in HCT. Type I error rate control is critical for establishing treatment efficacy and remains a key requirement for regulatory approval. At the same time, achieving substantial power gain (e.g., a 10\% increase) by borrowing EC is essential; without a clear advantage in power, RCT-only analyses may be preferred due to their well-established internal validity. Existing work has shown that “power gains by using external information in clinical trials are typically not possible when requiring strict type I error rate control” under Bayesian borrowing frameworks \citep{kopp2020power,kopp2024simulating}. Frequentist methods that rely on asymptotic inference may also inflate the type I error rate, as they assume large RCT sample sizes, which is often unrealistic in practice since limited sample size is typically the motivation for borrowing ECs. Moreover, both Bayesian dynamic and frequentist selective borrowing introduce selection uncertainty, which can further compromise type I error rate control. While some recent work explores alternative criteria beyond the traditional type I error rate \citep{best2024beyond,gao2025control}, the type I error rate remains the prevailing benchmark in practice.

To address this third challenge, we propose using the Fisher randomization test (FRT) \citep{fisher1935} to ensure strict type I error rate control and use CSB as the test statistic to enable power gain. The validity of FRT relies solely on the randomization and holds for any test statistic as long as the “analyze as you randomize” principle is followed. FRTs are commonly used in small-sample clinical trials due to their exact finite-sample validity and model-free nature \citep{zheng2008multi,ji2017,wang2023randomization}, and are often recommended as a backup option in adaptive designs \citep{Simon2011,plamadeala2012sequential,carter2024regulatory}. We develop a valid FRT for HCT by permuting only within the RCT and keeping the assignment of ECs fixed. Using the proposed CSB estimator as the test statistic, we show that power gain is possible under strict type I error rate control when some ECs are unbiased and others exhibit detectable bias.

\subsection{Motivation Example: CALGB 9633 Trial with External Control from NCDB}\label{sec:data}
The challenges and concerns discussed above are motivated by the following real-world scientific problems.
Cancer and Leukemia Group B (CALGB) 9633 is an RCT targeting patients with Stage IB Non-Small Cell Lung Cancer (NSCLC) \citep{Strauss2008}. Their primary objective is to study the effectiveness of adjuvant chemotherapy on the overall survival compared to observation only after surgical resections. From 1996 to 2003, a total of $\nrct=335$ patients were recruited in CALGB 9633, with $n_1=167$ randomized to the adjuvant chemotherapy (treated) group and $n_0=168$ to the observation (control) group. The measured pre-treatment covariates include gender, age, ethnicity, performance status, weight loss, indicator of symptoms, duration of symptoms, tumor size in diameter, histology records, tumor differentiation, indicator of mediastinoscopy, type of surgical procedure, and extent of resection. 

From previous studies, the overall survival for adjuvant chemotherapy did not statistically significantly outperform the overall survival of the observation with $p$-value of 0.125 and Hazard Ratio (HR) of 0.83. The limited trial size was criticized as underpowered when evaluating the effectiveness of adjuvant chemotherapy \citep{Khan2018}. Therefore, incorporating EC data may enrich the dataset and support the evaluation of treatment effect. 

National Cancer Database (NCDB) is an oncology outcomes database that collects information on roughly $70\%$ new invasive cancer diagnoses across the U.S. Between 2004 and 2016, a total of 16217 patients were diagnosed with NSCLC and received either adjuvant chemotherapy or solely observation after the surgery \citep{NCDB}. We extracted 11700 patients as the source of EC. 
Although NCDB and CALGB 9633 collect similar types of information, their covariate distributions exhibit noticeable differences, motivating the need for methods that ensure covariate balance before integrating NCDB with CALGB 9633. Moreover, the ECOG variable is available in CALGB 9633 but is missing from NCDB, raising concerns about hidden bias due to unmeasured confounding. As shown in Figure \ref{fig:caseHBall}(A), there is also evidence suggesting the presence of potential outcome incomparability in NCDB. 

To address these challenges, we develop a suite of methods in this paper. Section~\ref{sec:framework} introduces the causal framework and RCT-only benchmarks. Sections~\ref{sec:covshift} and~\ref{sec:outcomeShift} address covariate and outcome incomparability through doubly robust estimation and conformal selective borrowing, respectively. Section~\ref{sec:FRT} introduces randomization inference for type I error control. Sections~\ref{sec:simstudy} and~\ref{sec:casestudy} provide the simulation study and real-data case study, and Section~\ref{sec:conclusion} concludes with practical implications and future directions.

\section{Causal Inference Framework for Hybrid Controlled Trials}\label{sec:framework} 

\subsection{Problem Setup}
We consider the RCT population ($S=1$) as the target due to its strong internal validity and role as the regulatory gold standard for drug approval and labeling, where $S$ indicates the data source and $S=0$ corresponds to EC data.  
Define the expectation of the potential outcome in the RCT population as  
$$
\theta_a = \mathbb{E}\{Y(a) \mid S=1\},
$$  
where $Y(a)$ is the potential outcome under treatment level $a\in\{0,1\}$.
We define estimands by contrasting $\theta_1$ and $\theta_0$, including risk difference (RD), risk ratio (RR), and odds ratio (OR):
$$
\taurd = \theta_1 - \theta_0, \quad 
\tau_{\text{RR}} = \theta_1 / \theta_0, \quad
\tau_{\text{OR}} = \frac{\theta_1 / (1 - \theta_1)}{\theta_0 / (1 - \theta_0)}.
$$  
Let $\nrct$ denote the number of RCT participants $\mathcal{R} = \{i : S_i = 1\}$, with $n_1$ and $n_0$ randomized to the treatment and control groups, respectively. We borrow $\nec$ EC participants $\mathcal{E} = \{i : S_i = 0\}$ to form the hybrid controlled trial with total sample size $n = \nrct + \nec$. 
The sampling score for enrolling in the RCT is $\pi(x) = \mathbb{P}(S = 1 \mid X = x)$. Within the RCT, $A$ is the treatment assignment, with $A = 1$ for treated and $A = 0$ for control; the propensity score for being assigned to treatment is $e(x) = \mathbb{P}(A = 1 \mid X = x, S = 1)$, which is known. $Y$ is the observed outcome. 
The data in the RCT are denoted as $\{Y_i, A_i, X_i, S_i = 1\}$, and data in the EC as $\{Y_i, A_i = 0, X_i, S_i = 0\}$.  
All observed data are denoted as $O_i = (Y_i, A_i, X_i, S_i)$ for $i = 1, \dots, n$.

Under the following Assumption \ref{as1}, which holds under the RCT design, $\theta_a$ and the above estimands are identifiable using only the RCT data.
\begin{assumption}\label{as1}
(i) (Consistency) $Y=A\cdot Y(1)+(1-A)\cdot Y(0)$
(ii) (Overlap) $0<e(x)<1$ with probability 1 for all $x$ $s.t.$ $f_{X|S}(X=x|S=1)>0$, where $f_{X|S}(X=x|S=1)$ is the conditional p.d.f. of $X$ given $S=1$.
(iii) (Unconfoundedness) $\{Y(1),Y(0)\}\indep A|(X,S=1)$
\end{assumption}

\subsection{Semiparametric Efficient RCT-only Estimators}\label{sec:framework_noBorrow} 

We introduce RCT-only estimators for $\theta_a$ in HCT, which serve as building blocks. The most straightforward estimator is \texttt{No Borrow Unadj}, which relies solely on the RCT data and does not do covariate adjustment between the treated and control groups: $\hat{\theta}_{a,\text{Unadj}}=n_a^{-1}\sum_{i=1}^n S_i\cdot\mathbb{I}(A_i=a)\cdot Y_i$.
Corresponding plug-in estimators for RD, RR, and OR are:
$$
\hat{\tau}_{\text{RD,Unadj}}=\hat{\theta}_{1,\text{Unadj}} -\hat{\theta}_{0,\text{Unadj}},\quad
\hat{\tau}_{\text{RR,Unadj}}=\hat{\theta}_{1,\text{Unadj}} / \hat{\theta}_{0,\text{Unadj}},\quad \hat{\tau}_{\text{OR, Unadj}}=\frac{\hat{\theta}_{1,\text{Unadj}}/(1-\hat{\theta}_{1,\text{Unadj}})}{\hat{\theta}_{0,\text{Unadj}}/(1-\hat{\theta}_{0,\text{Unadj}})}.
$$

To better evaluate the efficiency gain from EC borrowing, we use the RCT-only semiparametric efficient estimator \texttt{No Borrow CovAdj} as the benchmark, which adjusts for covariates within the RCT. The RCT-only efficient influence function (EIF) of $\theta_a$ is given by
\begin{align*}
    \ifrct(\theta_a) &= \frac{S}{\prct}\bigg[\underbrace{\frac{A^a(1-A)^{1-a}}{e(X)^a(1-e(X))^{1-a}}\{Y - \mu_a(X)\} + \mu_a(X)}_{\xi_a(O)}
- \theta_a\bigg]
\end{align*}
Let $\hat{\mu}_{a,\mathcal{R}}$ and $\hat{e}$ denote the corresponding estimators using only RCT data. Let $\hat\xi_a(O_i)$ denote the empirical version of $\xi_a(O)$ with estimated nuisance functions, that is,
\begin{align*}
    \hat\xi_a(O_i)\equiv
\frac{A_i^a(1-A_i)^{1-a}}{\hat{e}(X_i)^a(1-\hat{e}(X_i))^{1-a}}\{Y_i - \hat{\mu}_{a,\mathcal{R}}(X_i)\} + \hat{\mu}_{a,\mathcal{R}}(X_i).
\end{align*}
By solving empirical version of EIF $\sum_{i=1}^n(S_i/\prct)\{\hat\xi_a(O_i)-\theta_a\}=0$, we obtain
\begin{align}
\label{eq:rct}
\hat{\theta}_{a,\text{CovAdj}}=\frac{1}{\nrct}\sum_{i=1}^n S_i\hat\xi_a(O_i).
\end{align}
After obtaining $\theta_a$ estimators as building blocks, estimators for $\taurd$, $\taurr$, and $\tauor$ are derived by plug-in, with explicit forms and asymptotic inference in Section~\ref{Supp:formula_estimator_RCTonly} of the Supplementary Material.

\section{Doubly Robust Borrowing for Addressing Covariate Incomparability of ECs} \label{sec:covshift}

Borrowing EC can improve the efficiency of RCT-only analysis, but it may also introduce covariate incomparability between the EC and RCT populations. To address this, we introduce Doubly Robust Borrowing estimators by solving the empirical version of the EIF.
\begin{assumption}\label{as2}
    (Mean exchangeability) $\mathbb{E}\{Y(0)|X, S=1\}=\mathbb{E}\{Y(0)|X, S=0\}$.
\end{assumption}

For binary outcomes, Assumption \ref{as2} implies that the potential outcomes for the EC and RCT control groups share the same \textit{conditional distribution}, which will be relaxed in Section \ref{sec:outcomeShift}. Under Assumptions \ref{as1} and \ref{as2}, $\theta_0$ can be estimated using both EC and RCT data, while the EIF and estimator for $\theta_1$ remain the same as in the RCT-only analysis, since no external data are borrowed for the treatment group.
Let $\prct=\nrct/n$ denote the sample ratio of RCT data.
The EC Borrowing EIF of $\theta_0$ is given by \citep{li2023improving}:  
\begin{align*}  
\mathbb{IF}(\theta_0) = \underbrace{\frac{\pi(X)}{\prct}\frac{S(1-A) + (1-S)r(X)}{\pi(X)\{1-e(X)\} + \{1-\pi(X)\}r(X)}\{Y - \mu_0(X)\} + \frac{S}{\prct}\mu_0(X)}_{\phi_0(O)}
-\frac{S}{\prct} \theta_0.
\end{align*}
Let $\hat{\mu}_{0,\mathcal{R}+\mathcal{E}}(X)$ denote the outcome model fitted by both RCT control and EC. 
Let $\hat\pi(X)$ denote the fitted sampling model $\pi(X)=\mathbb{E}\{S=1|X\}$.
Let $\hat{r}(X)$ denote the fitted variance ratio model for $r(X)=\mathbb{V} \{Y(0)|X,S=1\}/\mathbb{V}\{Y(0)|X,S=0\}$.
Let $\hat\phi_0(O_i)$ denote the empirical version of $\phi_0(O)$ with estimated nuisance functions, that is,
$$
\hat\phi_0(O_i)\equiv
\frac{\hat\pi(X_i)}{\prct}\frac{S_i(1-A_i) + (1-S_i)\hat r(X_i)}{\hat\pi(X_i)\{1-\hat e(X_i)\} + \{1-\hat\pi(X_i)\}\hat r(X_i)}\{Y_i - \hat\mu_{0,\mathcal{R}+\mathcal{E}}(X_i)\} + \frac{S_i}{\prct}\hat\mu_{0,\mathcal{R}+\mathcal{E}}(X_i)
.
$$
By solving empirical version of EIF $\sum_{i=1}^n\{\hat\phi_0(O_i)-(S_i/\prct)\theta_0\}=0$, we obtain
\begin{align*}
\hat{\theta}_{0,\text{AIPW}}
=\frac{1}{\nrct}\sum_{i=1}^n\bigg[
\hat\pi(X_i)\frac{S_i(1-A_i) + (1-S_i)\hat r(X_i)}{\hat\pi(X_i)\{1-\hat e(X_i)\} + \{1-\hat\pi(X_i)\}\hat r(X_i)}\{Y_i - \hat\mu_{0,\mathcal{R}+\mathcal{E}}(X_i)\} + {S_i}\hat\mu_{0,\mathcal{R}+\mathcal{E}}(X_i)
\bigg].
\end{align*}
For the treatment arm, since no external information is borrowed, we use $\hat{\theta}_{1,\text{AIPW}} = \hat{\theta}_{1,\text{CovAdj}}$ as defined in \eqref{eq:rct}.
The \texttt{Borrow AIPW} estimators for $\taurd$, $\taurr$, and $\tauor$ are obtained by plug-in using $\hat{\theta}_{a,\text{AIPW}}$, with explicit forms and asymptotic inference in Section~\ref{Supp:formula_estimator_EC} of the Supplementary Material.

In addition to \texttt{Borrow AIPW}, we consider five alternative EC Borrowing approaches, with explicit formulas provided in Section~\ref{Supp:alternative_estimator} of the Supplementary Material. \texttt{Borrow Na\"{i}ve} pools RCT controls and ECs without adjusting for covariate shift and is thus not recommended. Inspired by covariate balancing techniques from the causal inference literature for observational studies, the remaining methods are adapted to the HCT setting to balance covariates between the RCT and EC populations \citep{colnet2024causal}. These include Inverse Probability Weighting (\texttt{Borrow IPW}), Calibration Weighting (\texttt{Borrow CW}), Outcome Modeling (\texttt{Borrow OM}), and Augmented Calibration Weighting (\texttt{Borrow ACW}).

\section{Conformal Selective Borrowing for Addressing Outcome Incomparability of ECs} \label{sec:outcomeShift} 
Although \texttt{Borrow AIPW} addresses covariate incomparability, it cannot account for outcome incomparability. 
\highlight{When hidden bias is present and manifests as incomparability in outcome patterns, Assumption \ref{as2} no longer holds for the full borrowing sample. To address this issue, we propose Conformal Selective Borrowing (CSB), whose goal is to identify a subset of EC subjects for whom Assumption \ref{as2} is more plausible. To this end, CSB leverages conformal inference \citep{vovk2005algorithmic} to perform individual-level exchangeability testing using flexible similarity measures, while enjoying finite-sample validity.}
\highlight{Using the RCT control data $\mathcal{C}$ as a benchmark, CSB aims to assess whether each EC subject is compatible with the outcome-covariate pattern observed among randomized controls. The quantity $b_j=Y_j-\mathbb{E}\{Y(0)\mid X=X_j,S=1\}$ can be viewed as a conceptual measure of individual-level outcome drift, but it is not directly identifiable from observed data. In practice, CSB assesses such incompatibility indirectly through conformal $p$-values that compare each EC subject with the RCT controls under an exchangeability-based criterion.} 
In the following, we introduce (i) how to better measure comparability for binary outcomes and calibrate it into a working conformal $p$-value, (ii) how to determine which EC subjects to borrow based on the conformal $p$-values and construct the resulting estimator, and (iii) how to adaptively determine the borrowing amount through MSE minimization.

\subsection{Conformal Score and Conformal p-value}\label{sec:outcomeShift_conformalscore}
Conformal scores measure how far the ECs deviate from the RCT controls. For continuous outcomes, commonly used conformal scores include the absolute residual score, scaled absolute residual score, and conformal quantile regression \citep{angelopoulos2024theoretical}. \highlight{However, these conformal scores can behave poorly in binary settings. To illustrate this, consider the residual-based score $s_j=|Y_j-\hat{\mu}(X_j)|$, where $\hat{\mu}(X_j)$ denotes the predicted probability. For EC subjects with the same observed outcome (e.g., $Y_j=1$), the score is then determined by $\hat{\mu}(X_j)$. Consequently, the procedure tends to favor EC subjects with more extreme predicted probabilities, rather than subjects more comparable to the RCT controls.}

Thus, we consider the nearest-neighbor (NN) conformal score \citep{Shafer2008}, which directly compares covariate similarity among subjects with the same observed outcome and therefore provides a more reasonable assessment of EC comparability under binary outcomes. Specifically, for each EC $j$ and training set $\mathcal{N}$, the conformal score is defined as the distance to its nearest neighbor with the same outcome, i.e., $s_j=\min\{d(X_j,X_l):Y_l=Y_j,\; l\in \mathcal{N}\}$, where $X_j$ and $X_k$ denote the covariate vectors for subjects $j$ and $k$, respectively, and $d(\cdot,\cdot)$ denotes the Euclidean distance, although other distance metrics can be used in general.
\highlight{In implementation, continuous covariates should be centered and scaled and categorical covariates should be dummy-coded before Euclidean distances are computed; alternatively, other clinically justified metrics, such as Mahalanobis distance or propensity-score-based distances, may be used when appropriate.}

To compute the conformal $p$-value, we employ the cross-validation+ (CV+) procedure \citep{Barber2021}. Specifically, we randomly split the RCT control set $\mathcal{C}$ into $K$ disjoint folds such that $\mathcal{C}=\bigcup_{k=1}^K\mathcal{C}_k$. Each fold $\mathcal{C}_k$ is treated in turn as the calibration set, while the remaining subjects $\mathcal{C}\backslash\mathcal{C}_k$ form the training set. 
For any EC subject $j\in\mathcal{E}$, we first compute its conformal score $s_j^{(i)}=\min\{d(X_j,X_l):Y_l=Y_j,\; l\in \mathcal{C}\backslash\mathcal{C}_{k(i)}\}$, where $k(i)\in\{1,\ldots,K\}$ is a function indicating that $i\in\mathcal{C}_k$. Note that for $i\neq i^\prime$ with $k(i)=k(i^\prime)$, we have $s_j^{(i)}=s_j^{(i^\prime)}$. We then compute conformal scores $s_i=\min\{d(X_i,X_l):Y_l=Y_i,\; l\in \mathcal{C}\backslash\mathcal{C}_{k(i)}\}$ for all RCT controls. The conformal $p$-value for EC subject $j$ is then defined as
$$
p_{j}^{\text{NN}}=
\frac{\sum_{i\in \mathcal{C}}\mathbb{I}(s_i\geq s_j^{(i)})+1}{|\mathcal{C}|+1}.
$$

Additionally, inspired by label-conditional coverage guarantees \citep{vovk2012conditional}, we further propose a label-conditional conformal $p$-value based on the NN conformal score, denoted by LC-NN. Specifically, when computing the conformal $p$-value for an EC subject $j$, LC-NN restricts the comparison to calibration subjects with the same observed outcome as $j$. Let $\mathcal{C}_{Y_j}=\{i\in\mathcal{C}: Y_i=Y_j\}$ denote the subset of RCT controls with outcome equal to $Y_j$. The conformal $p$-value for EC subject $j$ is then defined as
$$
p_{j}^{\text{LC-NN}}=
\frac{\sum_{i\in \mathcal{C}_{Y_j}}\mathbb{I}(s_i\geq s_j^{(i)})+1}{|\mathcal{C}_{Y_j}|+1}.
$$

\subsection{Selected ECs and Treatment Effect Estimator}\label{sec:outcomeShift_conformalP}
\highlight{Given a threshold $\gamma\in[0,1]$, which will be determined later, we define the selected EC subset as $\hat{\mathcal{E}}(\gamma)=\{j\in \mathcal{E}:p_j^*>\gamma\}$, for which subjects are more comparable to the RCT controls and Assumption \ref{as2} is more plausible to hold relative to the full EC sample, where $*\in\{\text{NN},\text{LC-NN}\}$.} 

\highlight{Then, the estimator is constructed by replacing the entire EC set $\mathcal{E}$ in \texttt{Borrow AIPW} with the selected EC set $\hat{\mathcal{E}}(\gamma)$. {Specifically, the building-block estimators are}
\begin{align*}
\hat{\theta}_{1,\mathrm{CSB}}
&=
\frac{1}{\nrct}\sum_{i=1}^n
\left[
S_i\hat\mu_{1,R}(X_i)
+
\frac{S_iA_i}{\hat e(X_i)}
\left\{Y_i-\hat\mu_{1,R}(X_i)\right\}
\right], \\
\hat{\theta}_{0,\mathrm{CSB},\gamma}
&=
\frac{1}{\nrct}\sum_{i=1}^n
\left[
S_i\hat\mu_{0,R+\hat{\mathcal E}(\gamma)}(X_i)
+
\hat W
\left\{Y_i-\hat\mu_{0,R+\hat{\mathcal E}(\gamma)}(X_i)\right\}
\right],
\end{align*}
where
\begin{align*}
    \hat{W} =\hat{\pi}_{\hat{\mathcal{E}}(\gamma)}(X_i)\frac{ S_i(1 - A_i) + (1 - S_i) \mathbb{I}\{i\in\hat{\mathcal{E}}(\gamma)\}\hat{r}_{\hat{\mathcal{E}}(\gamma)}(X_i) }{\hat{\pi}_{\hat{\mathcal{E}}(\gamma)}(X_i) \{1 - \hat{e}(X_i)\} + \{1 - \hat{\pi}_{\hat{\mathcal{E}}(\gamma)}(X_i)\} \hat{r}_{\hat{\mathcal{E}}(\gamma)}(X_i)}.
\end{align*}
Finally, we obtain the treatment effect estimators
$$
\hat\tau_{\mathrm{CSB},\gamma}^{\mathrm{RD}}
= \hat{\theta}_{1,\mathrm{CSB}}
 -
\hat{\theta}_{0,\mathrm{CSB},\gamma},
\quad
\hat\tau_{\mathrm{CSB},\gamma}^{\mathrm{RR}}
= \hat{\theta}_{1,\mathrm{CSB}}
/
\hat{\theta}_{0,\mathrm{CSB},\gamma},
\quad
\hat\tau_{\mathrm{CSB},\gamma}^{\mathrm{OR}}
= \frac{\hat{\theta}_{1,\mathrm{CSB}}/(1-\hat{\theta}_{1,\mathrm{CSB}})}{\hat{\theta}_{0,\mathrm{CSB},\gamma}/(1-\hat{\theta}_{0,\mathrm{CSB},\gamma})}.
$$
Notably, \texttt{No Borrow CovAdj} and \texttt{Borrow AIPW} are special cases corresponding to $\gamma=1$ and $\gamma=0$, respectively.}

\subsection{Adaptive Selection Threshold}\label{sec:adaGamma}

\highlight{To determine $\gamma$, we use a data-adaptive procedure that minimizes the mean squared error (MSE) of the estimator $\hat{\tau}_{\mathrm{CSB},\gamma}^\dagger$, where $\dagger\in\{{\rm RD, RR, OR}\}$. For each estimand, we perform adaptive $\gamma$ selection separately.
The MSE decomposes into squared bias and variance:
$
\mathrm{MSE}(\gamma)
=
\mathbb{E}(\hat{\tau}_{\mathrm{CSB},\gamma}^\dagger-\tau_\dagger)^2
=
\{\mathbb{E}(\hat{\tau}^\dagger_{\mathrm{CSB},\gamma}-\tau_\dagger)\}^2
+
\mathbb{V}(\hat{\tau}^\dagger_{\mathrm{CSB},\gamma}).
$}

\highlight{Since $\tau_\dagger$ is unknown, we use the \texttt{No Borrow CovAdj} estimator $\hat{\tau}^\dagger_{\mathrm{CSB},1}$, which is consistent for $\tau_\dagger$, to approximate the squared bias. Thus,
$
\mathrm{MSE}(\gamma)
\approx
\mathbb{E}(\hat{\tau}^\dagger_{\mathrm{CSB},\gamma}-\hat{\tau}^\dagger_{\mathrm{CSB},1})^2
-
\mathbb{V}(\hat{\tau}^\dagger_{\mathrm{CSB},\gamma}-\hat{\tau}^\dagger_{\mathrm{CSB},1})
+
\mathbb{V}(\hat{\tau}^\dagger_{\mathrm{CSB},\gamma}),
$
where variances are estimated by bootstrap, and $\mathbb{E}(\hat{\tau}^\dagger_{\mathrm{CSB},\gamma}-\hat{\tau}^\dagger_{\mathrm{CSB},1})^2$ is approximated by $(\hat{\tau}^\dagger_{\mathrm{CSB},\gamma}-\hat{\tau}^\dagger_{\mathrm{CSB},1})^2$.
We then perform a grid search over $\gamma$ to minimize the estimated MSE.
A finer $\gamma$ grid and more bootstrap samples may improve threshold selection. The detailed algorithm is provided in Section~\ref{Supp:alg} of the Supplementary Material.}

\subsection{Summary of all methods}
Table \ref{tab:Sum_Methods} summarizes ten methods, including two No Borrow approaches, six Borrow approaches, and two CSB methods. Compared to the No Borrow approaches, our proposed CSB methods can incorporate information from the EC to improve power. In contrast to the Borrow methods, the CSB methods address covariate incomparability and mitigate outcome incomparability by identifying hidden bias. 

The applicability of these methods under asymptotic inference is limited. Asymptotic validity depends on both large sample sizes and correct model specification; thus, inference may be unreliable when either condition is violated. Among Borrow methods designed to address covariate incomparability, their consistency relies on the correct specification of the SM or OM. While \texttt{Borrow AIPW} and \texttt{Borrow ACW} have double robustness, they still require at least one nuisance model to be correctly specified, which is an assumption that may be hard to verify and hold in practice. Motivated by this limitation, we introduce the FRT as a randomization inference in the next section to provide a model-free alternative that ensures valid inference even in small samples and in the presence of outcome incomparability.
 
\begin{table}[!t]
\caption{Summary of all the methods to be discussed}
\label{tab:Sum_Methods}
\centering
\renewcommand{\arraystretch}{1}
\resizebox{\textwidth}{!}{
\begin{tabular}{lcccc}
\hline
Method & EC Borrow & Model Specification & \makecell{Adjust Covariate\\Incomparability of EC} &\makecell{Adjust Outcome\\Incomparability of EC}\\
\hline
No Borrow Unadj  &  $\times$ &  No need & - & - \\
No Borrow CovAdj  &  $\times$ & No need & - & - \\
Borrow Na\"ive & $\checkmark$ & Covariates perfectly balanced & $\times$ & $\times$ \\
Borrow IPW  & $\checkmark$ & SM$^a$ correct & $\checkmark$ & $\times$ \\
Borrow CW  & $\checkmark$ & SM correct/OM$^b$ linear & $\checkmark$ & $\times$ \\
Borrow OM  & $\checkmark$ & OM correct & $\checkmark$ & $\times$ \\
Borrow AIPW  & $\checkmark$ & SM/OM correct & $\checkmark$ & $\times$ \\
Borrow ACW  & $\checkmark$ & SM/OM correct & $\checkmark$ & $\times$ \\
CSB NN  & $\checkmark$ & SM/OM correct & $\checkmark$ & $\checkmark$ \\
CSB LC-NN  & $\checkmark$ & SM/OM correct & $\checkmark$ & $\checkmark$ \\
\hline
\end{tabular}}\\
\raggedright
\footnotesize{$^a$ Sampling model; $^b$ Outcome model}
\end{table}

\section{Randomization Inference for Type I Error Rate Control} \label{sec:FRT}
The validity of asymptotic inference relies on large-sample theory and doubly robust model specification. Full borrowing further requires outcome comparability of ECs, while selective borrowing introduces data-driven selection uncertainty that is not captured by standard asymptotic variance estimates \citep{Gao2024}. The FRT proposed in this section avoids these requirements by permuting treatment assignments within the RCT while holding EC assignments fixed, and recalculating the test statistic after reselecting ECs for each permutation. Thus, when CSB estimators are used as test statistics, FRT can account for selection uncertainty while selectively incorporating compatible ECs. 
FRT is flexible and can be adapted to any test statistic \citep{Rubin1980}, including the CSB estimators. Below, we describe its implementation and integration into our framework.

FRT is constructed on a sharp null hypothesis, i.e., $H_0:Y_i(0)=Y_i(1)$ for $\forall i\in \mathcal{R}$. This sharp null hypothesis implies that any units in RCT fail to reflect any treatment effect. Given $H_0$, the potential outcomes are equal and also equal to the observed outcome, i.e., $Y_i(0)=Y_i(1)=Y_i$ for $\forall i\in \mathcal{R}$. Let the treatment assignment vector be $\boldsymbol{A}\in\mathcal{A}$, where $\mathcal{A}$ is a set of possible treatment assignment vectors. The test statistic $T(\boldsymbol{A}, \boldsymbol{D})$ is defined on all observed data, where $\boldsymbol{D} = (\boldsymbol{Y}, \boldsymbol{X}, \boldsymbol{S})$. We separate $\boldsymbol{A}$ and $\boldsymbol{D}$ because, under the sharp null hypothesis, only $\boldsymbol{A}$ varies during randomization while $\boldsymbol{D}$ remains fixed.
For example, $T(\boldsymbol{A},\boldsymbol{D})$ can be $|\hat{\tau}_{\text{CovAdj}}|$, 
$|\hat{\tau}_{\text{Borrow AIPW}}|$, $|\hat{\tau}_{\text{CSB NN}}|$, or any of the estimators introduced in Section \ref{sec:covshift} and Section \ref{sec:outcomeShift}. The FRT $p$-value is defined as $p^\text{FRT}=\mathbb{P}_{\boldsymbol{A}^*}\{T(\boldsymbol{A}^*,\boldsymbol{D})\ge T(\boldsymbol{A},\boldsymbol{D})\}$, where $\boldsymbol{A}^*\in\mathcal{A}$ shares the same distribution but is independent of $\boldsymbol{A}$.

\highlight{With Theorem \ref{th1}, under the sharp null, the randomization distribution of the test statistic is obtained from the actual randomization process within the RCT. Therefore, finite-sample validity of the FRT $p$-value for the sharp null does not require Assumption \ref{as2} or correct specification of nuisance models, provided that the entire data-adaptive analysis procedure, including EC selection, is repeated under each randomized assignment. This point should not be interpreted as making EC comparability irrelevant: approximate validity of Assumption \ref{as2} within the selected EC subset remains important for the efficiency, power, and causal interpretation of CSB-based effect-size estimates.}

\begin{theorem}\label{th1}
    Under $H_0$, we have $\mathbb{P}_{\boldsymbol{A}}(p^\text{FRT}\leq\alpha)\leq \alpha$ for all $\alpha\in (0,1)$, where $\mathbb{P}_{\boldsymbol{A}}$ denotes the probability taken over the distribution of ${\boldsymbol{A}}$. If we assume that $T({\boldsymbol{A}},\boldsymbol{D})$ varies with ${\boldsymbol{A}}\in \mathcal{A}$, we have $\mathbb{P}_{\boldsymbol{A}}(p^\text{FRT}\leq\alpha)=\lfloor \alpha |\mathcal{A}| \rfloor/|\mathcal{A}| > \alpha - 1/|\mathcal{A}|$, where $\lfloor x\rfloor$ is largest integer $\leq x$.
\end{theorem}

To calculate the FRT $p$-value in practice, we apply Monte Carlo to perform repeated shuffling based on the observed $A^\text{obs}\in\mathcal{A}$, which comes from the actual randomization in RCT. Therefore, for each observation $i\in \mathcal{R}$, treatment will be updated, while outcome $Y$ and covariates $X$ will not change. For $i\in \mathcal{E}$, they always have $A_i\equiv0$. During each shuffling of treatment assignment, observations are reassigned new treatment labels. Let $\boldsymbol{A}_b=\{A_{b1},\cdots, A_{bn}\}$ denote the new treatment assignment vector for $b$-th sampling. Therefore, for $b$-th sampling, a test statistic $T(\boldsymbol{A}_b,\boldsymbol{D})$ can be calculated based on the newly arranged treatment vector and the newly selected ECs when using CSB estimators. Estimating test statistics relies on the estimator with which the FRT is integrated. After repeated sampling for $B$ times, the estimated FRT $p$-value is 
\begin{align*}
    \hat{p}^{\text{FRT}}=\frac{\sum_{b=1}^B\mathbb{I}\{T(\boldsymbol{A}_b,\boldsymbol{D})\ge T(\boldsymbol{A}^\text{obs},\boldsymbol{D})\}+1}{B+1}.
\end{align*}

\highlight{For interpreting hypothesis testing under the sharp null (i.e., no individual treatment effect in the RCT), rejection of the sharp null in a one-sided test implies that at least some patients benefit from treatment. This can be particularly meaningful in precision medicine settings where treatment effects are heterogeneous across individuals and may motivate further investigation rather than simply concluding a zero average treatment effect.}

\highlight{Regarding which subgroups benefit from treatment, when subgroups are prespecified, practitioners can apply conditional FRT within each subgroup to test subgroup-specific effects \citep{li2026selective}. For non-prespecified subgroups, one may leverage selective randomization tests for subgroup discovery with valid inference \citep{gao2025selective}. Our framework can naturally be combined with those approaches while additionally borrowing ECs to improve power.}

\highlight{For treatment effect magnitude and clinically interpretable estimation, the proposed \texttt{CSB AIPW} estimator in Section~\ref{sec:outcomeShift} serves a complementary role in addition to FRT and directly targets estimands such as risk difference, risk ratio, and odds ratio. This provides interpretable effect size estimates together with asymptotic inference under double robustness conditions and selection consistency of the CSB procedure.}


\section{Simulation Studies}\label{sec:simstudy}
\subsection{Setup}\label{sec:simstudy_setp}
In the simulation study, we combine FRT with the proposed methods in Sections~\ref{sec:covshift} and~\ref{sec:outcomeShift} to demonstrate how FRT provides more robust inference in small HCTs under both scenarios, i.e., whether hidden bias exists or not. Four scenarios of model specification regarding sampling model (SM) and outcome model (OM) are considered. 

The data-generating processes have eight magnitudes of hidden bias, ranging from 0 to 14, and four model specification scenarios, including SM Correct OM Correct, SM Correct OM Wrong, SM Wrong OM Correct, and SM Wrong OM Wrong. The sample sizes are $\{\nrct,n_1,n_0,\nec\}=\{75,50,25,150\}$, where RCT sample is with $\nrct=75$ ($n_1=50$ and $n_0=25$) and EC sample is with $\nec=150$. The observed covariates $X=\{X_1, X_2, X_3\}$ is with $p=3$ dimension, where $X_1, X_2, X_3\sim U(-2,2)$. 
The sampling indicator $S\sim \text{Bernoulli}(\pi(X))$, where $\pi(X)=\{1+\exp(\eta_0+X^T\eta)\}^{-1}$, where $\eta=(2,2,2)$ and $\eta_0$ is used to adjust $P(S=1)$ such that $P(S=1)=\nrct/(\nec+\nrct)$. For RCT sample ($S=1$), we generate the treatment assignment $A$ by $A\sim \text{Bernoulli}(n_1/(n_1+n_0))$. For EC($S=0$), treatment assignment $A=0$. 

In terms of the outcome, for RCT sample ($S=1$), the potential outcomes are generated by $Y(a)\sim \text{Bernoulli}(\mu_a(X))$ with $\mu_a(X)=\{1+\exp(\beta_{a0}+X^T\beta_a)\}^{-1}$, where $a\in\{0,1\}$, $\{\beta_0, \beta_1\}=\{(1,1,1), (2,2,2)\}$, $\beta_{00}$ ensures the $P(Y(0)=1)=0.3$, and $\beta_{10}$ ensures the $P(Y(1)=1)=0.4$. For the EC sample ($S=0$), two scenarios are considered: (i) no hidden bias; (ii) the partial EC sample is biased. For scenario (i), the generating process of potential outcomes for EC follow the same way as RCT's. For scenario (ii), we randomly select $\rho = 50\%$ ECs to be biased. The potential outcomes for these selected ECs are generated by $Y(0)\sim \text{Bernoulli}(\mu_{00}(X))$ with $\mu_{00}(X)=\{1+\exp(\beta_{00}+X^T\beta_0-b/20)\}^{-1}$, where $b$ is the magnitude of hidden bias and $b=\{2,4,6,8,10,12,14\}$. The remaining $(1-\rho)$ ECs retain the same distribution of $Y(0)$ as the RCT sample. Under the alternative hypothesis, the observed outcome is given by $Y=A\cdot Y(1)+(1-A)\cdot Y(0)$, following Assumption \ref{as1}. Under the null hypothesis, the observed outcome is $Y=Y(0)$. When SM is misspecified, the covariates used to generate the sampling indicator $S$ are transformed as $X^*=e^X+10\cdot\sin{X}\cdot \cos{X}$, affecting $\pi(X)$. Similarly, when the OM is misspecified, the covariates used in generating potential outcomes are transformed into $X^*$, impacting $\mu_a(X)$. 

\highlight{To approximate FRT $p$-values, we use $B_{\mathrm{FRT}}=2000$ Monte Carlo randomizations for each simulated dataset. Each simulation scenario consists of 1000 Monte Carlo replicates. For the CSB approach, the data-adaptive threshold $\hat\gamma$ is selected by MSE minimization using $B_\gamma=200$ bootstrap resamples, with the candidate grid $\Gamma=\{0,0.05,\ldots,0.95,1\}$. Sensitivity analyses for $B_\gamma\in\{100,150,200\}$ are reported in Section~\ref{Supp:sim_sens_nsize} of the Supplementary Material. Two conformal score functions are considered: NN and LC-NN. For each conformal score function, the adaptive selection process selects the $\gamma$ value that minimizes estimated MSE. CV+ uses 10 folds for conformal $p$-values.}

Section~\ref{sec:simstudy_threeEstimands} presents simulation results for all three estimands: RD, RR, and OR, under both scenarios with and without hidden bias. Among the methods, \texttt{No Borrow CovAdj} and \texttt{Borrow AIPW} are representative examples of No Borrow and Borrow approaches, respectively, and two CSB methods are also included. In addition, Section~\ref{sec:simstudy_diffTrue} evaluates method performance across a range of true values for RD and RR. Section~\ref{Supp:sim_allBorrow} of the Supplementary Material provides comprehensive simulation results for all methods listed in Table~\ref{tab:Sum_Methods}, focusing on the RD estimand under various scenarios.

\subsection{Simulation results among three binary estimands}\label{sec:simstudy_threeEstimands}
In this section, we evaluate the performance of the proposed methods when targeting different estimands, as defined in Sections \ref{sec:framework_noBorrow} and \ref{sec:covshift}. In addition to the two proposed conformal scores, we consider the standardized absolute residual (SAR) as a conformal score within the CSB framework. Section~\ref{Supp:sim_SAR} of the Supplementary Material provides simulation results by adding SAR.

\subsubsection{Covariate incomparability}\label{sec:simstudy_threeEstimands_b0}

In this section, we compare FRT and asymptotic inference for No Borrow, Borrow, and Conformal Selective Borrow methods under no hidden bias, while covariate incomparability still exists. We are also interested in the point estimation performance of these methods. The primary focus is addressing covariate incomparability between the RCT and EC groups, as well as the impact of model misspecification.

\begin{figure}[!t]
    \centering
    \includegraphics[width=\linewidth]{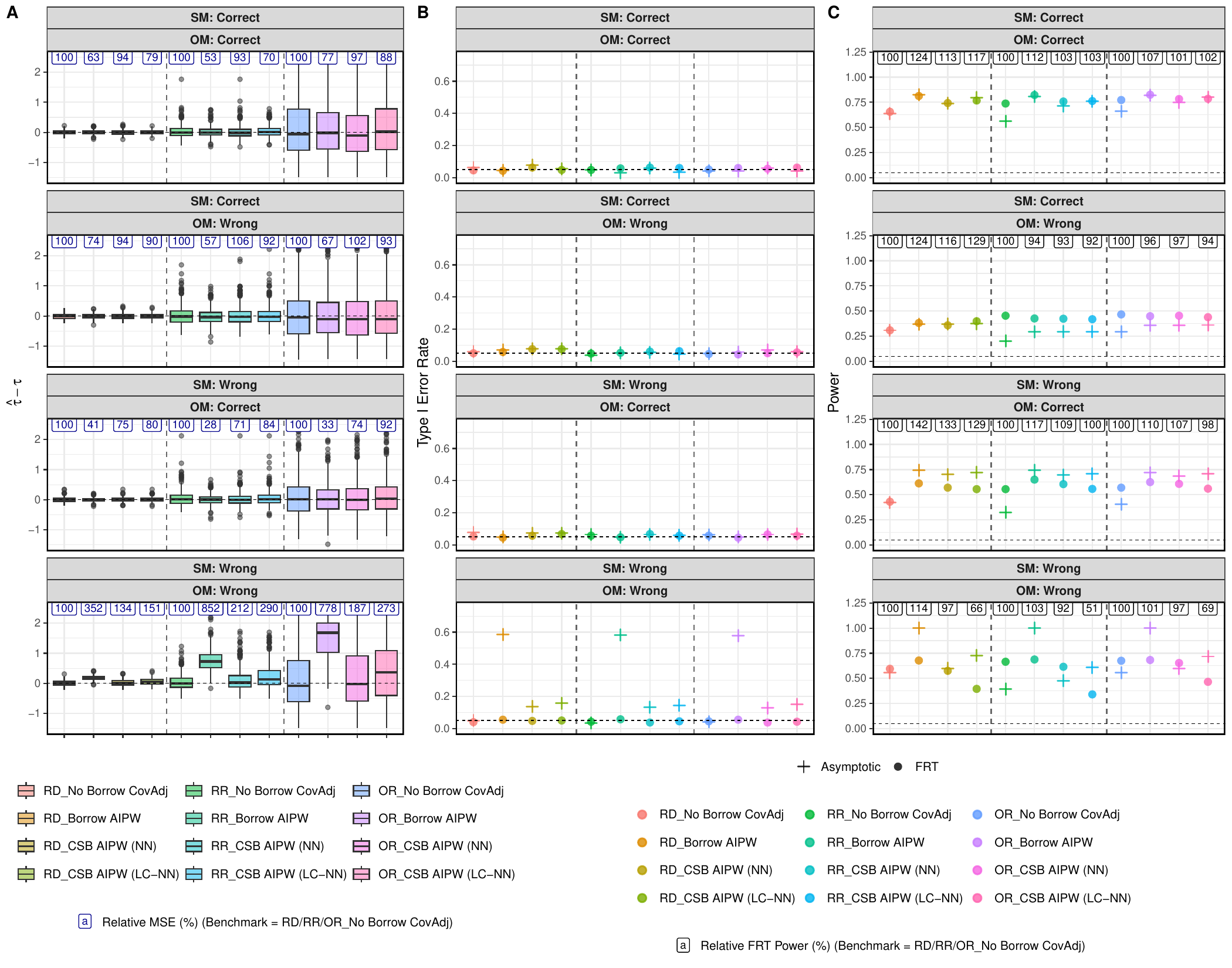}
    \vspace{-25pt}
    \caption{Simulation results for three different estimands ($b=0$)}
    \label{fig:sim_hb0_3estimands}
\end{figure}

Figure~\ref{fig:sim_hb0_3estimands} presents the estimation and inference results for RD, RR, and OR when using one No Borrow method (\texttt{No Borrow CovAdj}), one Borrow method (\texttt{Borrow AIPW}), and two Conformal Selective Borrow methods (\texttt{CSB NN} and \texttt{CSB LC-NN}). When $b=0$, all methods yield unbiased estimates across all three estimands, provided that at least one of the nuisance models is correctly specified. Among them, \texttt{CSB NN} shows greater robustness to misspecification of both models and remains valid for all estimands. In general, RD and RR exhibit lower variances due to different scales. Using \texttt{No Borrow CovAdj} as a benchmark within each estimand group, \texttt{Borrow AIPW} generally achieves lower MSE when at least one model is correctly specified, as no hidden bias exists in the EC. However, its performance deteriorates when both models are misspecified.

Figure \ref{fig:sim_hb0_3estimands} (B) shows that asymptotic inference fails to control the type I error rate when both nuisance models are misspecified, whereas FRT consistently maintains control across all methods and estimands, even under dual misspecification. Figure \ref{fig:sim_hb0_3estimands} (C) uses \texttt{No Borrow CovAdj} with RD as the benchmark. Their power is comparable since FRT targets the same sharp null across RD, RR, and OR. RR and OR generally yield greater power than RD across all scenarios. Without hidden bias, \texttt{Borrow AIPW} achieves the highest power derived from FRT, while \texttt{CSB NN} also performs relatively well even under model misspecification. Since asymptotic inference fails to control the type I error rate, the corresponding power estimates are invalid and should be interpreted cautiously.

\subsubsection{Outcome incomparability}\label{sec:simstudy_threeEstimands_b6}
\begin{figure}[ht]
    \centering
    \includegraphics[width=\linewidth]{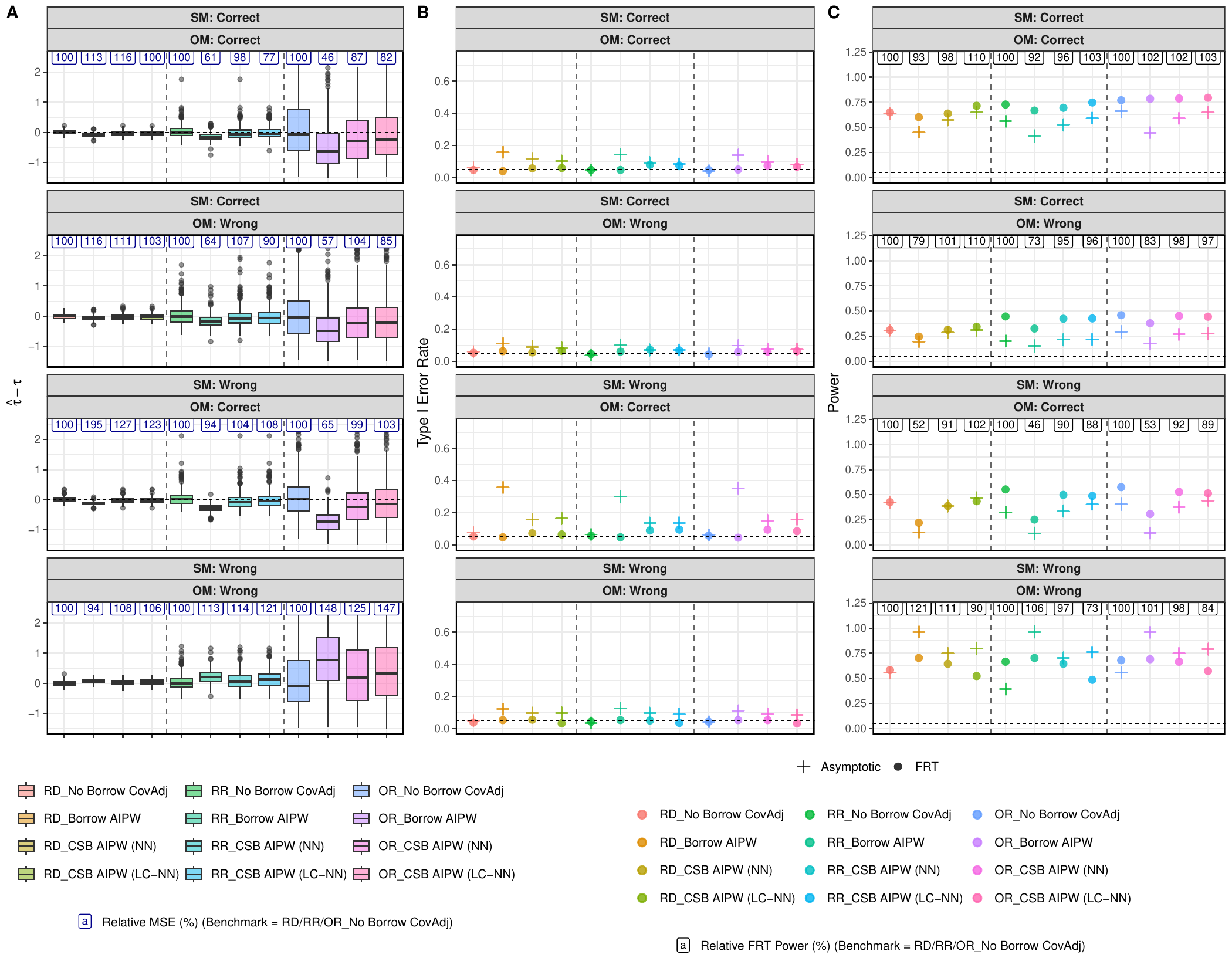}
    \vspace{-25pt}
    \caption{Simulation results for three different estimands ($b=6$)}
    \label{fig:sim_hb6_3estimands}
\end{figure}

When outcome incomparability exists, i.e., hidden bias is present ($b=6$), as indicated in Figure \ref{fig:sim_hb6_3estimands}, \texttt{Borrow AIPW} produces biased estimates across all three estimands, even when both models are correctly specified. By contrast, CSB methods maintain bias around zero, particularly for RD and RR, where the bias magnitudes are notably smaller than for OR. Regarding MSE, \texttt{CSB LC-NN} achieve comparable MSE to \texttt{No Borrow CovAdj} in most settings.

Asymptotic inference results in inflated type I error rates across all scenarios and estimands, while FRT consistently controls the type I error rate, even under large hidden bias. RR and OR continue to have higher FRT power than RD. However, unlike in the no hidden bias setting, \texttt{Borrow AIPW} no longer provides power gains over \texttt{No Borrow}. \texttt{CSB LC-NN} demonstrates robustness to both hidden bias and model misspecification, maintaining stable power gains in most settings.

\subsection{Simulation results across varying magnitudes of hidden bias}\label{sec:sim_mag_hb}
\begin{figure}[!t]
    \centering
    \includegraphics[width=\linewidth]{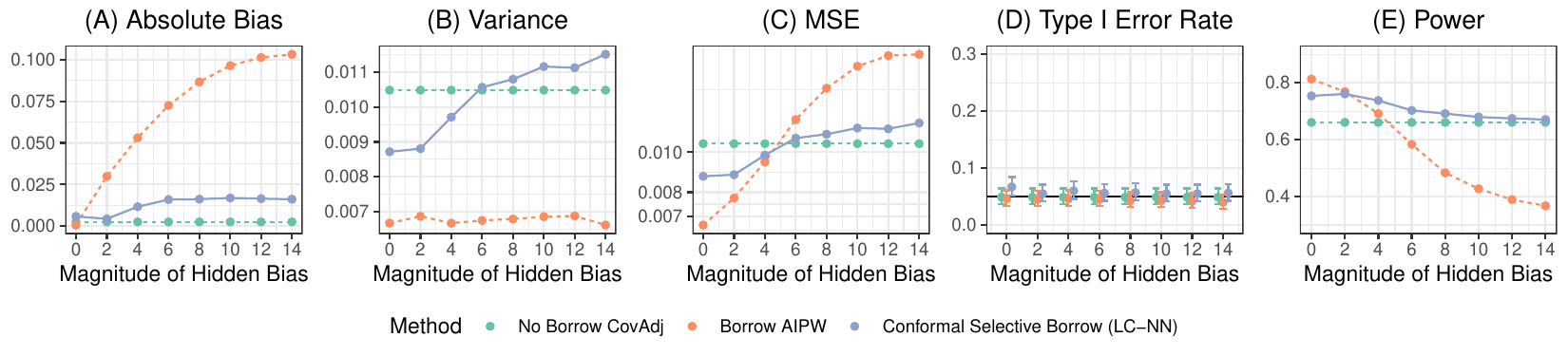}
    \vspace{-25pt}
    \caption{Simulation results across different magnitudes of hidden bias}
    \label{fig:sim_HB_c1_mt}
\end{figure}

In this section, we examine a range of hidden bias magnitudes ranging from 0 to 14, with results for RD are presented in Figure \ref{fig:sim_HB_c1_mt}. FRTs with all test statistics effectively control the type I error rate under varying degrees of hidden bias.
When no hidden bias is present, \texttt{Borrow AIPW} achieves the highest power, but its power drops sharply and becomes lower than that of \texttt{No Borrow CovAdj} as hidden bias increases. In contrast, \texttt{CSB LC-NN} consistently achieves higher power than \texttt{No Borrow CovAdj}, with up to a 15\% improvement when $b=0$.
Moreover, the CSB method retains absolute bias below 0.025, whereas \texttt{Borrow AIPW} shows bias exceeding 0.1.
These results suggest that the CSB method is more robust to hidden bias in EC data. 
Additional simulation results, including those for \texttt{CSB NN} and other scenarios, are provided in Section~\ref{Supp:sim_seqHiddenBias} of the Supplementary Material. We found that \texttt{CSB NN} may lose power when hidden bias is difficult to detect, partially aligning with the conclusion of no uniform power gain in \citet{kopp2020power}, but it can improve power when there is no hidden bias or when bias is detectable. Importantly, FRT effectively controls the type I error rate regardless of bias detectability. Moreover, \texttt{CSB NN} demonstrates greater robustness in EC selection and power gain when both models are misspecified.

\subsection{Simulation results across varying true estimands}\label{sec:simstudy_diffTrue}
This section examines a sequence of true estimand values to explore the proposed approaches' performance under varying conditions. Both scenarios, with and without hidden bias, are considered. For simplicity, we focus on the FRT power under the first model specification scenario. 

\begin{figure}[ht]
    \centering
    \includegraphics[width=\linewidth]{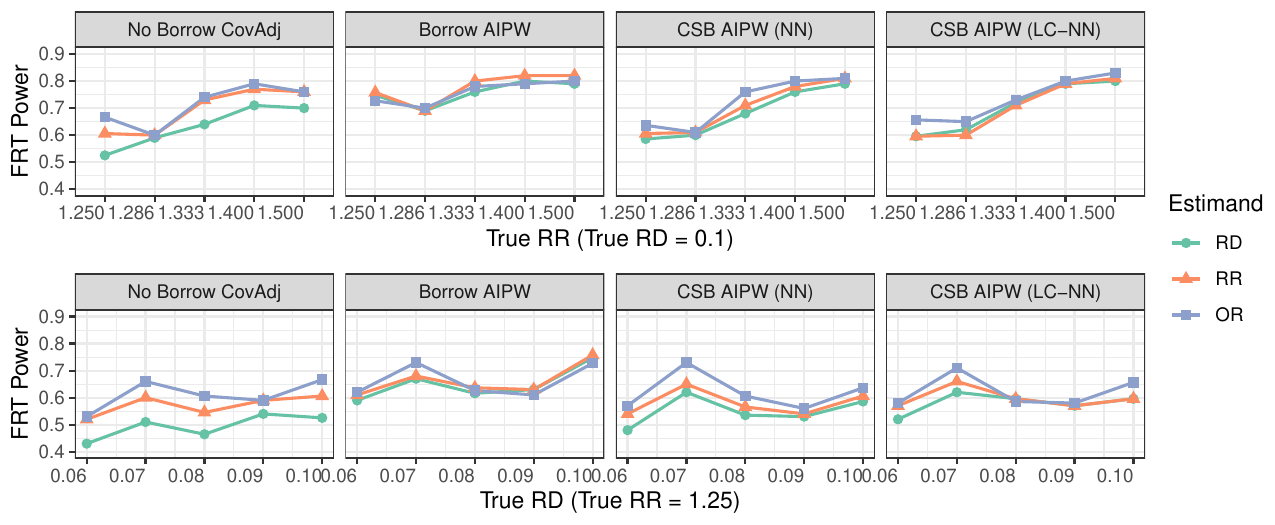}
    \vspace{-10pt}
    \caption{Power curves across different $\tau_\text{RD}$ and $\tau_\text{RR}$ ($b=0$)}
    \label{fig:powercurve_hb0_main}
\end{figure}

When no hidden bias is present ($b=0$) and true RD is fixed, power increases as $\tau_\text{RR}$ increases (Figure \ref{fig:powercurve_hb0_main}). Conversely, when true RR is fixed, power remains relatively stable with slight fluctuations as $\tau_\text{RD}$ increases. RR and OR estimands achieve greater FRT power than RD, especially under \texttt{No Borrow CovAdj}. With Borrow and CSB methods, power is more consistent across estimands.
When hidden bias exists ($b=6$), a similar increasing trend in power is observed as RR increases in the power curve plot provided in Section~\ref{Supp:sim_trueEstimand} of the Supplementary Material. However, unlike the flat trend in the bias-free setting, FRT power rises with the true RD. In addition, differences in FRT power across estimands become more noticeable. Even under Borrow and CSB methods, the choice of estimand significantly affects power, distinguishing from the no hidden bias case, where power levels are more uniform. \highlight{Additional sensitivity analyses, including RCT sample size and matching ratio, computational burden, higher covariate dimensionality, and the weak-null scenario, are provided in Sections~\ref{Supp:sim_sens_nsize}, \ref{Supp:sim_compu_burden}, \ref{Supp:sim_HD_p10}, and~\ref{Supp:sim_weak_null} of the Supplementary Material.}

\section{Application to the CALGB 9633 Trial with External Control from NCDB}\label{sec:casestudy} 

To assess the performance of FRT in controlling type I error rate in practice, we apply it to the estimators in Table \ref{tab:Sum_Methods} using real datasets introduced in Section \ref{sec:data}, with data preparation detailed in Section \ref{sec:dataPre}. The primary data analysis is based on the CALGB 9633, using all RCT observations ($\nrct=335$) and evaluating all three estimands. A representative example with 1:1 matching ($\nec=\nrct$) is provided in Section \ref{sec:caseRes1}. We further explore how $p$-values change under other common practical scenarios, including allocation ratios, RCT sample size (small and moderate), and EC sizes. For simplicity, this exploration focuses on RD, with results shown in Section~\ref{sec:caseRes2} of the Supplementary Material. 

\subsection{Data Preparation}\label{sec:dataPre}
CALGB 9633 and NCDB share five common pre-treatment covariates: gender, age, race, histology, and tumor size. These covariates capture key baseline characteristics and serve as the basis for addressing covariate incomparability in the analysis. Missing tumor size values in CALGB 9633 are imputed using the median of observed values. 

\begin{figure}[!t]
\centering
\includegraphics[width=0.9\textwidth]{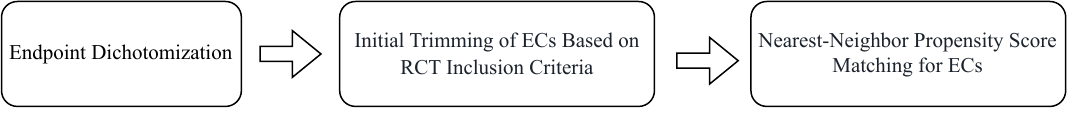}
\vspace{-10pt}
\caption{Three-stage data pre-processing}
\label{fig:flowchart}
\end{figure}

Figure \ref{fig:flowchart} outlines the three-stage data pre-processing: endpoint dichotomization, initial selection, and nearest-neighbor matching. The primary endpoint is time-to-event outcome and is dichotomized at 3 years following \cite{Strauss2008}, where a success indicator is coded as 1 if survival time exceeds 3 years and 0 otherwise. After this transformation, the conclusions from \texttt{No Borrow CovAdj} method for both overall survival and the subgroup with tumor size $>4$ cm remain consistent with original findings \citep{Strauss2008}. Consequently, for the RD estimand, the point estimate is interpreted as the difference in 3-year survival probabilities between the treatment and control groups.

\begin{figure}[t]
\includegraphics[width=\textwidth]{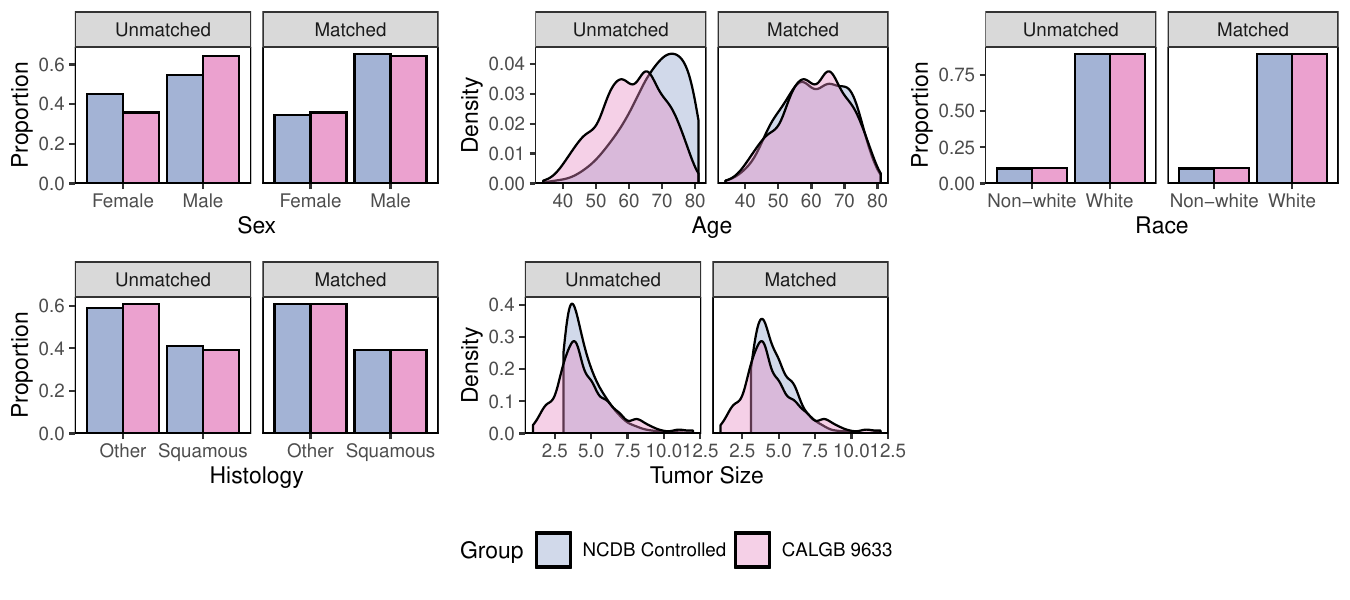}
\vspace{-20pt}
\caption{Covariate distribution before and after matching when $\nrct=335$ and $\nec=335$}
\label{fig:match}
\end{figure}

After dichotomization, we perform initial trimming based on the RCT’s inclusion/exclusion criteria and assess covariate balance between CALGB 9633 (RCT) and NCDB controls (the source of external controls). Primary disparities are observed in tumor size and age, where NCDB includes more elderly patients and has a wider age distribution, along with larger average tumor sizes (Figure~\ref{fig:match}). To improve comparability, NCDB controls with age or tumor size outside the CALGB 9633 range are excluded, aligning with the eligibility criteria of RCT.

To further enhance comparability across the covariates, we apply nearest-neighbor matching, which is commonly used in causal estimation \citep{stuart2008matching,qian2025matching,Qiu2025}. The distance between neighbors is measured using the \textit{Euclidean} distance computed from the covariates between EC and RCT subjects. \highlight{Before matching and conformal score calculation, continuous covariates were standardized and categorical covariates were coded as indicator variables so that the Euclidean distance was not dominated by variables with larger numerical scales.} For example, keeping all CALGB 9633 participants ($\nrct=335$) in a 1:1 matching design, we select $\nec=335$ NCDB subjects as EC. Figure \ref{fig:match} demonstrates that matching improves covariate balance, although some imbalances remain, which are addressed by methods introduced in Section \ref{sec:covshift} other than \texttt{Borrow Naive}. Baseline summary table in Section~\ref{Supp:case_Table1s} of the Supplementary Material confirms that the covariates are well balanced across the three arms.

Determining how much external information to borrow is critical in HCTs \citep{FDA2023}. Borrowing more EC data can increase bias, while too little may minimize efficiency gains. Given the limited guidance on EC sample size determination, we explore how $p$-values change under different matching ratios in Section~\ref{sec:caseRes2} of the Supplementary Material. 

\subsection{Primary Analysis Results}\label{sec:caseRes1}
In this section, we will discuss the result of primary analysis, which explores the entire RCT dataset (CALGB 9633) and thus corresponds to the moderate HCT with equal allocation ratio. Although we are motivated by type I error rate issues arising from small HCTs, the asymptotic inference also depends on the nuisance model specifications. Even for approaches with double robustness properties, such as AIPW and ACW, their validity still requires at least one of the nuisance models is correctly specified. However, this requirement is not guaranteed in real practice. Therefore, it is meaningful to investigate whether FRT can also benefit moderate HCTs.

\begin{figure}[t]
\includegraphics[width=\textwidth]{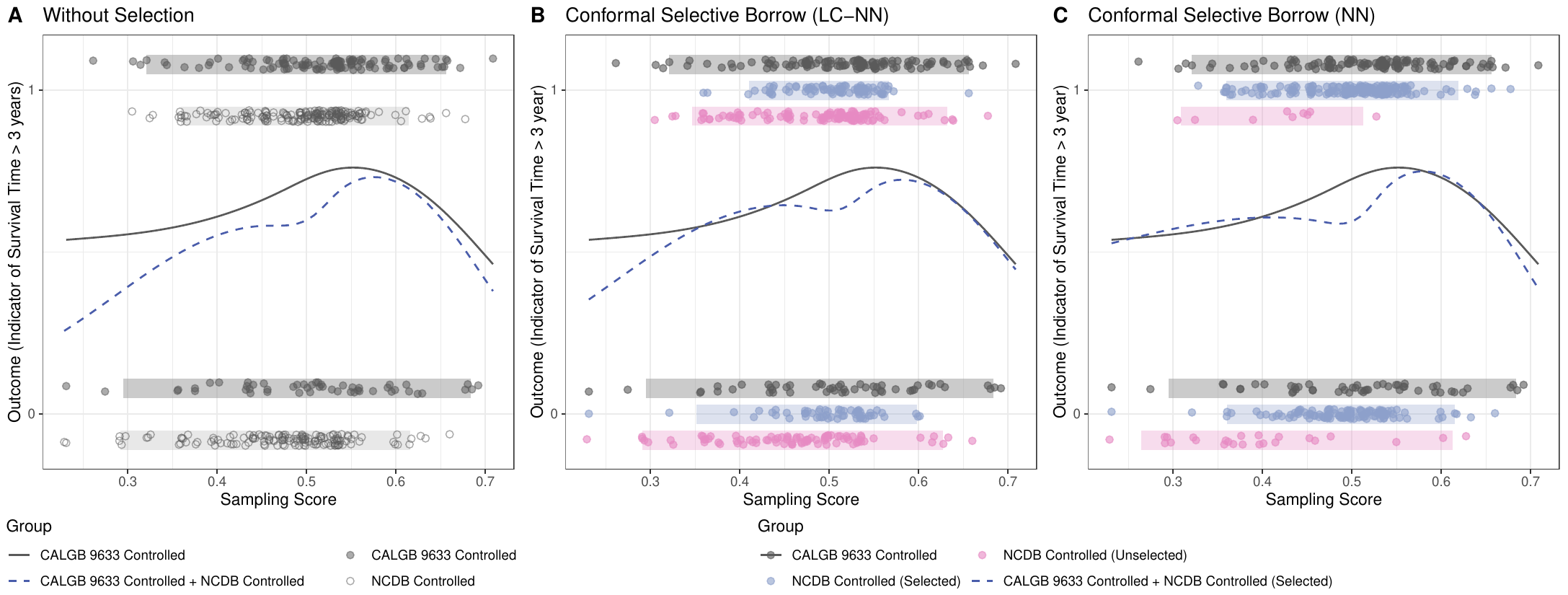}
\vspace{-10pt}
\caption{Sampling Score Distribution ($n_1=167, n_0=168, \nec=335$)}
\label{fig:caseHBall}
\end{figure}

In Figure \ref{fig:caseHBall} (A), some NCDB control observations fall outside the 95\% quantile range of the estimated sampling scores for CALGB 9633 controls. These outliers indicate potential hidden bias, further supported by the separation between the smoothed sampling-score curves for the NCDB and CALGB 9633 controls. Including all EC subjects in the hybrid control arm lowers the average probability of survival beyond three years, potentially overestimating the treatment effect. In contrast, as shown in Figure \ref{fig:caseHBall} (B-C), the smooth curves under \texttt{CSB LC-NN} and \texttt{CSB NN} more closely align with RCT control curve. The distribution of selected ECs is more concentrated and strictly follows RCT control distribution. Compared to \texttt{CSB LC-NN}, \texttt{CSB NN} is less strict with the selection and retains more EC subjects, while \texttt{CSB LC-NN} prioritizes those within the intersection of intervals for both NCDB and CALGB 9633 controls.

\highlight{As an additional diagnostic assessment of outcome comparability, we compared a reduced covariate-adjusted logistic regression model $Y\sim X$ with a full model $Y\sim X+S+X\times S$ among subjects with $A=0$. The full EC set showed evidence against outcome comparability with the RCT controls ($p=0.017$), whereas the CSB-selected subset yielded a larger $p$-value ($p=0.158$), suggesting improved comparability after selective borrowing. This diagnostic should not be interpreted as a formal post-selection test of Assumption~\ref{as2}, but it provides empirical support that the selected subset is more compatible with the RCT controls than the full EC sample.}

For each approach, we use FRT to obtain robust $p$-values and compare them with asymptotic inference across target estimands. In Table~\ref{tab:resC1_rd}, all six Borrow methods yield asymptotic $p$-values below 0.001, whereas FRT gives more conservative values around 0.05. A similar pattern appears for CSB methods: the asymptotic $p$-values decrease sharply, while FRT $p$-values decrease more conservatively. The \texttt{CSB NN} and \texttt{CSB LC-NN} estimates lie between No Borrow and Borrow methods. Since no ground truth exists in this case study, the goal is not simply to obtain smaller or significant $p$-values. Instead, the results suggest that, even under strict type I error control, methods with FRT $p$-values below 0.05 support that chemotherapy significantly improves the probability of surviving beyond 3 years compared with observation in CALGB 9633.

\highlight{In this case study, \texttt{CSB NN} selects 302 EC subjects while \texttt{CSB LC-NN} selects 144, although LC-NN does not always borrow less than NN in general. Since the optimal amount of borrowing depends on the unknown ground truth, it is difficult to determine which selection is objectively better. However, based on simulation studies, we generally prefer LC-NN because conditioning on the outcome label provides a more reliable assessment of comparability under hidden bias. When one outcome class is very rare, however, LC-NN may become less stable due to an insufficient calibration set, in which case the simpler NN score may be preferable. Importantly, despite the substantial difference in selected sample size, the two methods produce very similar treatment effect estimates in this application (0.130 vs.\ 0.134), suggesting that the proposed framework is relatively robust to the choice of conformal score in this example.}

\highlight{Additionally, we note that the computational burden of CSB with FRT was acceptable in the motivating application. The runtimes in Table~\ref{tab:resC1_rd} are obtained from the full analysis pipeline, including conformal selective borrowing and FRT procedure for valid inference, which are approximately 57.5 seconds for \texttt{CSB NN} and 64.5 seconds for \texttt{CSB LC-NN}. This suggests that, despite the need to re-run selection within each permutation to account for selection uncertainty, the proposed procedure is feasible for many moderate-sized HCT applications.} 

\begin{table}[!t]
\caption{Results of case study (RD, $n_1=167, n_0=168, \nec=335$)}
\label{tab:resC1_rd}
\centering
\resizebox{\linewidth}{!}{
\renewcommand{\arraystretch}{1}
\begin{tabular}{llllrrrrr}
\hline
 & \multicolumn{1}{c}{} & \multicolumn{3}{c}{Asymptotic Inference} & \multicolumn{1}{c}{FRT} &  &  \\ \cline{3-5}
Method & Point Est. & SE & 95\% CI & $p$-value & $p$-value & Num. of EC & ESS of EC & FRT Runtime (s)\\
\hline
No Borrow Unadj & 0.076 & 0.048 & (-0.018, 0.169) & 0.110 & 0.120 & 0 & 0 & 0.001 \\
No Borrow CovAdj & 0.081 & 0.049 & (-0.015, 0.178) & 0.097 & 0.096 & 0 & 0 & 22.026\\
Conformal Selective Borrow NN & 0.134 & 0.040 & (0.058, 0.212) & $<$0.001 & 0.062 & 302 & 294 & 57.478\\
Conformal Selective Borrow LC-NN & 0.130 & 0.043 & (0.046, 0.215) & 0.002 & 0.036 & 144 & 138 & 64.492\\
Borrow Na\"ive  & 0.151 & 0.039 & (0.075, 0.227) & $<$0.001 & 0.056 & 335 & 335 & 36.916\\
Borrow IPW  & 0.143 & 0.039 & (0.066, 0.220) & $<$0.001 & 0.058 & 335 & 315 & 18.059\\
Borrow CW  & 0.142 & 0.039 & (0.067, 0.218) & $<$0.001 & 0.055 & 335 & 313 & 23.003 \\
Borrow OM  & 0.148 & 0.038 & (0.073, 0.223) & $<$0.001 & 0.044 & 335 & 335 & 29.474\\
Borrow AIPW  & 0.148 & 0.039 & (0.071, 0.225) & $<$0.001 & 0.047 & 335 & 315 & 37.312 \\
Borrow ACW  & 0.145 & 0.039 & (0.068, 0.222) & $<$0.001 & 0.046 & 335 & 313 & 44.060\\
\hline
\end{tabular}
}
\end{table}
These findings are consistent across RD, RR, and OR, as shown in Section~\ref{Supp:case_summaryTables_prim} of the Supplementary Material, indicating that the choice of estimand does not substantially impact the conclusions.
Supplementary analyses under varying EC sizes, allocation ratios, and small-sample RCT settings are presented in Section~\ref{sec:caseRes2} of the Supplementary Material. These results suggest that power gains depend on EC quality, and FRT with CSB helps prevent over-reliance on biased external data.

\section{Discussion}\label{sec:conclusion}

In this paper, we propose (i) doubly robust borrowing estimators for three estimands in HCTs with binary outcomes to address covariate incomparability of ECs; (ii) CSB methods using nearest-neighbor-based conformal scores to address binary outcome incomparability of ECs; and (iii) randomization inference to control the type I error rate while enhancing power when combined with the proposed methods that address both covariate and outcome incomparability. 

We evaluated the finite-sample performance of the proposed approach through extensive simulation studies, demonstrating its robustness in both estimation and inference. CSB methods can adaptively select comparable ECs even when some exhibit hidden bias, outperforming full-borrowing approaches.
Partly echoing the conclusion in \citet{kopp2020power}, we observe that the power gain from EC borrowing is not uniform; in fact, power can be compromised when hidden bias is complex or hard to detect. This emphasizes the need to identify high-quality ECs to ensure power improvement. \highlight{Nonetheless, FRT consistently protects the sharp-null type I error under the randomized experiment, even when EC quality affects the efficiency and interpretability of the CSB-based test statistic.} 
Moreover, the result of \citet{kopp2020power} relies on settings in which UMP tests exist under parametric assumptions, which we do not assume here.
We applied our method, along with alternative borrowing estimators, to the CALGB 9633 trial with ECs from the NCDB, improving upon the original underpowered analysis while mitigating bias from EC borrowing.

\textbf{Alternative target population and outcome.}
We consider the RCT population ($S=1$) as the target population in this work. Future work may explore alternative targets using weighting methods \citep{lee2023improving}, including the external control population ($S=0$), the pooled population ($S=0$ and $S=1$), and the overlapping population \citep{wang2025integrating}.
Beyond binary outcomes, HCTs with survival outcomes are of great interest \citep{kwiatkowski2024case}. \citet{gao2024survival} propose modeling bias using a DR-learner and penalizing the estimated bias to guide selective borrowing for survival outcomes. A promising direction for future work is to apply conformalized survival analysis \citep{candes2023conformalized} to test the individual exchangeability of ECs.

\textbf{Extension of FRT.} By using studentized test statistics, FRT can also be valid for common weak null hypotheses, such as the average treatment effect being zero, though this validity holds only in the asymptotic sense \citep{wu2021randomization}. \citet{caughey2023} show that FRTs can also be valid under bounded nulls, where individual treatment effects are all non-positive (or all non-negative). \citet{ding2016randomization} use FRT to test treatment effect heterogeneity by taking the maximum p-value over a confidence set of nuisance parameters \citep{berger1994p}. Extending these approaches to the HCT setting remains an important direction for future work. 

\highlight{\textbf{Multiple external databases.} From a statistical implementation perspective, since CSB evaluates EC comparability at the \textit{individual level}, the framework can be naturally extended to settings with multiple external sources by assessing each EC subject separately regardless of its originating database.
From a practical data perspective, consistent with FDA guidance emphasizing that external control data should be ``\textit{fit for purpose}'' and comparable to the trial population \citep{FDA2023}, the selected database should align closely with the RCT in terms of study population, baseline covariates, and outcome definitions. Databases with richer and more compatible covariate information are particularly important for CSB, since the proposed framework relies on covariates to assess comparability and guide borrowing decisions. Outcome definitions should also be aligned across the external database and the RCT, as differences in how outcomes are defined, measured, or timed may introduce substantial bias and limit interpretability. In addition, temporal alignment, data quality, data completeness, and pre-specified database selection are all important considerations.}

\highlight{\textbf{Computational cost.} Because the conformal selection procedure evaluates EC subjects individually and is repeated within each FRT permutation, the computational cost increases approximately linearly with both the number of EC subjects and the number of permutations. However, these computations are naturally parallelizable. Our \texttt{intFRT} R package includes \textit{parallel implementations} to substantially improve computational efficiency in practice.
Importantly, the HCT setting is typically motivated by scenarios where the RCT sample size is limited and additional external information is needed, while at the same time the EC sample should not overwhelm or dominate the randomized evidence from the RCT. Therefore, when the initial EC dataset is large, we recommend first applying propensity score matching to identify an EC subset that is more comparable to the RCT in terms of covariate distributions before applying CSB. As a byproduct, this step also reduces the computational burden of the subsequent CSB and FRT procedures. Finally, from an application perspective, the CALGB 9633 trial in our case study required approximately 12 years because of slow accrual. Reliable EC borrowing methods may therefore help accelerate drug development, making the additional computational cost practically worthwhile.}

\highlight{\textbf{Trade-off between practical gains and methodological complexity.} When the RCT is already sufficiently large, accrual is timely, and enrolling concurrent controls is feasible, a standard RCT-only analysis may be preferable because of its simplicity. However, in settings such as rare diseases, oncology studies with slow accrual, or situations where reducing enrollment to the control arm is ethically or practically desirable, an RCT-only analysis may be underpowered or difficult to implement. In such settings, carefully borrowing external controls can provide meaningful practical gains by improving efficiency and potentially accelerating drug development. The goal of proposed methods is therefore not to replace simple RCT-only analyses in all settings, but to provide a principled option for settings where the practical value of borrowing justifies the added methodological complexity. To facilitate implementation, we also provide the R package \texttt{intFRT} with user-friendly functions, parallel computation support, and detailed vignettes.}

\section*{Acknowledgments}
The authors thank the Associate Editor and two anonymous reviewers for their constructive comments and valuable suggestions.

\section*{Funding Statement}
This project is supported by the Food and Drug Administration (FDA) of the U.S. Department of Health and Human Services (HHS) as part of a financial assistance award U01FD007934, \$2,556,429 over three years, funded by FDA/HHS. It is also supported by the National Institute on Aging of the National Institutes of Health under Award Number R01AG06688, totaling \$1,565,763 over four years. The contents are those of the authors and do not necessarily represent the official views of, nor an endorsement by, the FDA/HHS, the National Institutes of Health, or the U.S. Government.

\section*{Data Availability Statement}
Code used to reproduce the simulation studies and data analyses reported in the Supplementary Material. The NCDB data may be made available upon reasonable request, whereas the CALGB 9633 data cannot be shared due to restricted access.

\bibliographystyle{agsm}
\bibliography{ref}

\appendix

\setcounter{equation}{0}
\renewcommand{\theequation}{S\arabic{equation}}
\setcounter{table}{0}
\renewcommand{\thetable}{S\arabic{table}}
\setcounter{figure}{0}
\renewcommand{\thefigure}{S\arabic{figure}}
\setcounter{theorem}{0}
\renewcommand{\thetheorem}{S\arabic{theorem}}
\setcounter{lemma}{0}
\renewcommand{\thelemma}{S\arabic{lemma}}
\setcounter{remark}{0}
\renewcommand{\theremark}{S\arabic{remark}}

\newpage
\begin{center}
{\large\bf SUPPLEMENTARY MATERIAL}
\end{center}

\newtheorem{definition}{Definition}
\newtheorem{condition}{Condition}

\newcounter{rmk}
\newcommand\rmk[1]{\vspace*{1mm} \par \stepcounter{rmk}{\noindent \bf Remark \thermk}. {#1}\vspace*{1mm}}

Section~\ref{Supp:formula_estimator} provides details about how to construct estimators for risk difference (RD), risk ratio (RR), and odds ratio (OR), including both \texttt{No Borrow CovAdj} and \texttt{Borrow AIPW} as examples. This way of constructing estimators for three estimands can be generalized to the alternative estimators provided in Section~\ref{Supp:alternative_estimator}. Section \ref{Supp:alg} provides the algorithm of adaptive selection of threshold $\gamma$. Section~\ref{Supp:sim} focuses on the simulation study. Specifically, Section~\ref{Supp:sim_allBorrow} presents the simulation results for all Borrow and No Borrow methods under both the presence and absence of hidden bias, with RD as the estimand of interest. Section~\ref{Supp:sim_SAR} includes results using SAR as the conformal score, a commonly used approach for continuous outcomes. Section~\ref{Supp:sim_seqHiddenBias} further provides simulation figures under varying levels of hidden bias across alternative scenarios. In Section~\ref{Supp:sim_trueEstimand}, we present power curves for varying true estimands under a fixed hidden bias magnitude ($b=6$). Section~\ref{Supp:sim_senAna} consists of several sensitivity analyses to provide additional discussion. Section~\ref{Supp:sim_compu_burden} discusses the computational burden under different settings. Section~\ref{Supp:sim_HD_p10} presents simulation results under higher dimensional covariates.
Section~\ref{Supp:case_Table1s} provides the baseline covariate summary after preprocessing. Additional case study analyses under varying RCT sample sizes, EC sizes, and allocation ratios are presented in Section~\ref{sec:caseRes2}, with primary and supplementary result tables in Sections~\ref{Supp:case_summaryTables_prim} and~\ref{Supp:case_summaryTables}. Further analyses of Full Borrow and CSB methods and diagnostic assessments are provided in Sections~\ref{Supp:case_sizeECchange} and~\ref{Supp:case_diag_outcome}, respectively.


\section{Semiparametric Efficient Estimators and Asymptotic Inference}\label{Supp:formula_estimator}
In this section, we provide detailed formulations of the estimators corresponding to the estimation approaches discussed in the main text, using the consistent notation as defined in main text.

\subsection{RCT-only Analysis}\label{Supp:formula_estimator_RCTonly}

Let $z_{1-\alpha/2}$ denote the lower $1-\alpha/2$ quantile of standard normal distribution.

\begin{example}[Risk Difference, No Borrow CovAdj]
\label{ex:rd}
The RCT-only plug-in estimator for $\taurd$ is 
$$
\htaurdR=\hthetaR{1}-\hthetaR{0}.
$$
The RCT-only EIF of $\taurd$ is 
\begin{align*}
\ifrct(\taurd)&=\ifrct(\theta_1-\theta_0)=\ifrct(\theta_1)-\ifrct(\theta_0)\\
&=\frac{S}{\prct}\big\{\xi_1(O)-\xi_0(O)- \taurd\big\}.    
\end{align*}
The variance estimator for $\sqrt{n} \htaurdR$ is
\begin{align*}
\hat{V}_{\rm RD,\mathcal{R}}&=\frac1n\sum_{i=1}^n\bigg[\frac{S_i}{\prct}\big\{\hat\xi_1(O_i)-\hat\xi_0(O_i)- \htaurdR\big\}\bigg]^2\\
&=\frac{1}{n\prct^2}\sum_{i:S_i=1}\big\{\hat\xi_1(O_i)-\hat\xi_0(O_i)- \htaurdR\big\}^2.
\end{align*}
The asymptotic confidence interval is
$$
\bigg[
\htaurdR-z_{1-\alpha/2}\sqrt{\hat{V}_{\rm RD,\mathcal{R}}/n},\;
\htaurdR+z_{1-\alpha/2}\sqrt{\hat{V}_{\rm RD,\mathcal{R}}/n}
\bigg],
$$
For a sanity check, we see that the variance estimator for $\htaurdR$ is
\begin{align*}
\hat{V}_{\rm RD,\mathcal{R}}/n
&=\frac{1}{n^2\prct^2}\sum_{i:S_i=1}\big\{\hat\xi_1(O_i)-\hat\xi_0(O_i)- \htaurdR\big\}^2\\
&=\frac{1}{\nrct^2}\sum_{i:S_i=1}\big\{\hat\xi_1(O_i)-\hat\xi_0(O_i)- \htaurdR\big\}^2.
\end{align*}
\end{example}

\begin{example}[Risk Ratio, No Borrow CovAdj]
\label{ex:rr}
The RCT-only plug-in estimator for $\taurr$ is 
$$
\htaurrR=\hthetaR{1}/\hthetaR{0}.
$$
The RCT-only EIF of $\taurr$ is 
\begin{align*}
\ifrct(\taurr)&=\ifrct\Bigg(\frac{\theta_1}{\theta_0}\Bigg)=\frac{\ifrct(\theta_1)}{\theta_0}-\frac{\ifrct(\theta_0)}{\theta_0}\Bigg(\frac{\theta_1}{\theta_0}\Bigg)\\
&=\frac{S}{\prct}\frac{1}{\theta_0}\big\{\xi_1(O)-\xi_0(O)\taurr\big\}.
\end{align*}
The variance estimator for $\sqrt{n} \htaurrR$ is
\begin{align*}
\hat{V}_{\rm RR,\mathcal{R}}
&=\frac1n\sum_{i=1}^n
\bigg[
\frac{S_i}{\prct}\frac{1}{\hthetaR{0}}\big\{\hat\xi_1(O_i)-\hat\xi_0(O_i)\htaurrR\big\}
\bigg]^2\\
&=\frac{1}{n\prct^2}\sum_{i:S_i=1}\bigg[
\frac{1}{\hthetaR{0}}\big\{\hat\xi_1(O_i)-\hat\xi_0(O_i)\htaurrR\big\}
\bigg]^2.
\end{align*}
The asymptotic confidence interval is
\begin{align*}
    &\exp\bigg[
\log(\htaurrR)-z_{1-\alpha/2}\sqrt{\hat{V}_{\rm log(RR),\mathcal{R}}/n},\;
\log(\htaurrR)+z_{1-\alpha/2}\sqrt{\hat{V}_{\rm log(RR),\mathcal{R}}/n}
\bigg]\\
\approx&{\bigg[
\htaurrR\cdot \exp\left(-z_{1-\alpha/2}\cdot\frac{\sqrt{\hat{V}_{\rm RR,\mathcal{R}}/n}}{\htaurrR}\right),\;
\htaurrR\cdot \exp\left(z_{1-\alpha/2}\cdot\frac{\sqrt{\hat{V}_{\rm RR,\mathcal{R}}/n}}{\htaurrR}\right)
\bigg]}.
\end{align*}

\end{example}

\begin{example}[Odds Ratio, No Borrow CovAdj]
\label{ex:or}
The RCT-only plug-in estimator for $\tauor$ is 
$$
\htauorR=
\frac{\hthetaR{1}/(1-\hthetaR{1})}{\hthetaR{0}/(1-\hthetaR{0})}.
$$
The RCT-only EIF of $\tauor$ is 
\begin{align*}
\ifrct(\tauor)&=\ifrct\Bigg\{\frac{\theta_1/(1-\theta_1)}{\theta_0/(1-\theta_0)}\Bigg\}
=\frac{\ifrct\{\theta_1/(1-\theta_1)\}}{\theta_0/(1-\theta_0)}-
\frac{\ifrct\{\theta_0/(1-\theta_0)\}}{\theta_0/(1-\theta_0)}\Bigg\{\frac{\theta_1/(1-\theta_1)}{\theta_0/(1-\theta_0)}\Bigg\}\\
&=\frac{S/\prct}{\theta_0/(1-\theta_0)}\Bigg\{
\frac{\xi_1(O)-\theta_1}{(1-\theta_1)^2}
-
\frac{\xi_0(O)-\theta_0}{(1-\theta_0)^2}
\tauor 
\Bigg\},
\end{align*}
where the last equality is due to
$$
\ifrct\{\theta_a/(1-\theta_a)\}=\frac{\ifrct(\theta_a)}{(1-\theta_a)^2} \quad \text{and}\quad \ifrct(\theta_a)=(S/\prct)\{\xi_a(O)-\theta_a\}.
$$
The variance estimator for $\sqrt{n} \htauorR$ is
\begin{align*}
\hat{V}_{\rm OR, \mathcal{R}}
&=\frac{1}{n}\sum_{i=1}^n
\Bigg[
\frac{S_i/\prct}{\hthetaR{0}/(1-\hthetaR{0})}\Bigg\{
\frac{\hat\xi_1(O_i)-\hat\theta_1}{(1-\hthetaR{1})^2}
-
\frac{\hat\xi_0(O_i)-\hthetaR{0}}{(1-\hthetaR{0})^2}
\htauorR 
\Bigg\}
\Bigg]^2\\
&=\frac{1}{n\prct^2}\sum_{i:S_i=1}
\Bigg[
\frac{1}{\hthetaR{0}/(1-\hthetaR{0})}\Bigg\{
\frac{\hat\xi_1(O_i)-\hat\theta_1}{(1-\hthetaR{1})^2}
-
\frac{\hat\xi_0(O_i)-\hthetaR{0}}{(1-\hthetaR{0})^2}
\htauorR 
\Bigg\}
\Bigg]^2.
\end{align*}
The asymptotic confidence interval is
\begin{align*}
    &\exp\bigg[
\log(\htauorR)-z_{1-\alpha/2}\sqrt{\hat{V}_{\rm log(OR),\mathcal{R}}/n},\;
\log(\htauorR)+z_{1-\alpha/2}\sqrt{\hat{V}_{\rm log(OR),\mathcal{R}}/n}
\bigg]\\
\approx&{\bigg[
\htauorR\cdot \exp\left(-z_{1-\alpha/2}\cdot\frac{\sqrt{\hat{V}_{\rm OR,\mathcal{R}}/n}}{\htauorR}\right),\;
\htauorR\cdot \exp\left(z_{1-\alpha/2}\cdot\frac{\sqrt{\hat{V}_{\rm OR,\mathcal{R}}/n}}{\htauorR}\right)
\bigg]}.
\end{align*}
\end{example}

\subsection{EC Borrowing}\label{Supp:formula_estimator_EC}

\begin{continuedexample}{ex:rd}[Risk Difference, Borrow AIPW]
The plug-in estimator for $\taurd$ is 
$$
\htaurd=\hat{\theta}_1-\hat{\theta}_0.
$$
The EIF of $\taurd$ is 
\begin{align*}
\mathbb{IF}(\taurd)&=\mathbb{IF}(\theta_1-\theta_0)=\mathbb{IF}(\theta_1)-\mathbb{IF}(\theta_0)\\
&=\phi_1(O)-\phi_0(O)-\frac{S}{\prct} \taurd.    
\end{align*}
The variance estimator for $\sqrt{n} \htaurd$ is
$$
\hat{V}_{\rm RD}=\frac1n\sum_{i=1}^n\bigg\{\hat\phi_1(O_i)-\hat\phi_0(O_i)-\frac{S_i}{\prct} \htaurd\bigg\}^2.
$$
The asymptotic confidence interval is
$$
\bigg[
\htaurd-z_{1-\alpha/2}\sqrt{\hat{V}_{\rm RD}/n},\;
\htaurd+z_{1-\alpha/2}\sqrt{\hat{V}_{\rm RD}/n}
\bigg].
$$
\end{continuedexample}

\begin{continuedexample}{ex:rr}[Risk Ratio, Borrow AIPW]
The plug-in estimator for $\taurr$ is 
$$
\htaurr=\hat{\theta}_1/\hat{\theta}_0.
$$
The EIF of $\taurr$ is 
\begin{align*}
\mathbb{IF}(\taurr)&=\mathbb{IF}\Bigg(\frac{\theta_1}{\theta_0}\Bigg)=\frac{\mathbb{IF}(\theta_1)}{\theta_0}-\frac{\mathbb{IF}(\theta_0)}{\theta_0}\Bigg(\frac{\theta_1}{\theta_0}\Bigg)\\
&=\frac{1}{\theta_0}\big\{\phi_1(O)-\phi_0(O)\taurr\big\}.
\end{align*}
The variance estimator for $\sqrt{n} \htaurr$ is
$$
\hat{V}_{\rm RR}=\frac1n\sum_{i=1}^n
\bigg[
\frac{1}{\hat\theta_0}\big\{\hat\phi_1(O_i)-\hat\phi_0(O_i)\htaurr\big\}
\bigg]^2.
$$
The asymptotic confidence interval is
\begin{align*}
    &\text{exp}\bigg[
\text{log}(\htaurr)-z_{1-\alpha/2}\sqrt{\hat{V}_{\rm log(RR)}/n},\;
\text{log}(\htaurr)+z_{1-\alpha/2}\sqrt{\hat{V}_{\rm log(RR)}/n}
\bigg]\\
\approx&{\bigg[
\htaurr\cdot \text{exp}\left(-z_{1-\alpha/2}\cdot\frac{\sqrt{\hat{V}_{\rm RR}/n}}{\htaurr}\right),\;
\htaurr\cdot \text{exp}\left(z_{1-\alpha/2}\cdot\frac{\sqrt{\hat{V}_{\rm RR}/n}}{\htaurr}\right)
\bigg]}.
\end{align*}
\end{continuedexample}

\begin{continuedexample}{ex:or}[Odds Ratio, Borrow AIPW]
The plug-in estimator for $\tauor$ is 
$$
\htauor=
\frac{\hat{\theta}_1/(1-\hat{\theta}_1)}{\hat{\theta}_0/(1-\hat{\theta}_0)}.
$$
The EIF of $\tauor$ is 
\begin{align*}
\mathbb{IF}(\tauor)&=\mathbb{IF}\Bigg\{\frac{\theta_1/(1-\theta_1)}{\theta_0/(1-\theta_0)}\Bigg\}
=\frac{\mathbb{IF}\{\theta_1/(1-\theta_1)\}}{\theta_0/(1-\theta_0)}-
\frac{\mathbb{IF}\{\theta_0/(1-\theta_0)\}}{\theta_0/(1-\theta_0)}\Bigg\{\frac{\theta_1/(1-\theta_1)}{\theta_0/(1-\theta_0)}\Bigg\}\\
&=\frac{1}{\theta_0/(1-\theta_0)}\Bigg\{
\frac{\phi_1(O)-(S/\prct)\theta_1}{(1-\theta_1)^2}
-
\frac{\phi_0(O)-(S/\prct)\theta_0}{(1-\theta_0)^2}
\tauor 
\Bigg\},
\end{align*}
where the last equality is due to
$$
\mathbb{IF}\{\theta_a/(1-\theta_a)\}=\frac{\mathbb{IF}(\theta_a)}{(1-\theta_a)^2} \quad \text{and} \quad \mathbb{IF}(\theta_a)=\phi_a(O)-(S/\prct)\theta_a.
$$
The variance estimator for $\sqrt{n} \htauor$ is
$$
\hat{V}_{\rm OR}=\frac1n\sum_{i=1}^n
\Bigg[
\frac{1}{\hat\theta_0/(1-\hat\theta_0)}\Bigg\{
\frac{\hat\phi_1(O_i)-(S_i/\prct)\hat\theta_1}{(1-\hat\theta_1)^2}
-
\frac{\hat\phi_0(O_i)-(S_i/\prct)\hat\theta_0}{(1-\hat\theta_0)^2}
\htauor 
\Bigg\}
\Bigg]^2.
$$
The asymptotic confidence interval is
\begin{align*}
    &\text{exp}\bigg[
\text{log}(\htauor)-z_{1-\alpha/2}\sqrt{\hat{V}_{\rm log(OR)}/n},\;
\text{log}(\htauor)+z_{1-\alpha/2}\sqrt{\hat{V}_{\rm log(OR)}/n}
\bigg]\\
\approx&{\bigg[
\htauor\cdot \text{exp}\left(-z_{1-\alpha/2}\cdot\frac{\sqrt{\hat{V}_{\rm OR}/n}}{\htauor}\right),\;
\htauor\cdot \text{exp}\left(z_{1-\alpha/2}\cdot\frac{\sqrt{\hat{V}_{\rm OR}/n}}{\htauor}\right)
\bigg]}.
\end{align*}
\end{continuedexample}

\section{Alternative Estimators for EC Borrowing}\label{Supp:alternative_estimator}


\subsection{Borrow Na\"ive}
\texttt{Borrow Na\"ive} method pools the RCT and EC data by using AIPW estimator to adjust covariates imbalance between treatment and control groups but ignores the source indicator $S$. Both the outcome mean function $\mu_a(X)=E[Y(a) \mid X]$ and propensity score function $e(X)=P(A=1 \mid X)$ are estimated by RCT and EC data. The average is taken over all subjects participating in hybrid controlled trials. 
\begin{align*}
    \widehat{\tau}_{\text{Na\"ive}}=&\frac{1}{n} \sum_{i\in\mathcal{R}\cup\mathcal{E}}\Big[\frac{A_i Y_i}{e\left(X_i ; \widehat{\alpha}\right)}-\frac{\left(1-A_i\right) Y_i}{1-e\left(X_i ; \widehat{\alpha}\right)}\\
    &+\left\{1-\frac{A_i}{e\left(X_i ; \widehat{\alpha}\right)}\right\} \mu_1\left(X_i ; \widehat{\beta}_1\right)
    -\left\{1-\frac{1-A_i}{1-e\left(X_i ; \widehat{\alpha}\right)}\right\} \mu_0\left(X_i ; \widehat{\beta}_0\right) \Big]. 
\end{align*}

\subsection{Borrow IPW}
\texttt{Borrow IPW} method pools the RCT and EC data by using IPW estimator and does not involve outcome model. Both EC and RCT data assist in the estimation of sampling score function $\pi(X)=P(S=1\mid X)$ and the ratio $r(X)=\operatorname{var}\{Y(0) \mid X, S=1\}/\operatorname{var}\{Y(0) \mid X, S=0\}$. However, only RCT data is used in estimating propensity score function $e(X)=P(A=1 \mid X,S=1)$. The average is taken over only participants in RCT. 
\begin{align*}
    \widehat{\tau}_{\text{IPW}}=\frac{1}{\nrct} \sum_{i\in\mathcal{R}\cup\mathcal{E}}\left[\frac{S_i A_i}{e(X_i ; \hat{\alpha})} Y_i-\widehat{W}Y_i\right],
\end{align*}
where
\begin{align*}
     \widehat{W} =\frac{S_i(1-A_i)+(1-S_i) r(X_i ; \hat{\gamma})}{\pi(X_i ; \hat{\eta})\{1-e(X_i ; \hat{\alpha})\}+\{1-\pi(X_i ; \hat{\eta})\} r(X_i ; \hat{\gamma})} \pi(X_i ; \hat{\eta}) .
\end{align*}

\subsection{Borrow CW}
\texttt{Borrow CW} method replaces the propensity score $\pi(X)$ in \texttt{Borrow IPW} with the calibration weight $q(X)=\pi(X)/\{1-\pi(X)\}$ and updates the estimator as
\begin{align*}
    \widehat{\tau}_{\text{CW}}=\frac{1}{\nrct} \sum_{i\in\mathcal{R}\cup\mathcal{E}}
    \left[\frac{S_i A_i}{e(X_i ; \hat{\alpha})} Y_i-\widehat{W}Y_i\right],
\end{align*}
where
\begin{align*}
    \widehat{W} =\frac{S_i(1-A_i)+(1-S_i) r(X_i ; \hat{\gamma})}{\hat{q}(X_i)\{1-e(X_i ; \hat{\alpha})\}+ r(X_i ; \hat{\gamma})} \hat{q}(X_i).
\end{align*}
The calibration weight $q(X)$ is estimated as $\hat{q}(X)=\pi(X_i ; \hat{\eta})/\{1-\pi(X_i ; \hat{\eta})\}$, where $\pi(X_i ; \hat{\eta})$ is derived solely from RCT data, following the same approach as in the \texttt{Borrow IPW} method.

\subsection{Borrow OM}
\texttt{Borrow OM} method models the outcome mean function $\mu_a(X)=E[Y(a) \mid X]$ using both RCT and EC data but only use RCT data for outcome model estimator.
\begin{align*}
    \widehat{\tau}_{\text{OM}}=&\frac{1}{\nrct} \sum_{i\in\mathcal{R}}\Big[ \mu_1\left(X_i ; \widehat{\beta}_1\right)
    -\mu_0\left(X_i ; \widehat{\beta}_0\right) \Big].
\end{align*}

\subsection{Borrow ACW}
\texttt{Borrow ACW} method uses both RCT and EC data to model outcome mean function $\mu_a(X)=E[Y(a) \mid X]$ and ratio $r(X)=\operatorname{var}\{Y(0) \mid X, S=1\}/\operatorname{var}\{Y(0) \mid X, S=0\}$, but only RCT is used for modeling propensity score function $e(X)=P(A=1 \mid X,S=1)$. The calibration weight is $q(X)=\pi(X)/\{1-\pi(X)\}$, which replaces the propensity score $\pi(X)$ in \texttt{Borrow AIPW} and updates the estimator as
\begin{align*}
    \widehat{\tau}_{\text{ACW}}=\frac{1}{\nrct} \sum_{i\in\mathcal{R}\cup\mathcal{E}}\left[S_i \widehat{\Delta}+\frac{S_i A_i}{e(X_i ; \hat{\alpha})} \widehat{R}_1-\widehat{W}\widehat{R}_0\right],
\end{align*}
where $\widehat{\Delta}=\mu_1(X_i ; \hat{\beta}_1)-\mu_0(X_i ; \hat{\beta}_0)$, $\widehat{R}_a=$ $Y_i-\mu_a(X_i ; \hat{\beta}_a)$ and
\begin{align*}
     \widehat{W} =\frac{S_i(1-A_i)+(1-S_i) r(X_i ; \hat{\gamma})}{\hat{q}(X_i)\{1-e(X_i ; \hat{\alpha})\}+ r(X_i ; \hat{\gamma})} \hat{q}(X_i).
\end{align*}

The estimate of calibration weight $q(X)$ is $\hat{q}(X)=\pi(X_i ; \hat{\eta})/\{1-\pi(X_i ; \hat{\eta})\}$, where the sampling score $\pi(X_i ; \hat{\eta})$ is estimated solely by RCT data.

\section{Algorithm of Adaptive Selection Threshold}\label{Supp:alg}

\begin{algorithm}[H]
\caption{Adaptive Selection Threshold}
\textbf{Input:} Threshold grid $\Gamma = \{0, 0.05, \ldots, 1\}$, number of bootstrap samples $K = 200$.\\

\For{$\gamma \in \Gamma$ \hfill}{
    Compute $\hat{\tau}_\gamma$ from the original sample. \\
    \For{$k = 1, \ldots, K$ \textbf{do}}{
        Compute $\hat{\tau}_\gamma^{(k)}$ from the $k$-th bootstrap sample.
    }
}

\For{$\gamma \in \Gamma \setminus \{1\}$}{
    Compute 
    $\widehat{\text{Var}}(\hat{\tau}_\gamma - \hat{\tau}_1) = (K - 1)^{-1} \sum_{k=1}^{K} \left\{ 
    (\hat{\tau}_\gamma^{(k)} - \hat{\tau}_1^{(k)}) - K^{-1} \sum_{k'=1}^{K} (\hat{\tau}_\gamma^{(k')} - \hat{\tau}_1^{(k')}) 
    \right\}^2.$
    $\widehat{\text{Var}}(\hat{\tau}_\gamma) = (K - 1)^{-1} \sum_{k=1}^{K} \left( \hat{\tau}_\gamma^{(k)} - K^{-1} \sum_{l'=1}^{K} \hat{\tau}_\gamma^{(k')} \right)^2.$
    $\widehat{\text{MSE}}(\gamma) = (\hat{\tau}_\gamma - \hat{\tau}_1)^2 - \widehat{\text{Var}}(\hat{\tau}_\gamma - \hat{\tau}_1) + \widehat{\text{Var}}(\hat{\tau}_\gamma).$

}

Compute 
$\widehat{\text{MSE}}(1) = (K - 1)^{-1} \sum_{k=1}^{K} \left( \hat{\tau}_1^{(k)} - K^{-1} \sum_{k'=1}^{K} \hat{\tau}_1^{(k')} \right)^2.$

Find $\hat{\gamma} = \arg\min_{\gamma \in \Gamma} \widehat{\text{MSE}}(\gamma)$. \\

\textbf{Output:} $\hat{\gamma}$.
\end{algorithm}

\section{Additional Simulation Results}\label{Supp:sim}
\subsection{Simulation results for all the methods}\label{Supp:sim_allBorrow}
To provide a comprehensive simulation study, we study the performance of all the methods that aforementioned under the scenarios of both no hidden bias and hidden exists. 

\subsubsection{No hidden bias}\label{sec:simstudy_covshift}
In this section, we compare FRT and asymptotic inference for all the No Borrow, Borrow, and Conformal Selective Borrow methods under the assumption of no hidden bias. For simplicity, the results for RD is discussed in detail, but the results are consistent for RR and OR. 

\begin{figure}[ht]
\includegraphics[width=1\textwidth]{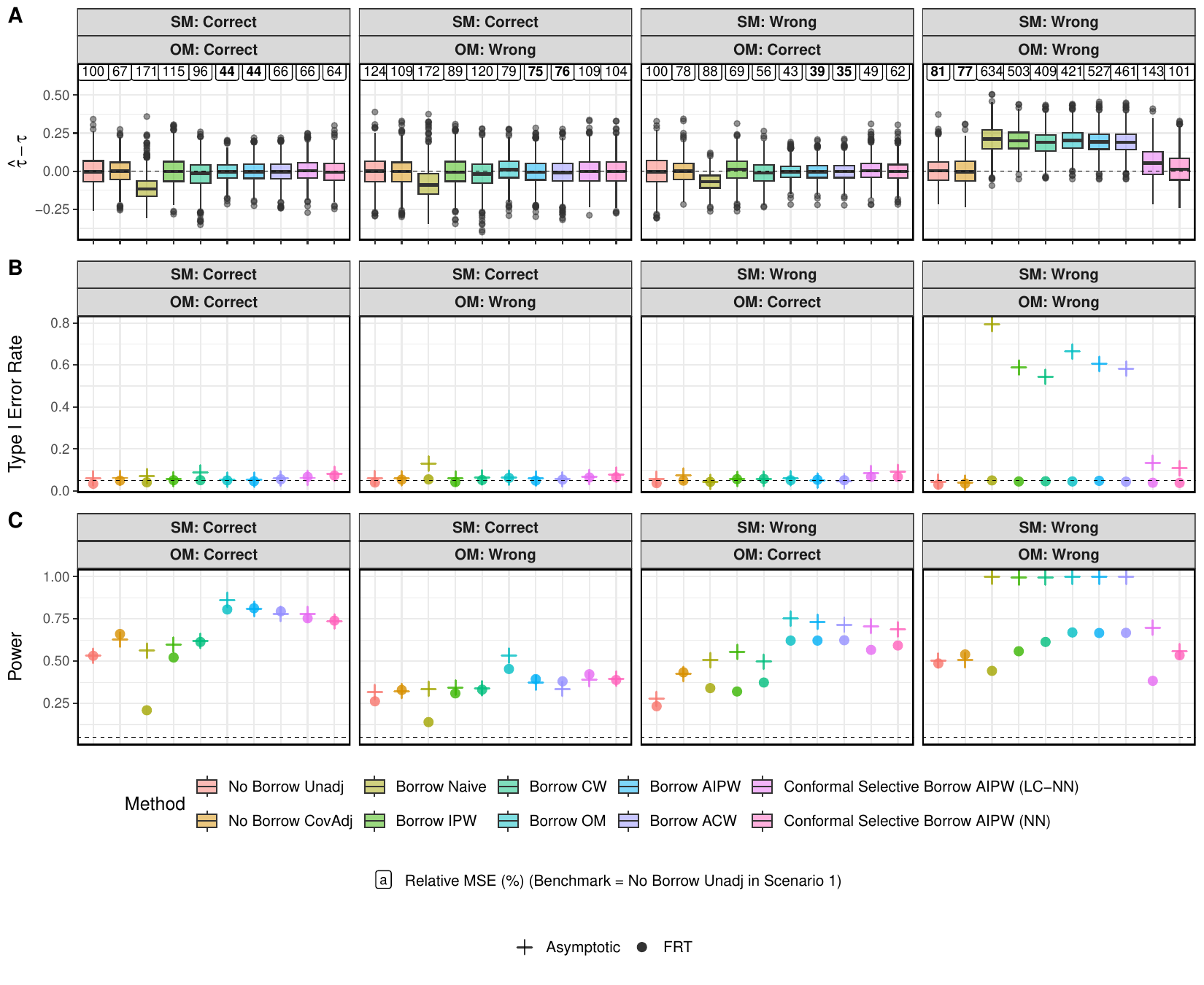}
\caption{Simulation results under no hidden bias $b=0$}
\label{fig:sim_est_inf_main_0}
\end{figure}

Figure~\ref{fig:sim_est_inf_main_0} summarizes the estimation and inference results for all No Borrow, Borrow, and CSB estimators. When at least one nuisance model is correctly specified, all estimators are nearly unbiased except for \texttt{Borrow Na\"ive}. \texttt{Borrow OM}, \texttt{Borrow ACW}, \texttt{Borrow AIPW}, \texttt{CSB NN}, and \texttt{CSB LC-NN} exhibit similar and relatively low variances. Consistent with variance, Relative MSE values (shown above each box) indicate that \texttt{Borrow OM}, \texttt{Borrow ACW}, and \texttt{Borrow AIPW} achieve the lowest Relative MSE under partial model correctness when  compared to \texttt{No Borrow Unadj} in Scenario 1. Notably, when both the SM and OM are misspecified, \texttt{CSB NN} and \texttt{CSB LC-NN} still yields almost unbiased estimates.

The testing results in Figure~\ref{fig:sim_est_inf_main_0} show that FRT consistently controls the type I error rate across all scenarios, while asymptotic inference exhibits inflation, especially when both models are misspecified. \texttt{Borrow OM}, \texttt{Borrow ACW}, and \texttt{Borrow AIPW} achieve higher power than \texttt{No Borrow} regardless of model specification. Under the no hidden bias setting, the two CSB estimators perform comparably to the Borrow estimators in the first three scenarios; however, the Borrow methods yield greater FRT power gains by enriching the RCT with additional information without introducing bias.


\subsubsection{Hidden bias exists}\label{sec:simstudy_hiddenbias}

In this section, we explore the performance of FRT-inference-based estimators in the presence of hidden bias. Since Conformal Selective Borrow approaches can identify potential hidden bias and select subjects from the EC that closely resemble those in the RCT, these estimators are of particular interest.
\begin{figure}[ht]
    \centering
    \includegraphics[width=\linewidth]{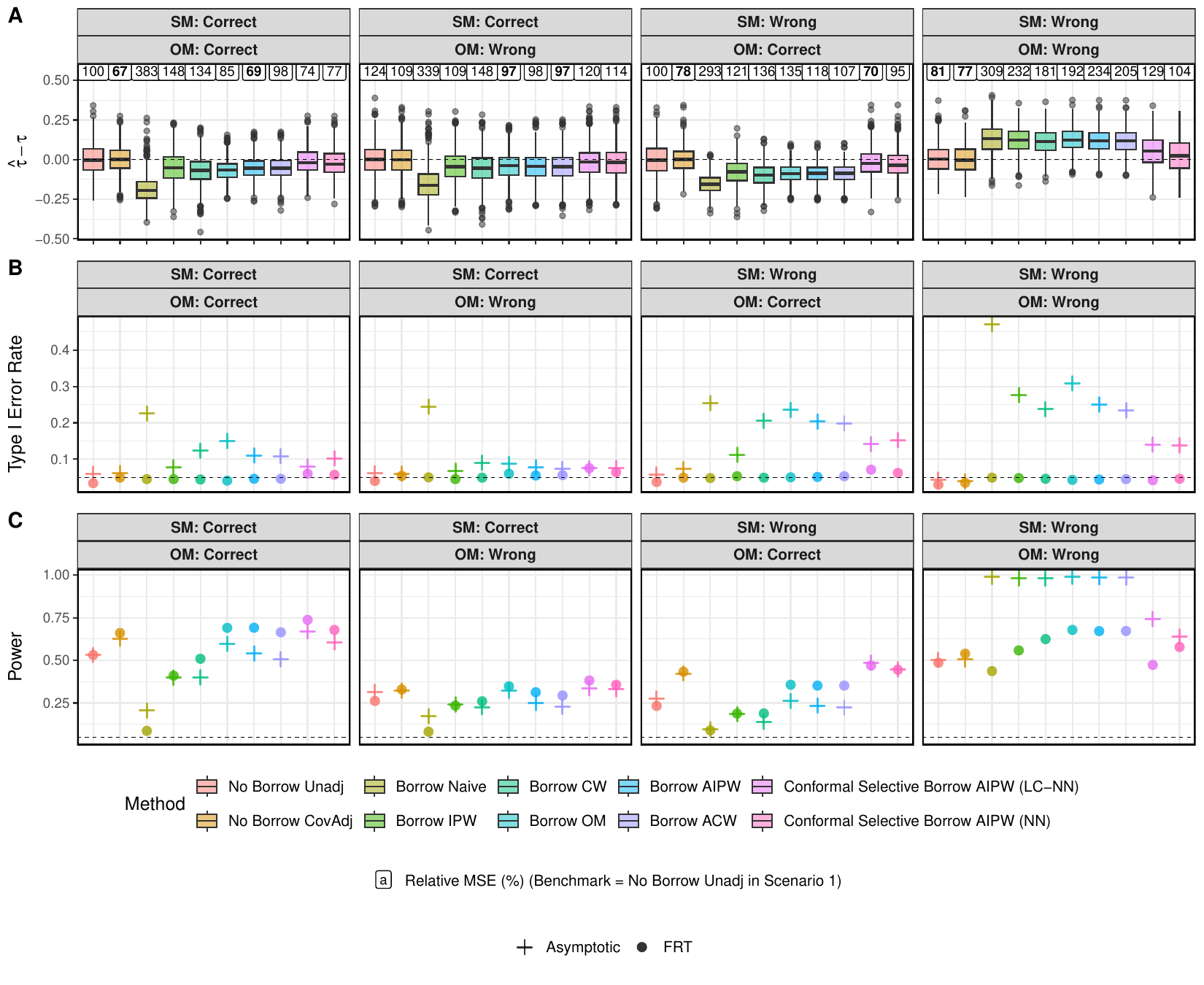}
    \caption{Simulation results under hidden bias $b=4$}
    \label{fig:sim_est_inf_main_4}
\end{figure}
In Figure \ref{fig:sim_est_inf_main_4}, under hidden bias $b=4$, where outcome incomparability exists, all six Borrow methods lead to biased estimates even when both SM and OM are correctly specified. In comparison, the \texttt{CSB NN} and \texttt{CSB LC-NN} maintain bias near zero across all scenarios. The figure also shows that asymptotic inference leads to inflated type I error rates for all methods, regardless of model specification, while FRT continues to strictly control type I error rates, consistent with the no hidden bias setting. In general, the power based on FRT is larger than that based on asymptotic inference. 
When at least one model is correctly specified, Borrow estimators fail to achieve power gains over \texttt{No Borrow} under FRT, whereas CSB methods provide some power improvements. These results highlight that integrating FRT with CSB estimators offers both valid type I error control and improved power in the presence of hidden bias.



\subsection{Simulation results with SAR conformal score}\label{Supp:sim_SAR}
\begin{figure}[ht]
    \centering
    \includegraphics[width=\linewidth]{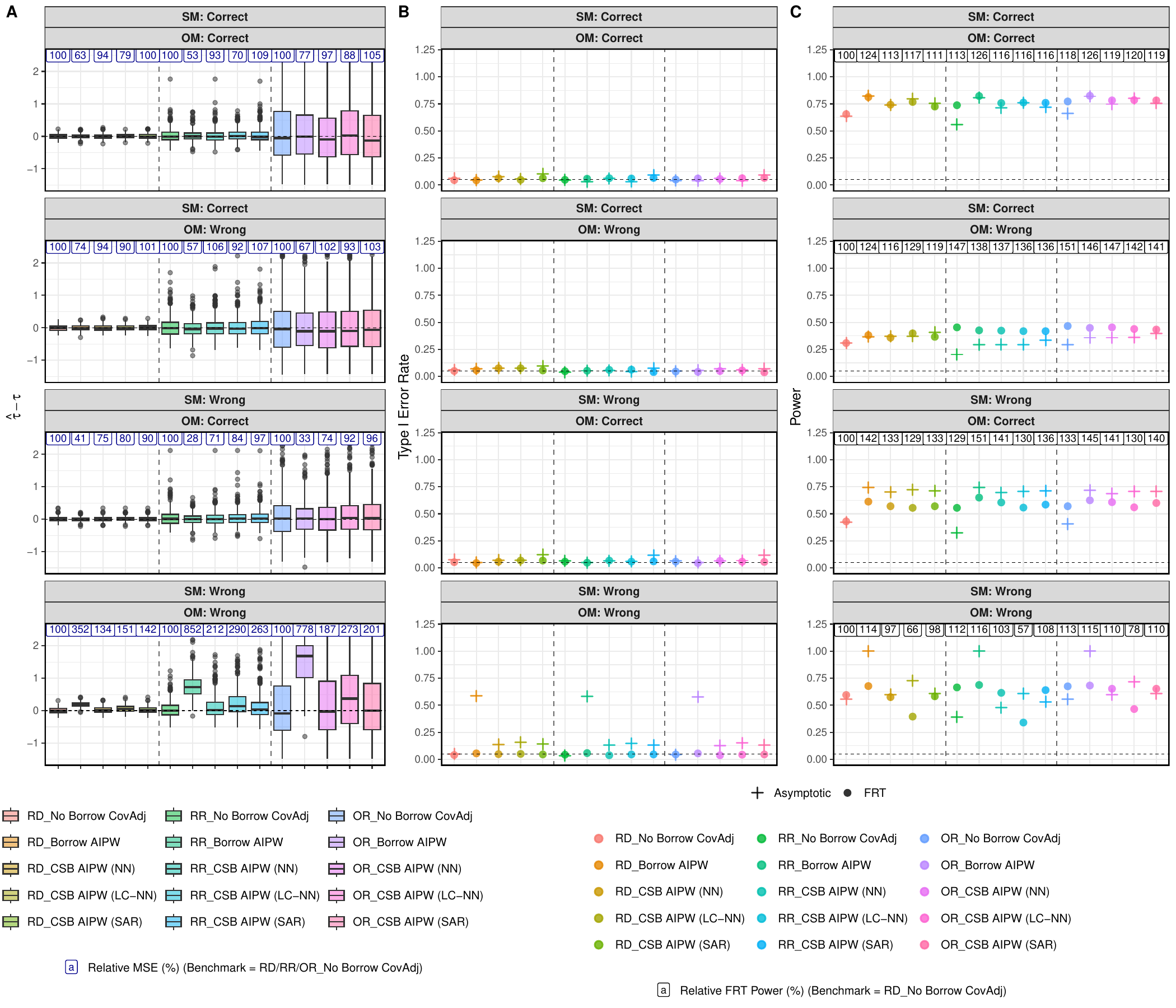}
    \caption{Simulation results for three different estimands with SAR ($b=0$)}
    \label{fig:sim_hb0_3estimands_SAR}
\end{figure}

Standardized absolute residial (SAR) is one of the most commonly used conformal score for continuous outcome. In this section, we additionally provided the simulation results when using SAR as the conformal scores. As shown in Figure \ref{fig:sim_hb0_3estimands_SAR}, under no hidden bias, similar estimation results can be observed when using NN as conformal score, although SAR leads to a generally larger Relative MSE. Similar to NN, SAR is also robust to both model misspecification. The FRT power of \texttt{CSB SAR} tends to be lower than that for \texttt{CSB NN}. 

\begin{figure}[ht]
    \centering
    \includegraphics[width=\linewidth]{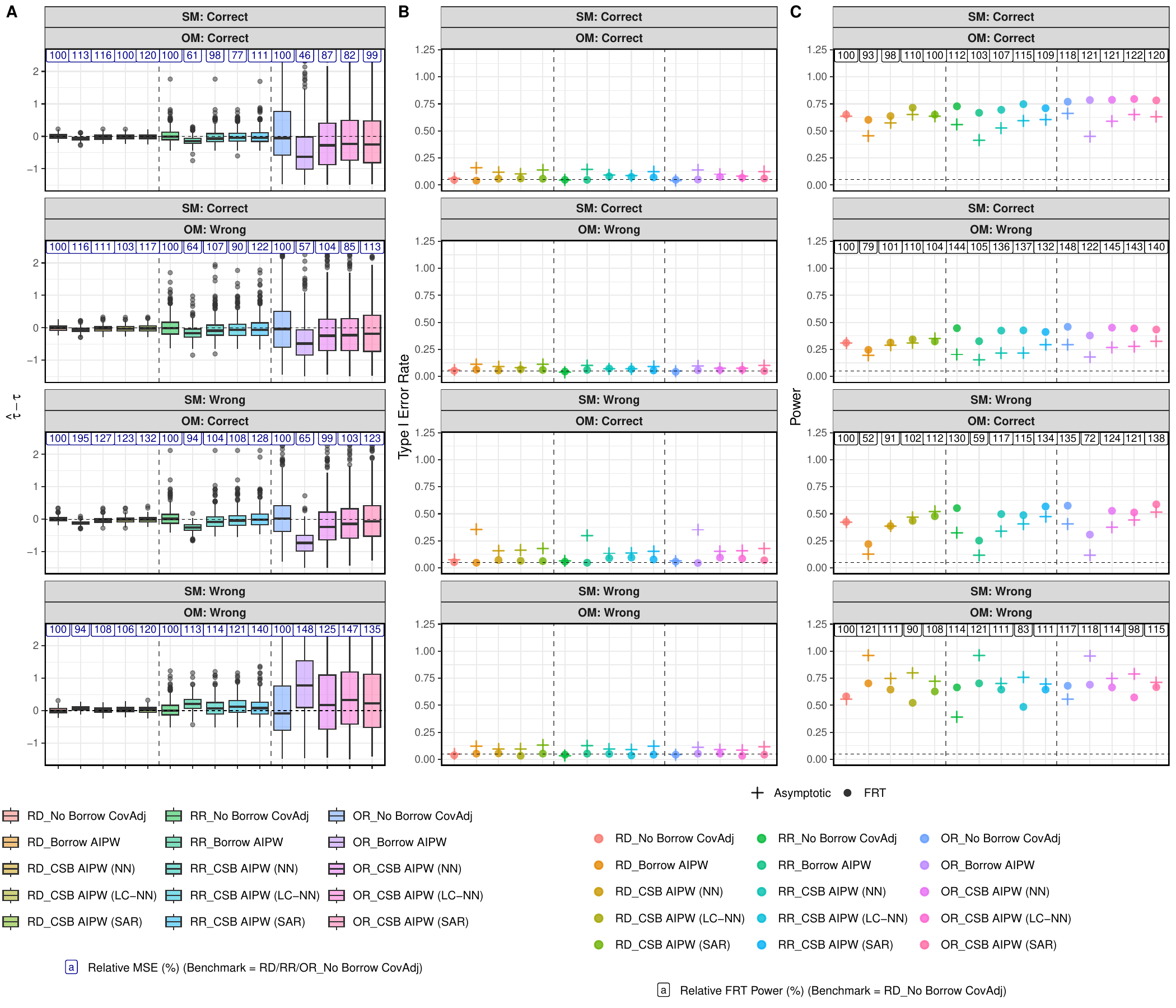}
    \caption{Simulation results for three different estimands with SAR ($b=6$)}
    \label{fig:sim_hb6_3estimands_SAR}
\end{figure}

Under hidden bias magnitude of 6, the comparison of \texttt{CSB SAR} and \texttt{CSB NN} is consistent with the comparison under no hidden bias. However, the variance of \texttt{CSB SAR} becomes more noticeably larger than \texttt{CSB NN}. 
\subsection{Simulation results under varying hidden bias}\label{Supp:sim_seqHiddenBias}

\begin{figure}[ht]
    \centering
    \includegraphics[width=1\linewidth]{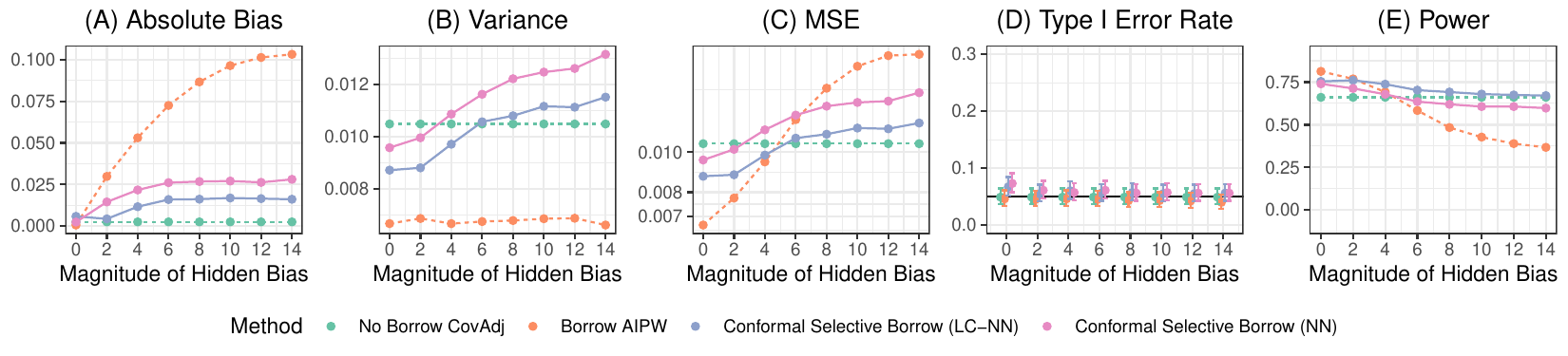}
    \caption{Simulation results across different magnitudes of hidden bias (SM Correct; OM Correct)}
    \label{fig:sim_HB_c1}
\end{figure}

\begin{figure}[ht]
    \centering
    \includegraphics[width=1\linewidth]{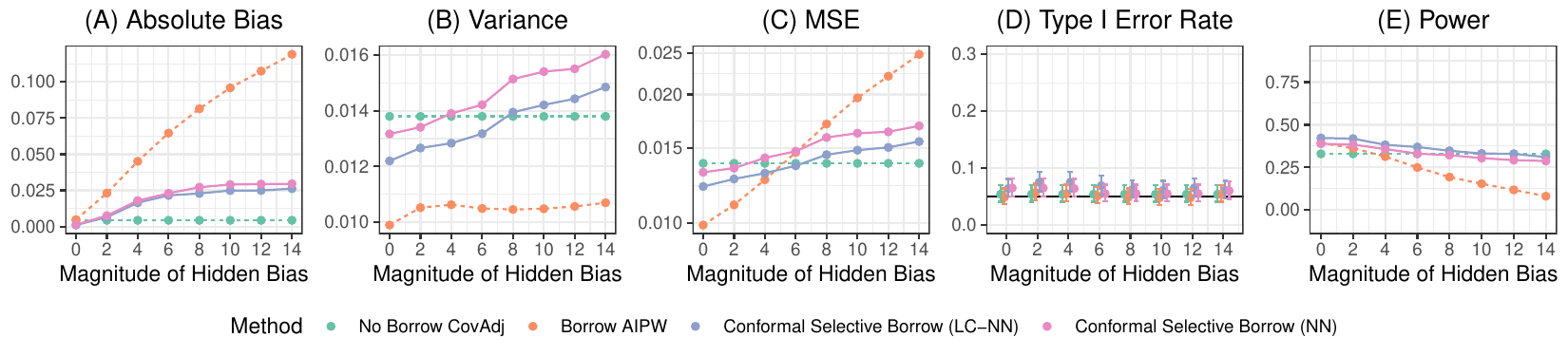}
    \caption{Simulation results across different magnitudes of hidden bias (SM Correct; OM Wrong)}
    \label{fig:sim_HB_c2}
\end{figure}

\begin{figure}[ht]
    \centering
    \includegraphics[width=1\linewidth]{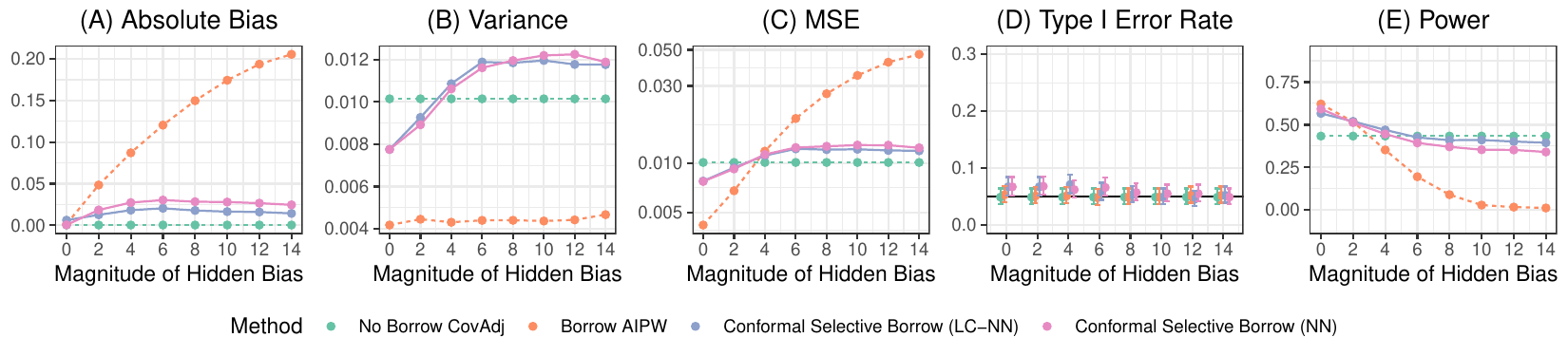}
    \caption{Simulation results across different magnitudes of hidden bias (SM Wrong; OM Correct)}
    \label{fig:sim_HB_c3}
\end{figure}

\begin{figure}[ht]
    \centering
    \includegraphics[width=1\linewidth]{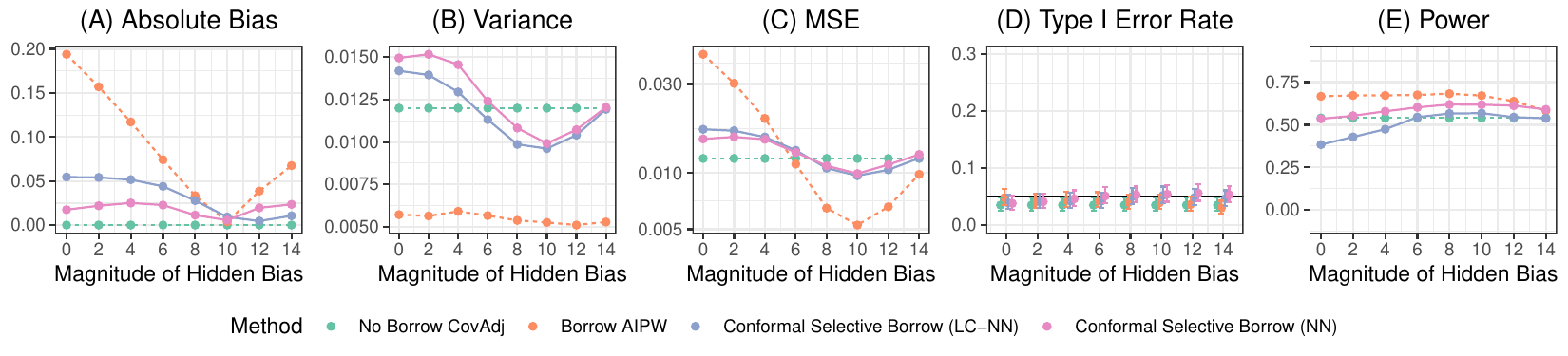}
    \caption{Simulation results across different magnitudes of hidden bias (SM Wrong; OM Wrong)}
    \label{fig:sim_HB_c4}
\end{figure}

In addition to the results under varying magnitudes of hidden bias provided in main text, we provide the results for other three scenarios in this section in Figure \ref{fig:sim_HB_c1} - \ref{fig:sim_HB_c4}. 

\subsection{Power curves across different true estimands (\textit{b=6})}\label{Supp:sim_trueEstimand}

Under hidden bias with $b=6$, we also provided the power curves based on FRT to explore how the FRT power changes as the true RD and true RR increases. 

\begin{figure}[ht]
    \centering
    \includegraphics[width=\linewidth]{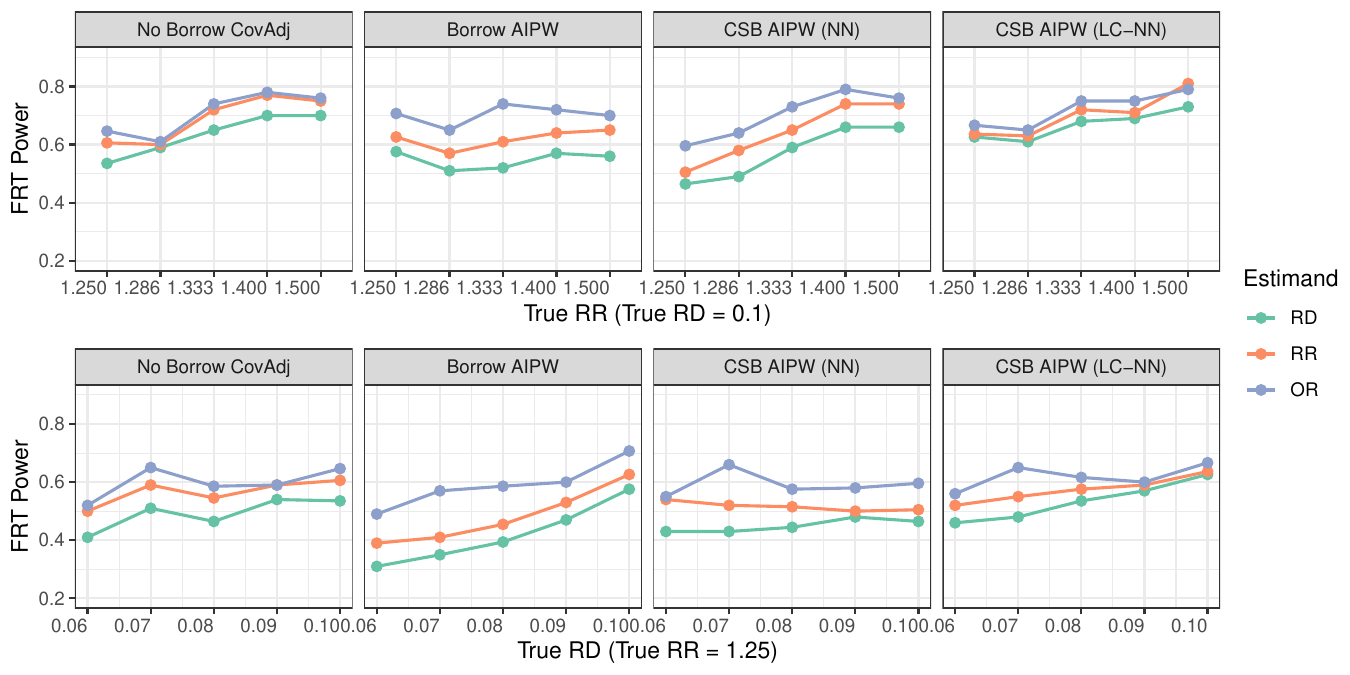}
    \caption{Power curves across different $\tau_{RD}$ and $\tau_{RR}$ ($b=6$)}
    \label{fig:powercurve_hb6_main}
\end{figure}

\subsection{Sensitivity analysis}\label{Supp:sim_senAna}
\subsubsection{Sensitivity analysis in terms of $\gamma$}\label{Supp:sim_senAna_gamma}
\begin{figure}[ht]
    \centering
    \includegraphics[width=0.55\linewidth]{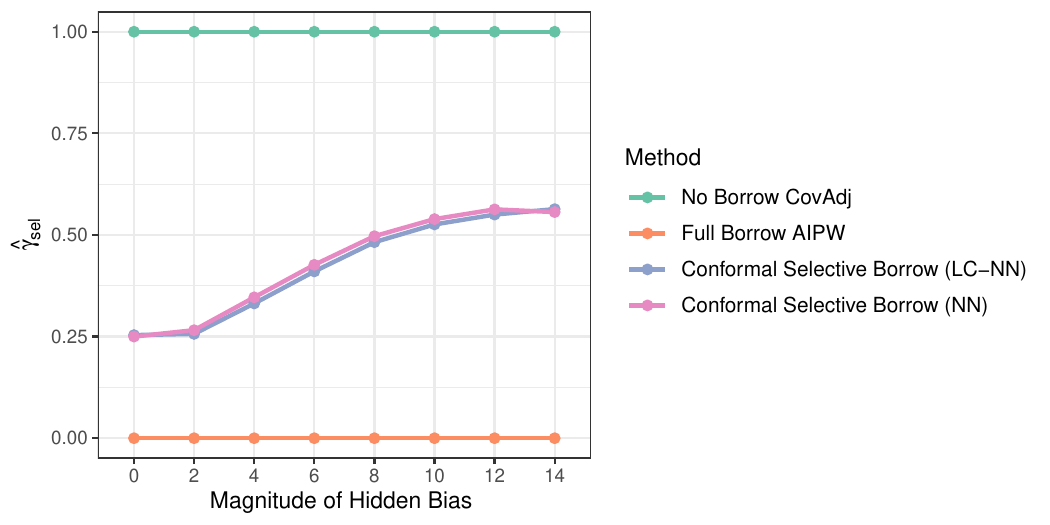}
    \vspace{-10pt}
    \caption{Adaptively selected $\gamma$ across different magnitudes of hidden bias}
    \label{fig:sim_HB_gamma}
\end{figure}
The threshold $\gamma$ plays a critical role in determining how much external control information is borrowed, and its optimal value is expected to vary with the magnitude of hidden bias. Therefore, we examine the relationship between the hidden bias magnitude $b$ and the selected threshold $\gamma$, where $\gamma$ is chosen by minimizing the MSE as described in Section~\ref{sec:adaGamma} of the main article. As shown in Figure \ref{fig:sim_HB_gamma}, both \texttt{CSB LC-NN} and \texttt{CSB NN} exhibit increasing values of $\gamma$ as the hidden bias magnitude increases. Since larger values of $\gamma$ correspond to less borrowing, this indicates that both selective borrowing methods adaptively reduce the amount of external information used when hidden bias becomes more severe. Consequently, our proposed methods are capable of detecting the presence of hidden bias and responding more conservatively when the quality of external control data deteriorates. In contrast, \texttt{No Borrow CovAdj} and \texttt{Borrow AIPW} can be viewed as two special cases of CSB, corresponding to $\gamma=1$ (no borrowing) and $\gamma=0$ (full borrowing), respectively.

\begin{figure}[ht]
    \centering    \includegraphics[width=\linewidth]{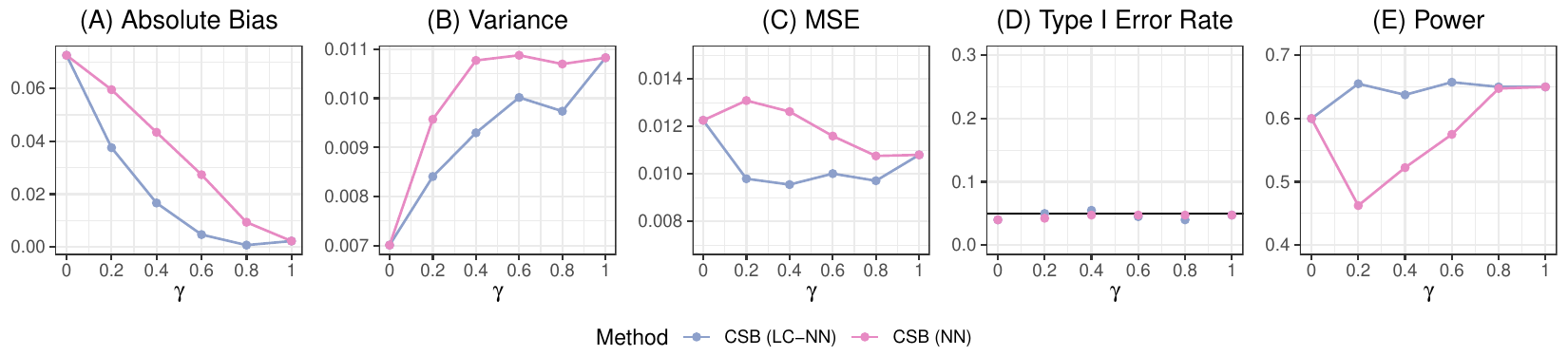}
    \vspace{-10pt}
    \caption{Sensitivity analysis of two CSB methods across different $\gamma$}
    \label{fig:sim_gamma_SenAna}
\end{figure}

To further understand the role of the selection threshold $\gamma$, we examine the performance of CSB methods across a range of fixed $\gamma$ values under hidden bias $b=6$. As shown in Figure \ref{fig:sim_gamma_SenAna}, the performance of both CSB methods varies nonlinearly with the selection threshold $\gamma$, which directly controls the strictness of external control borrowing. As $\gamma$ increases, the absolute bias decreases because fewer potentially incompatible external controls are retained, whereas the variance generally increases due to the reduced effective sample size. Consequently, both the MSE and power exhibit non-monotone behavior as functions of $\gamma$. As there is no universally optimal fixed choice of $\gamma$, we are motivated to propose the adaptive $\gamma$-selection procedure as discussed in Section~\ref{sec:adaGamma} of the main article. Rather than relying on an ad hoc threshold, our procedure selects $\gamma$ in a data-adaptive manner by minimizing the estimated MSE, thereby balancing the bias-variance trade-off.

Figure~\ref{fig:sim_gamma_SenAna} and Table~\ref{tab:sim_gamma_sensitivity_full} show the performance of \texttt{CSB NN} and \texttt{CSB LC-NN} in the sensitivity analysis across fixed values of $\gamma$. In terms of MSE, \texttt{CSB LC-NN} with $\gamma=0.2$--$0.8$ achieves lower MSE than both No Borrow ($\gamma=1$) and Full Borrow ($\gamma=0$), whereas \texttt{CSB NN} yields higher MSE than No Borrow across the considered values of $\gamma$. In terms of power, \texttt{CSB LC-NN} with $\gamma=0.2$ and $\gamma=0.6$ achieves comparable power compared with No Borrow and Full Borrow, whereas \texttt{CSB NN} yields lower power. In the primary analysis, however, $\gamma$ is selected adaptively and may therefore differ across simulation replicates. Consequently, its operating characteristics cannot be fully represented by any single fixed value of $\gamma$ considered here. Under the same simulation scenario, Figure~\ref{fig:sim_hb6_3estimands} in the main text shows that adaptive $\gamma$ selection with \texttt{CSB LC-NN} achieves approximately a 10\% power gain over No Borrow. Overall, these results further highlight that (i) choosing an appropriate conformal score is essential for achieving efficiency gains, supporting our recommendation of the LC-NN conformal score, and (ii) adaptive selection of $\gamma$ can provide greater power gains than those observed with individual fixed values of $\gamma$.

\begin{table}[t]
\centering
\caption{Simulation results across selective borrowing threshold $\gamma$, by two CSB methods}
\label{tab:sim_gamma_sensitivity_full}
\begin{tabular}{clccccc}
\toprule
Method & $\gamma$ & Abs. Bias & Variance & MSE & Type I Error & Power \\
\midrule
\multirow{6}{*}{CSB (LC-NN)} & 0 (Full Borrow) & 0.0726 & 0.0070 & 0.0123 & 0.0400 & 0.6000 \\
 & 0.2 & 0.0376 & 0.0084 & 0.0098 & 0.0500 & 0.6550 \\
 & 0.4 & 0.0166 & 0.0093 & 0.0096 & 0.0550 & 0.6375 \\
 & 0.6 & 0.0047 & 0.0100 & 0.0100 & 0.0450 & 0.6575 \\
 & 0.8 & 0.0007 & 0.0097 & 0.0097 & 0.0400 & 0.6500 \\
 & 1 (No Borrow) & 0.0022 & 0.0108 & 0.0108 & 0.0475 & 0.6500 \\
\midrule
\multirow{6}{*}{CSB (NN)} & 0 (Full Borrow) & 0.0726 & 0.0070 & 0.0123 & 0.0400 & 0.6000 \\
 & 0.2 & 0.0595 & 0.0096 & 0.0131 & 0.0425 & 0.4625 \\
 & 0.4 & 0.0434 & 0.0108 & 0.0126 & 0.0475 & 0.5225 \\
 & 0.6 & 0.0273 & 0.0109 & 0.0116 & 0.0475 & 0.5750 \\
 & 0.8 & 0.0093 & 0.0107 & 0.0108 & 0.0475 & 0.6475 \\
 & 1 (No Borrow) & 0.0022 & 0.0108 & 0.0108 & 0.0475 & 0.6500 \\
\bottomrule
\end{tabular}
\end{table}


\subsubsection{Sensitivity analysis in terms of $n_\mathcal{R}$, matching ratios, and bootstrap sizes}\label{Supp:sim_sens_nsize}
To further evaluate the robustness of the proposed procedure, we conducted additional sensitivity analyses under varying RCT sample sizes, external cohort sizes, and bootstrap sizes. Specifically, we considered two RCT sample sizes, $n_{\mathcal{R}}=75$ and $n_{\mathcal{R}}=350$, representing a relatively small trial and a moderate-sized trial, respectively; the latter is also close to the scale of our real-data application. To reflect the impact of external data size, we varied the matching ratio between the RCT and external cohort as 1:1, 1:2, and 1:4. In addition to $B_{\gamma}=200$ in primary simulation setting in main manuscript, we additionally explore two bootstrap sizes for adaptive $\gamma$ selection, $B_{\gamma}=\{100,150\}$. Following the setup of the main simulations, we fixed the hidden bias level at $b=6$ and focused on the risk difference estimand for simplicity, since the main text showed broadly similar performance patterns across the three estimands.

\begin{figure}[!ht]
    \centering
    \includegraphics[width=1\linewidth]{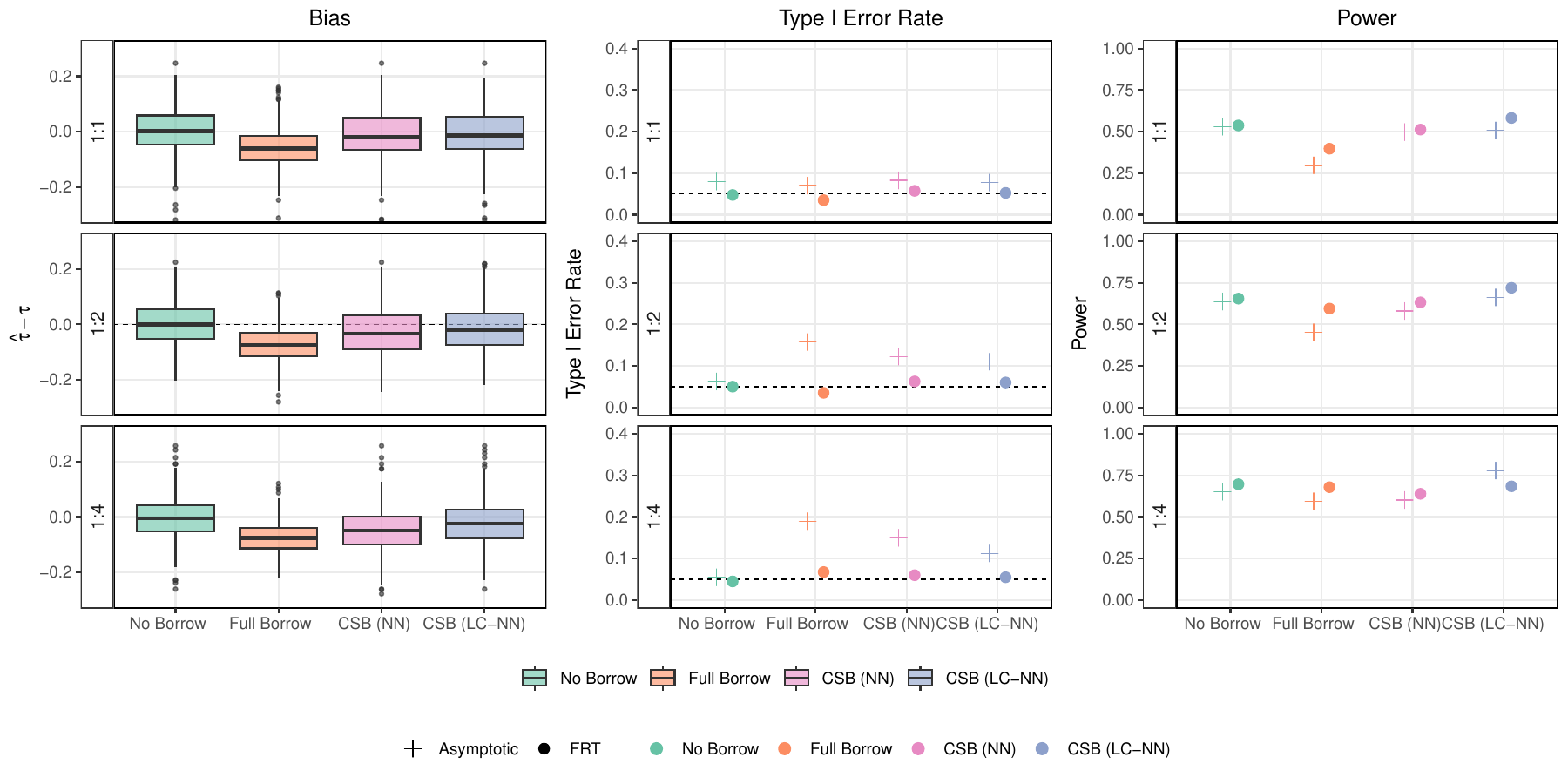}
    \caption{Simulation results when $B_{\gamma}=200$, $n_\mathcal{R}=75$, with correct SM and correct OM.}
    \label{fig:sim_bootG200_75_SMT_OMT}
\end{figure}

\begin{figure}[!ht]
    \centering
    \includegraphics[width=1\linewidth]{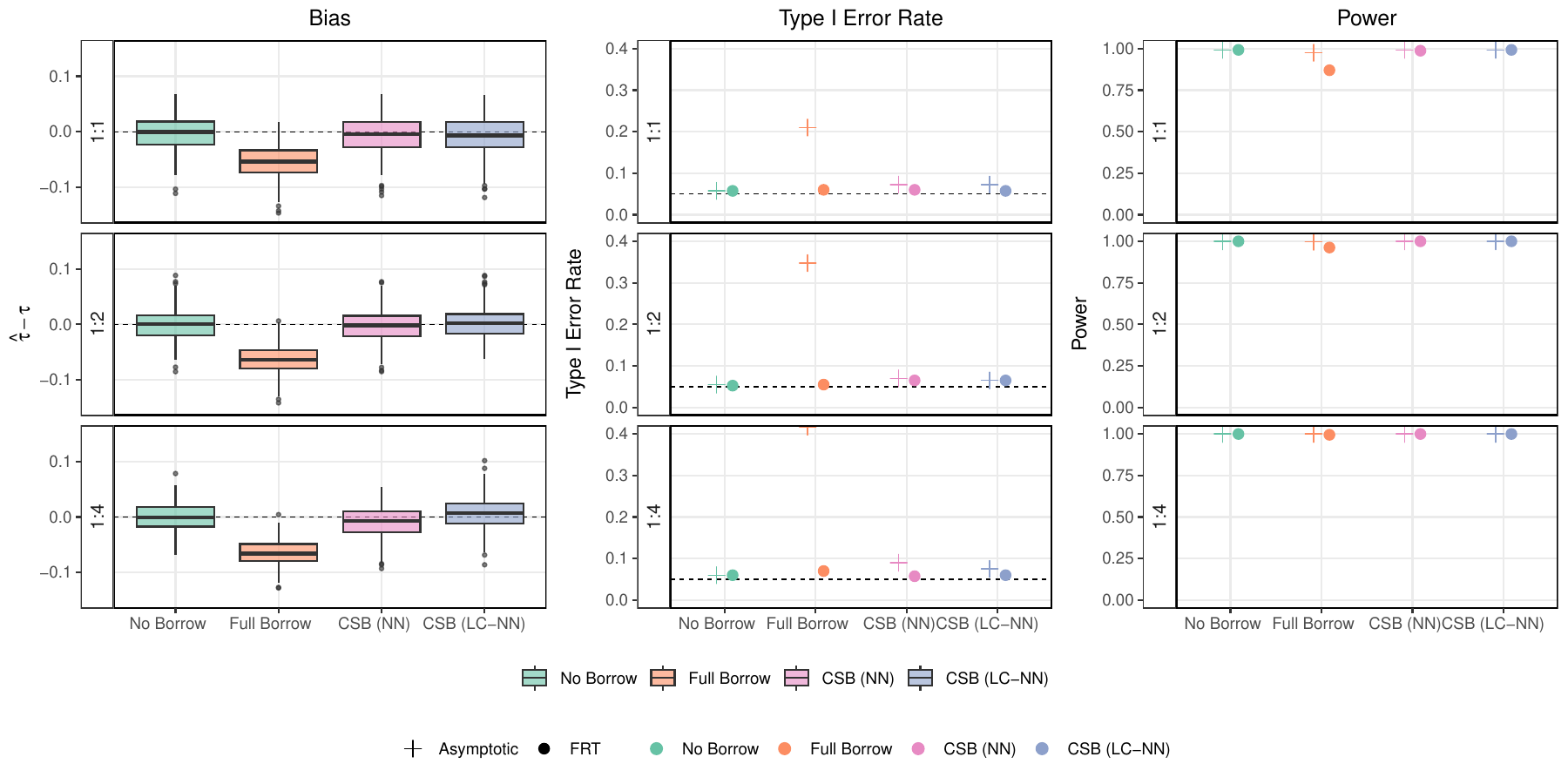}
    \caption{Simulation results when $B_{\gamma}=200$, $n_\mathcal{R}=350$, with correct SM and correct OM.}
    \label{fig:sim_bootG200_350_SMT_OMT}
\end{figure}

Figures \ref{fig:sim_bootG200_75_SMT_OMT} and \ref{fig:sim_bootG200_350_SMT_OMT} summarize the results when both nuisance models are correctly specified. When the RCT sample size is small ($n_{\mathcal{R}}=75$), full borrowing (\texttt{Borrow AIPW}) already introduces noticeable bias under hidden bias, and this bias becomes more pronounced as the matching ratio increases. In contrast, the proposed CSB procedures remain substantially more stable across the three matching ratios. The FRT-based procedures continue to maintain type I error close to the nominal 0.05 level. In terms of power, compared with \texttt{No Borrow}, the \texttt{CSB} procedures provide a clear gain under 1:1 and 1:2 matching, whereas this gain becomes less evident under 1:4 matching. This pattern suggests that, when the candidate external controls may still contain hidden bias, increasing the amount of borrowed data does not necessarily improve inference, because excessive borrowing may allow the external cohort to exert undue influence on the target trial analysis. Under these settings, a moderate matching ratio such as 1:2 appears to provide a more favorable balance between efficiency gain and protection against domination by external data.

When the internal RCT sample size increases to $n_{\mathcal{R}}=350$, the estimators generally become more stable, as reflected by the narrower bias distributions, and power is high across most methods even under 1:1 matching. This suggests that, when the internal trial itself is sufficiently informative, the incremental benefit of external borrowing becomes smaller and the analysis is less sensitive to the matching ratio. Nevertheless, even in this moderate-sample setting, the advantage of selective borrowing remains in its protection against harmful external information, as full borrowing still remains vulnerable once the working models are misspecified.

\begin{figure}[!ht]
    \centering
    \includegraphics[width=1\linewidth]{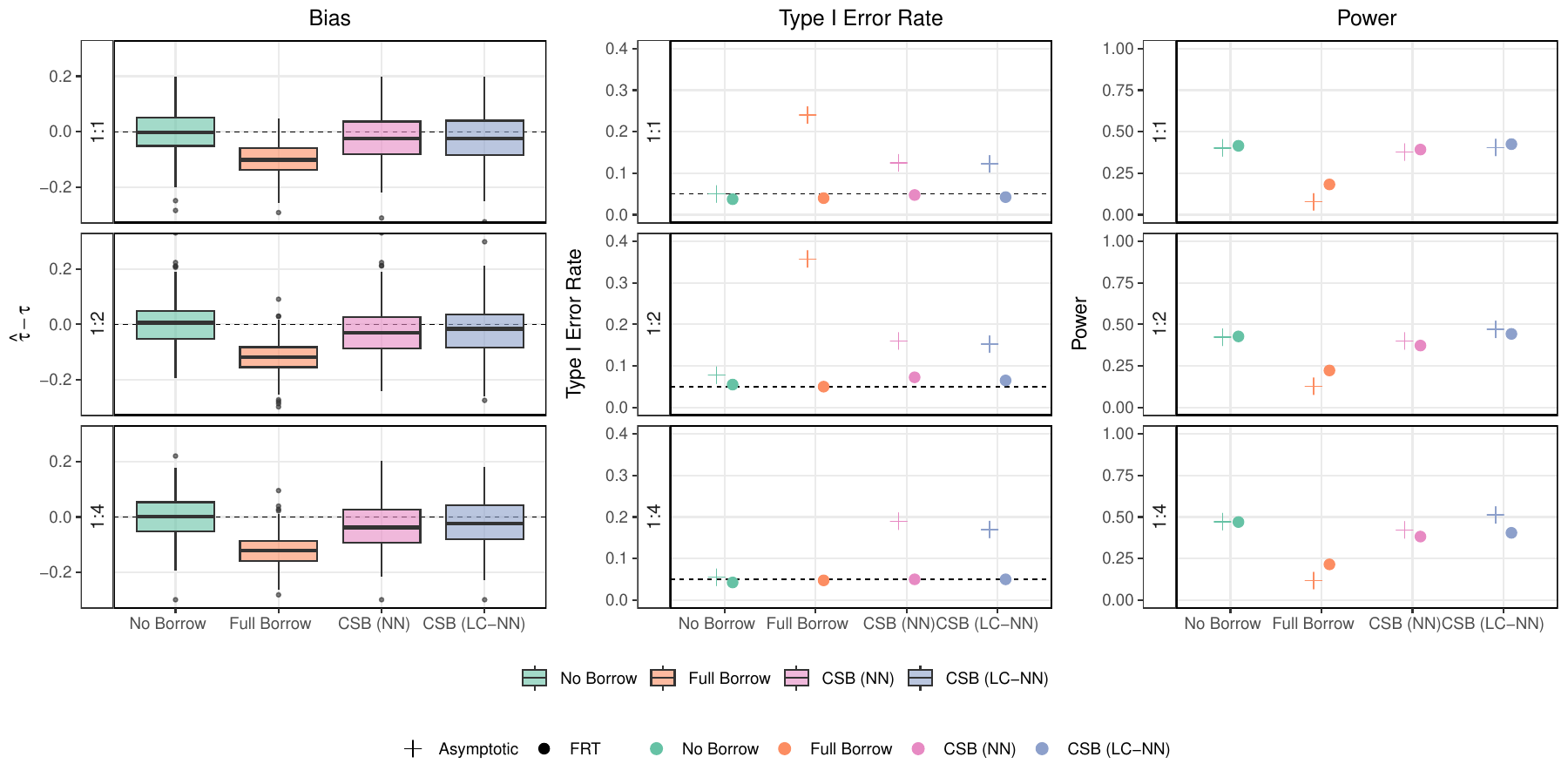}
    \caption{Simulation results when $B_{\gamma}=200$, $n_\mathcal{R}=75$, with wrong SM and correct OM.}
    \label{fig:sim_bootG200_75_SMF_OMT}
\end{figure}

\begin{figure}[!ht]
    \centering
    \includegraphics[width=1\linewidth]{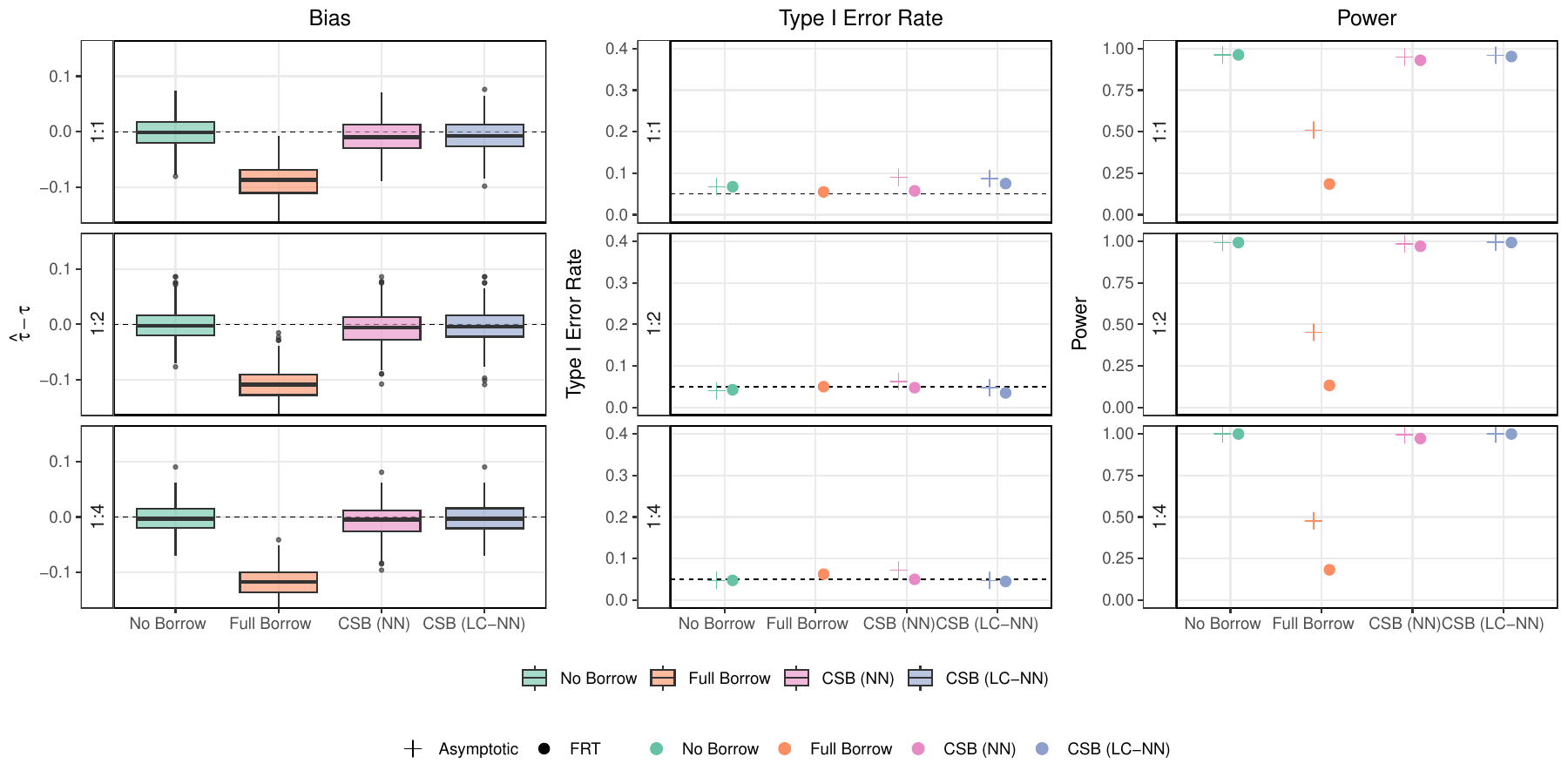}
    \caption{Simulation results when $B_{\gamma}=200$, $n_\mathcal{R}=350$, with wrong SM and correct OM.}
    \label{fig:sim_bootG200_350_SMF_OMT}
\end{figure}

Figures \ref{fig:sim_bootG200_75_SMF_OMT} and \ref{fig:sim_bootG200_350_SMF_OMT} present the corresponding results under sampling model misspecification. In both the small- and moderate-sample settings, \texttt{Borrow AIPW} exhibits clearly larger bias and reduced power, indicating that naive full borrowing is particularly vulnerable when hidden bias is combined with model misspecification. In contrast, \texttt{CSB} and \texttt{No Borrow} remain much more stable. Notably, when the RCT sample size is small, the CSB procedures can still achieve meaningful power improvement over \texttt{No Borrow} without the severe degradation observed under full borrowing. These results further illustrate that the benefit of selective borrowing is not simply borrowing more data, but borrowing external information in a way that is more robust to hidden bias and misspecification.

Taken together, these results suggest that the proposed \texttt{CSB+FRT} framework is reasonably robust across the sample size configurations considered here. Across both small and moderate RCT sizes, \texttt{CSB+FRT} generally achieves a better bias-power tradeoff than naive full borrowing while preserving appropriate type I error control. From a practical perspective, the results also suggest different recommendations depending on the strength of the internal trial. When the RCT sample size is small, selective borrowing can provide meaningful efficiency gains, but the borrowing ratio should still be chosen cautiously, since overly aggressive borrowing may allow the external data to dominate the analysis even under a selective borrowing mechanism. When the RCT sample size is sufficiently large, \texttt{No Borrow} is often already highly competitive, and CSB tends to perform similarly while still guarding against harmful borrowing. These findings reinforce a key practical point: more external data is not always better, and selective borrowing is most useful when it enriches the trial without overwhelming it.

\begin{figure}[!ht]
    \centering
    \includegraphics[width=1\linewidth]{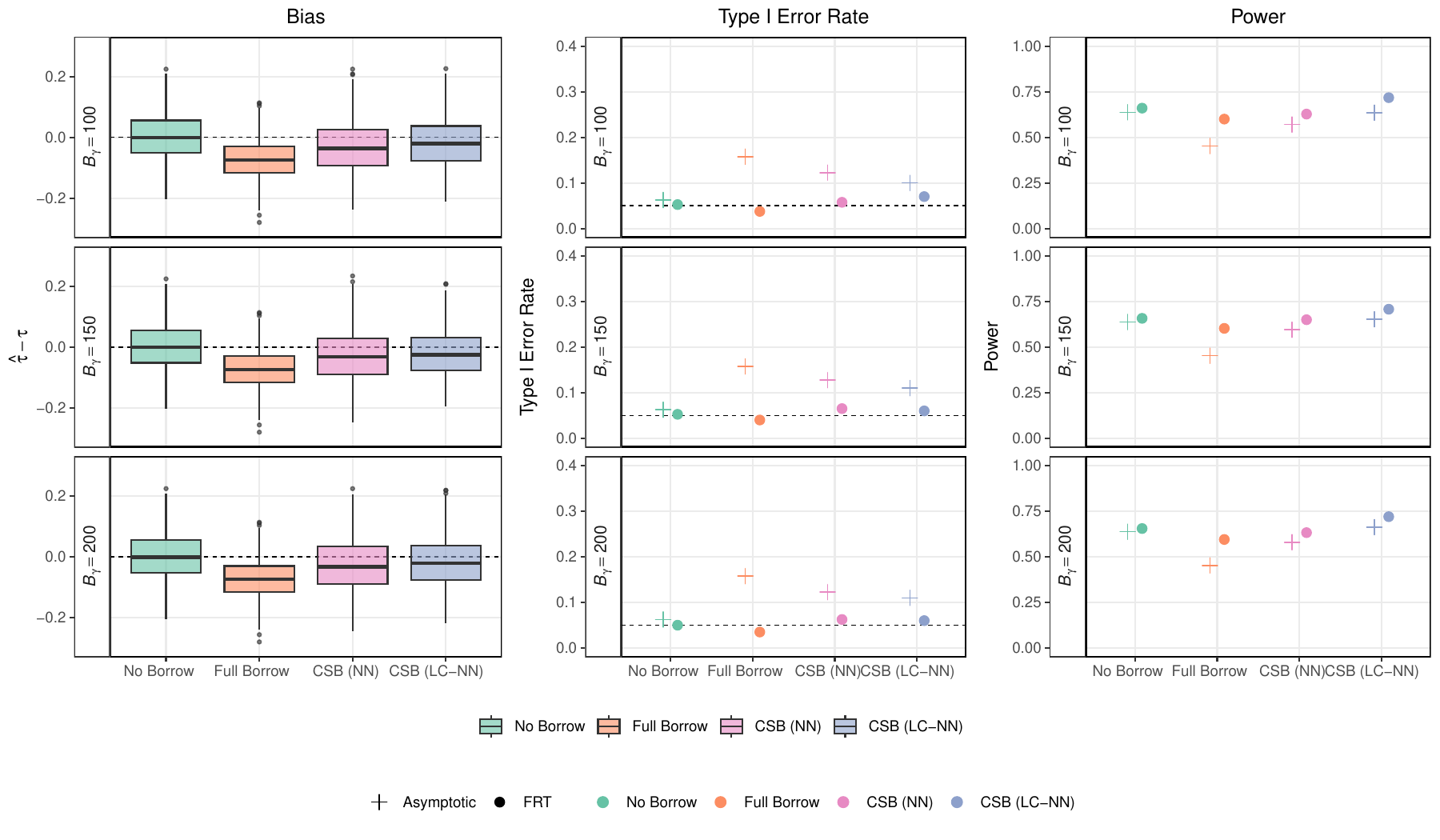}
    \caption{Simulation results across $B_{\gamma}=\{100,150,200\}$, with $n_\mathcal{R}=75$;  matching ratio 1:2; correct SM and correct OM.}
    \label{fig:sim_bootG_compare_SMT_OMT}
\end{figure}

\begin{figure}[!ht]
    \centering
    \includegraphics[width=1\linewidth]{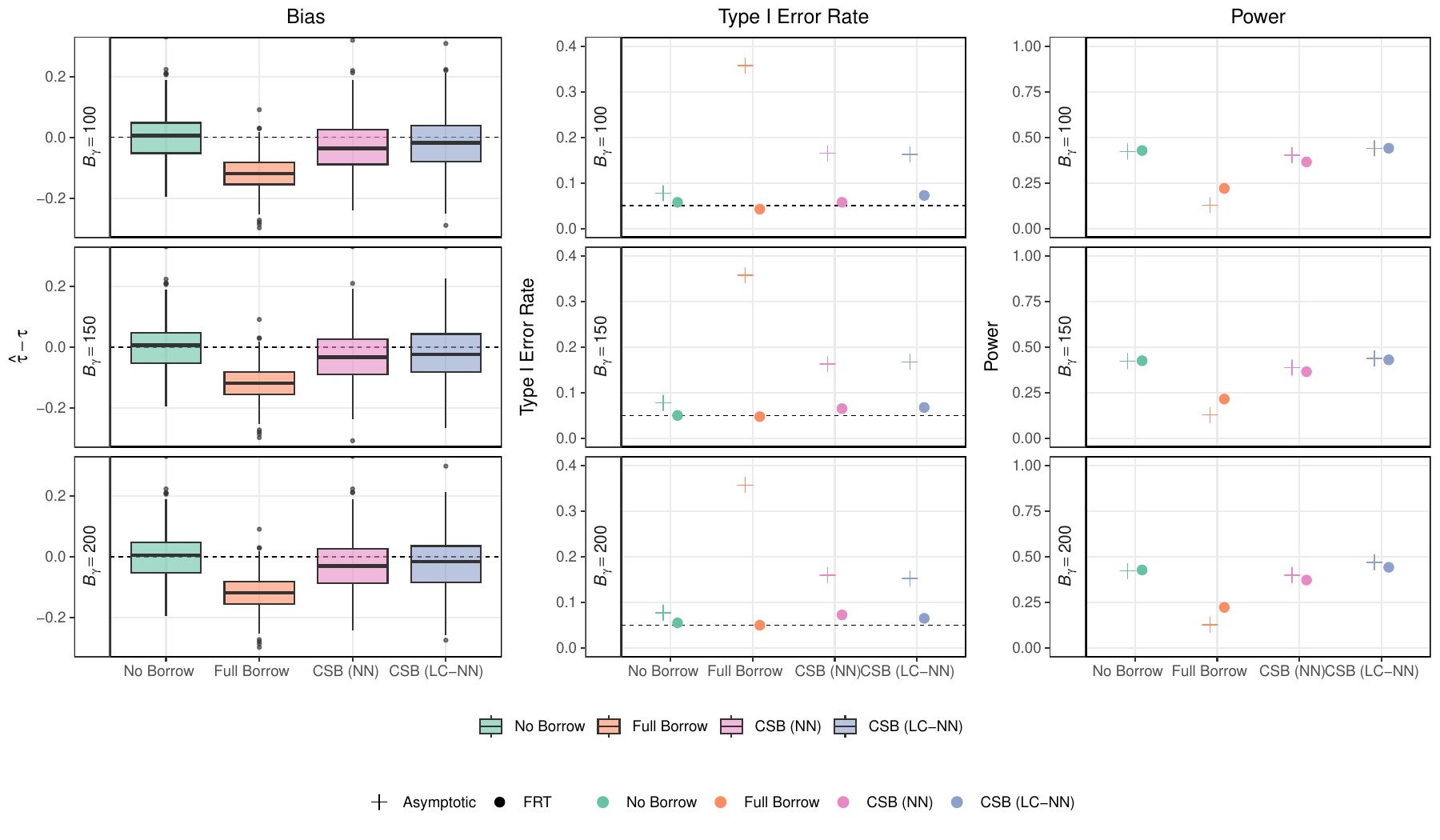}
    \caption{Simulation results across $B_{\gamma}=\{100,150,200\}$, with $n_\mathcal{R}=75$;  matching ratio 1:2; wrong SM and correct OM.}
    \label{fig:sim_bootG_compare_SMF_OMT}
\end{figure}

\begin{figure}[!ht]
    \centering
    \includegraphics[width=1\linewidth]{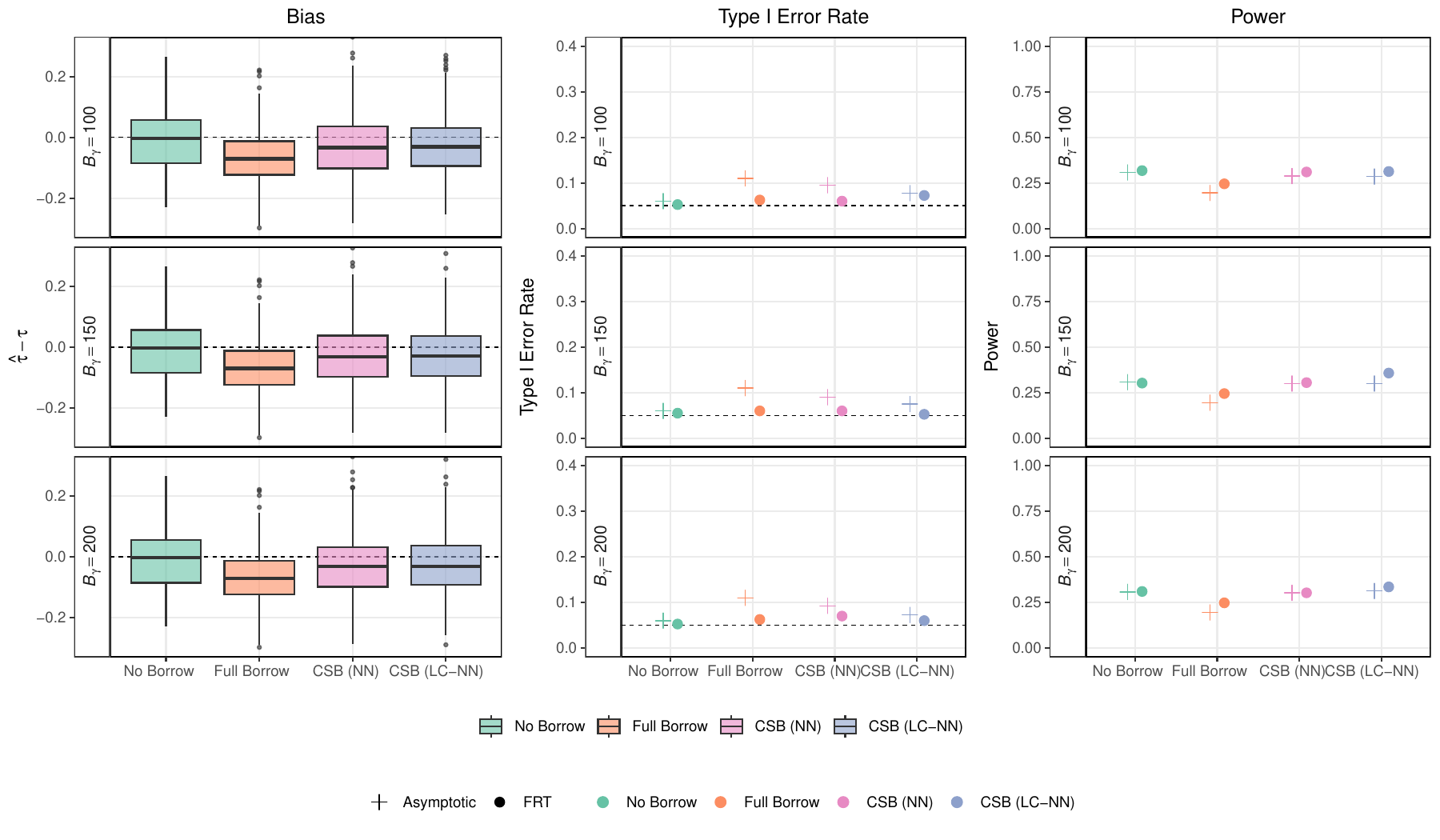}
    \caption{Simulation results across $B_{\gamma}=\{100,150,200\}$, with $n_\mathcal{R}=75$;  matching ratio 1:2; correct SM and wrong OM.}
    \label{fig:sim_bootG_compare_SMT_OMF}
\end{figure}

\begin{figure}[!ht]
    \centering
    \includegraphics[width=1\linewidth]{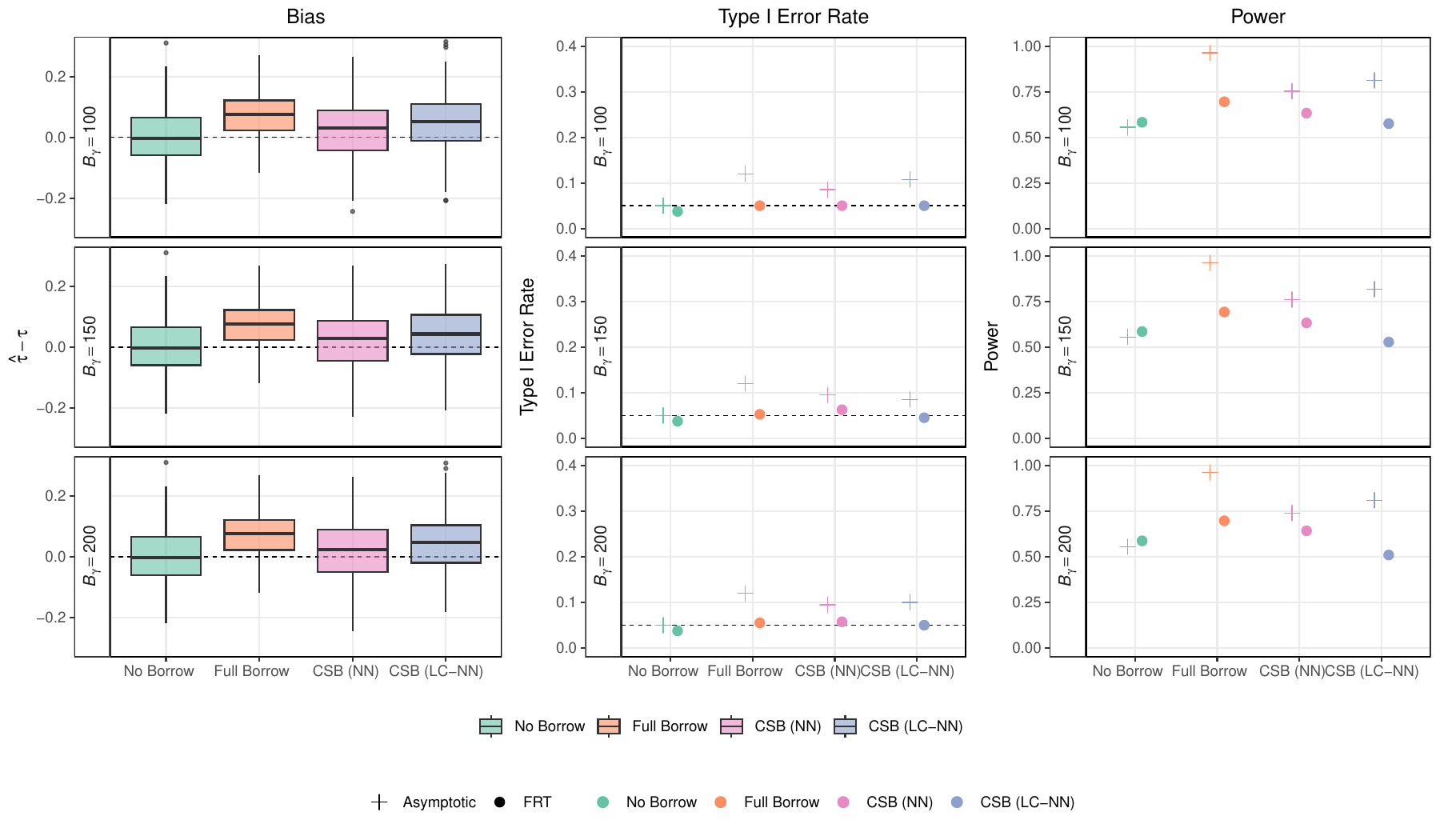}
    \caption{Simulation results across $B_{\gamma}=\{100,150,200\}$, with $n_\mathcal{R}=75$;  matching ratio 1:2; wrong SM and wrong OM.}
    \label{fig:sim_bootG_compare_SMF_OMF}
\end{figure}
To investigate the practical choice of bootstrap size, we additionally considered \(B_{\gamma}\in\{100,150\}\), alongside \(B_{\gamma}=200\), which was used in the main simulations reported in the manuscript. As shown in Figures \ref{fig:sim_bootG_compare_SMT_OMT}--\ref{fig:sim_bootG_compare_SMF_OMF}, the overall qualitative patterns remain similar across model-specification scenarios under this primary simulation setting as the bootstrap size for \(\gamma\)-selection increases. In particular, increasing \(B_{\gamma}\) from 100 to 150 or 200 does not materially change the conclusions regarding bias, type I error control, or power. Nevertheless, a modest improvement is observed when increasing \(B_{\gamma}\) from 100 to 150. For example, within a fixed model-specification scenario such as Figure \ref{fig:sim_bootG_compare_SMT_OMF}, the \texttt{CSB} methods exhibit somewhat greater power gains relative to \texttt{No Borrow} when \(B_{\gamma}\) increases from 100 to 150. In contrast, the additional change from 150 to 200 appears limited, with performance largely stabilizing. These findings suggest that, for the scenarios considered here, \(B_{\gamma}=150\) is sufficient to achieve stable inference, whereas a larger bootstrap size primarily increases computational cost without yielding meaningful practical improvement.

\begin{figure}[!ht]
    \centering
    \includegraphics[width=1\linewidth]{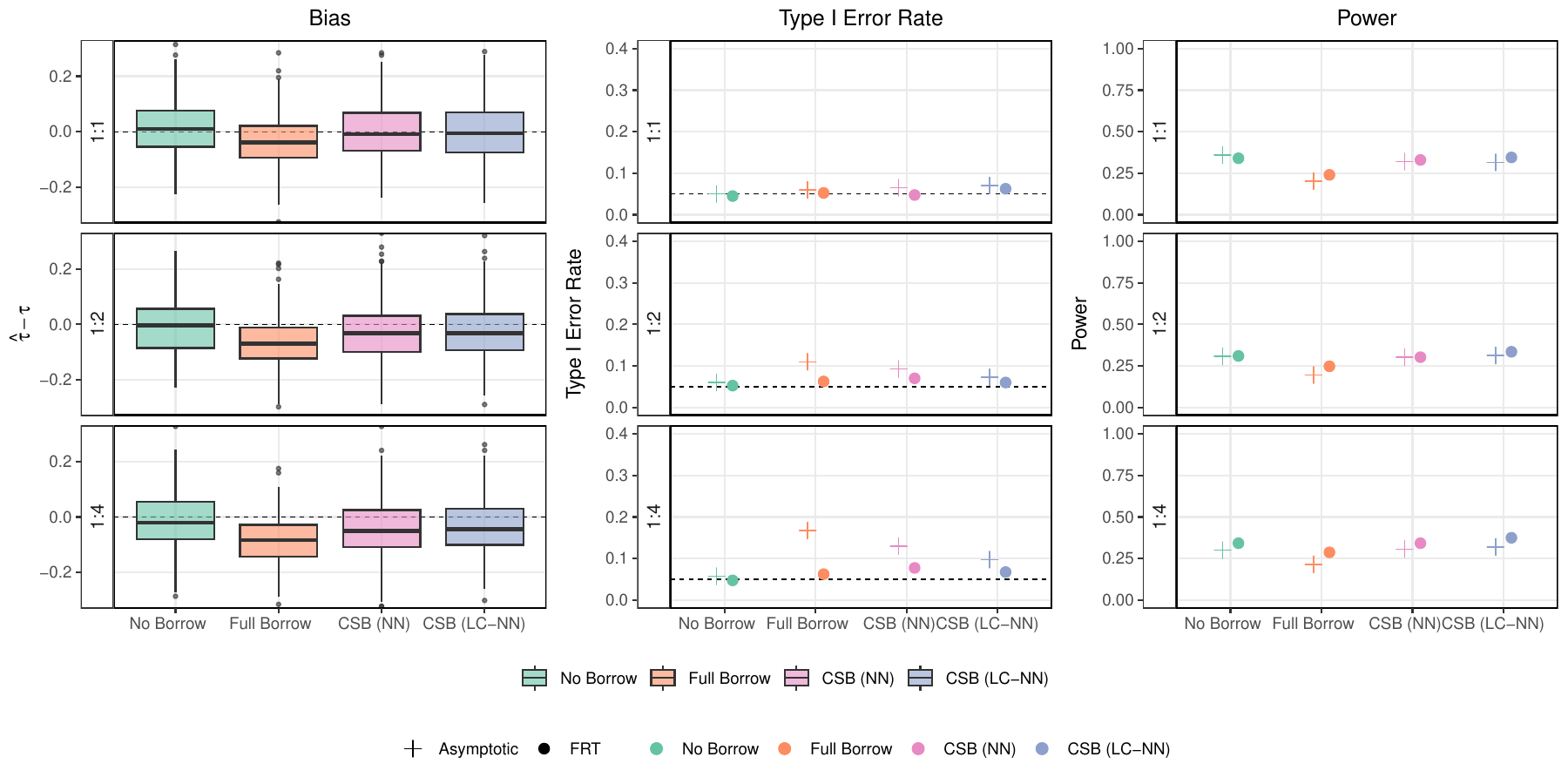}
    \caption{Simulation results when $B_{\gamma}=200$, $n_\mathcal{R}=75$, with correct SM and wrong OM.}
    \label{fig:sim_bootG200_75_SMT_OMF}
\end{figure}

\begin{figure}[!ht]
    \centering
    \includegraphics[width=1\linewidth]{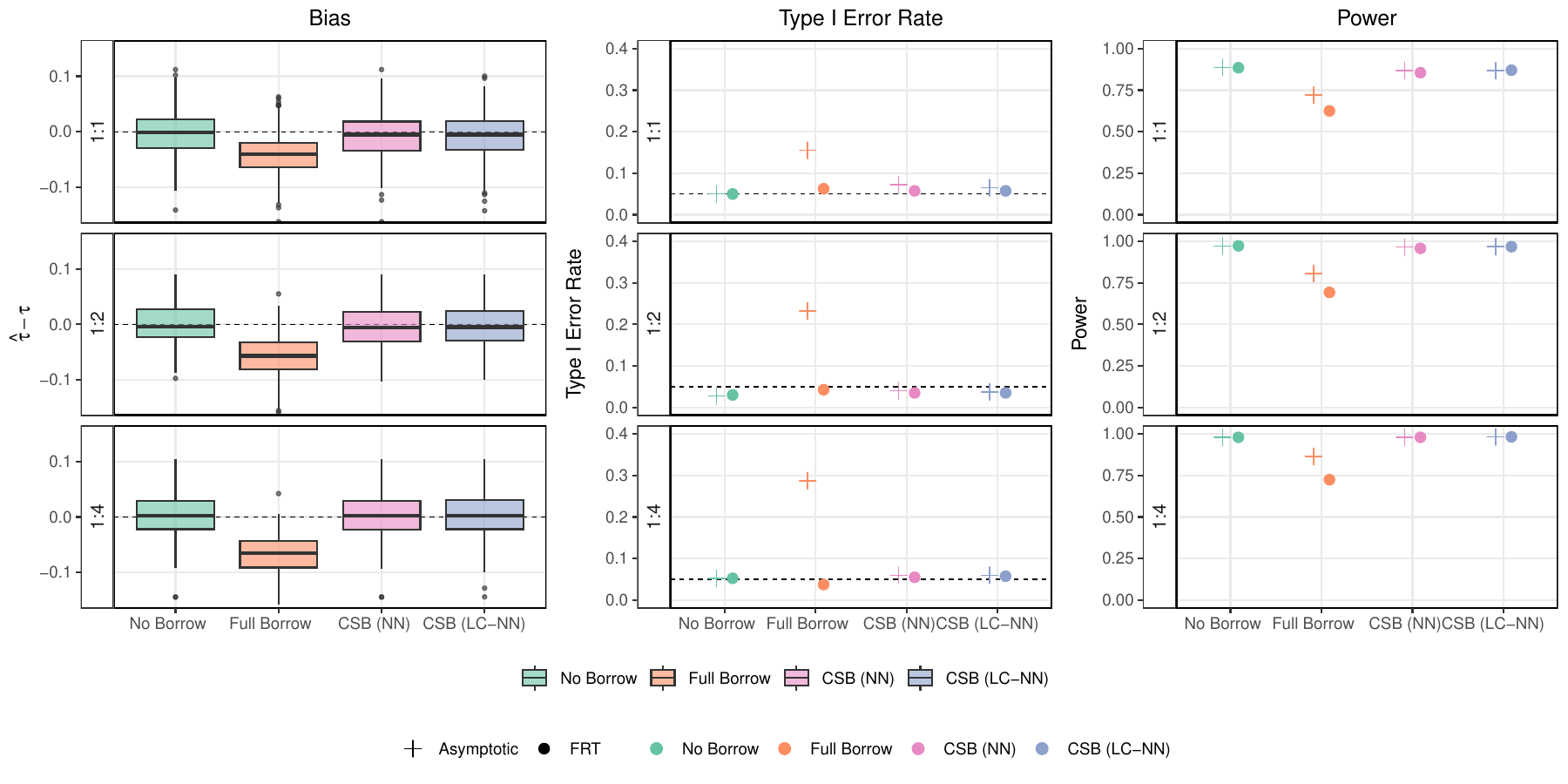}
    \caption{Simulation results when $B_{\gamma}=200$, $n_\mathcal{R}=350$, with correct SM and wrong OM.}
    \label{fig:sim_bootG200_350_SMT_OMF}
\end{figure}

\begin{figure}[!ht]
    \centering
    \includegraphics[width=1\linewidth]{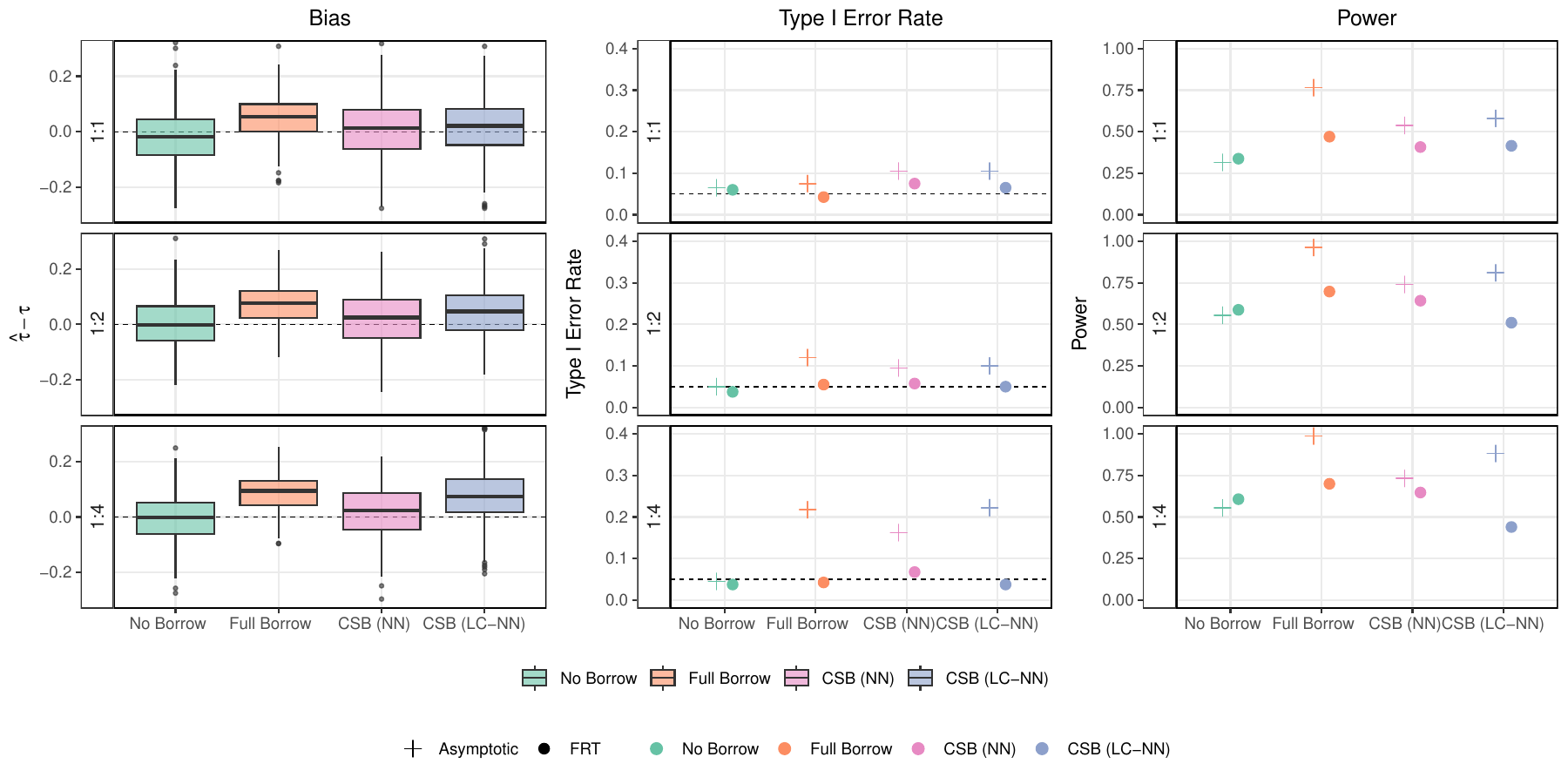}
    \caption{Simulation results when $B_{\gamma}=200$, $n_\mathcal{R}=75$, with wrong SM and wrong OM.}
    \label{fig:sim_bootG200_75_SMF_OMF}
\end{figure}

\begin{figure}[!ht]
    \centering
    \includegraphics[width=1\linewidth]{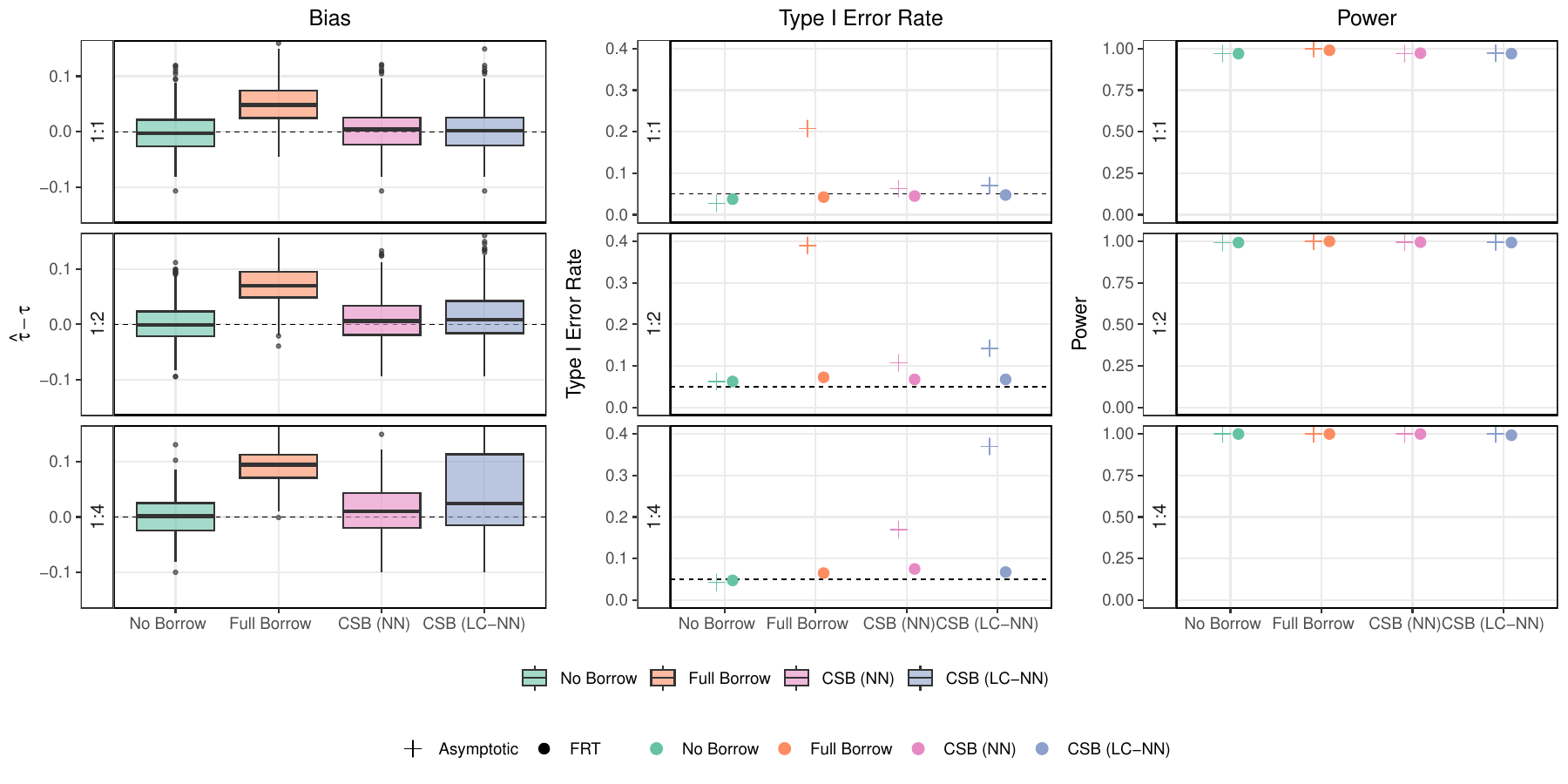}
    \caption{Simulation results when $B_{\gamma}=200$, $n_\mathcal{R}=350$, with wrong SM and wrong OM.}
    \label{fig:sim_bootG200_350_SMF_OMF}
\end{figure}

Results for the remaining scenarios are provided in Figures \ref{fig:sim_bootG200_75_SMT_OMF}--\ref{fig:sim_bootG200_350_SMF_OMF}.

\subsection{Discussion about computational burden}\label{Supp:sim_compu_burden}
To further assess the computational burden of the proposed $\texttt{CSB}$ with FRT framework, we recorded the runtime for each simulation scenario that studied in above Section \ref{Supp:sim_sens_nsize} and summarized the results in Table \ref{tab:runtime_boot}. Overall, runtime increases with the bootstrap size, the RCT sample size, and the amount of external data considered through the matching ratio. This pattern is expected, since larger bootstrap sizes require more repeated resampling, while larger internal and external sample sizes increase the computational cost of both the conformal selection step and the subsequent Fisher randomization test. Notably, these results correspond to the most computationally intensive version of our framework, as they include all core components and recommended procedures, including adaptive $\gamma$ selection. Therefore, they may be viewed as a conservative assessment, or practical upper bound, of the computational burden in routine use.

\begin{table}[htbp]
\centering
\caption{Average runtime (minutes) for FRT procedures under different bootstrap sizes, RCT sample sizes, and matching ratios.}
\label{tab:runtime_boot}
\resizebox{\linewidth}{!}{
\begin{tabular}{cccrrrr}
\toprule
\multirow{2}{*}{Bootstrap ($B_\gamma$)} & \multirow{2}{*}{Sample size ($n_\mathcal{R}$)} & \multirow{2}{*}{Matching ratio} & \multicolumn{4}{c}{Method} \\
\cmidrule(lr){4-7}
 &  &  & No Borrow & Full Borrow & CSB (NN) & CSB (LC-NN) \\
\midrule
\multirow{1}{*}{100}
& 75  & 1:2 & 0.91 & 1.12 & 5.63 & 6.11 \\
\midrule
\multirow{1}{*}{150}
& 75  & 1:2 & 0.96 & 1.19 & 5.92 & 6.38 \\
\midrule
\multirow{6}{*}{200}
& 75  & 1:1 & 0.95 & 1.23 & 6.11 & 6.10 \\
& 75  & 1:2 & 1.10 & 1.37 & 7.04 & 7.54 \\
& 75  & 1:4 & 1.18 & 1.43 & 7.36 & 8.35 \\
& 350 & 1:1 & 0.99 & 1.33 & 6.40 & 7.54 \\
& 350 & 1:2 & 1.07 & 1.58 & 7.50 & 8.74 \\
& 350 & 1:4 & 1.09 & 1.99 & 9.06 & 10.28 \\
\bottomrule
\end{tabular}
}
\end{table}

Despite these expected increases, the runtime remains practically manageable across all settings considered. As shown in Table \ref{tab:runtime_boot}, the \texttt{No Borrow} and \texttt{Full Borrow} procedures typically require about 1-2 minutes, whereas the proposed \texttt{CSB (NN)} and \texttt{CSB (LC-NN)} procedures generally require about 6-10 minutes. The longest runtime occurs in the most computationally demanding setting, namely \(B_\gamma=200\), \(n_{\mathcal{R}}=350\), and matching ratio 1:4, where \texttt{CSB (LC-NN)} takes approximately 10.28 minutes on average. Even in this setting, which combines a relatively large RCT with the most extensive external borrowing considered here, the total runtime remains within a practically reasonable range.

It is important to note that the additional computation required by $\texttt{CSB+FRT}$ is directly tied to its methodological goal. Simpler approaches that borrow all available external data are faster because they do not attempt to detect hidden bias or exclude less comparable external units. In contrast, the proposed framework uses subject-level conformal selection to identify and filter potentially biased external controls before borrowing. This feature is particularly valuable when the internal RCT sample size is limited, where carefully selected borrowing can improve power, whereas full borrowing $\texttt{Borrow AIPW}$ may amplify bias and degrade inferential performance. From this perspective, the additional runtime can be viewed as the computational cost of achieving more robust and selective borrowing, while remaining feasible for realistic applications.

\subsection{Additional simulation under higher-dimensional covariates}\label{Supp:sim_HD_p10}
Although high-dimensional settings may be less common in hybrid controlled trials because the overlapping covariates that are consistently measured and recorded in both the RCT and EC data sources are often limited, we conducted additional simulations with $p=5$ and $p=10$ covariates to evaluate the robustness of the proposed methods beyond the original low-dimensional setting. These settings are larger than what is typically encountered in many HCT applications, where the number of shared and harmonized baseline covariates is usually limited, but they serve as moderate stress tests for the proposed conformal score construction.

\begin{figure}[ht]
    \centering
    \includegraphics[width=0.9\linewidth]{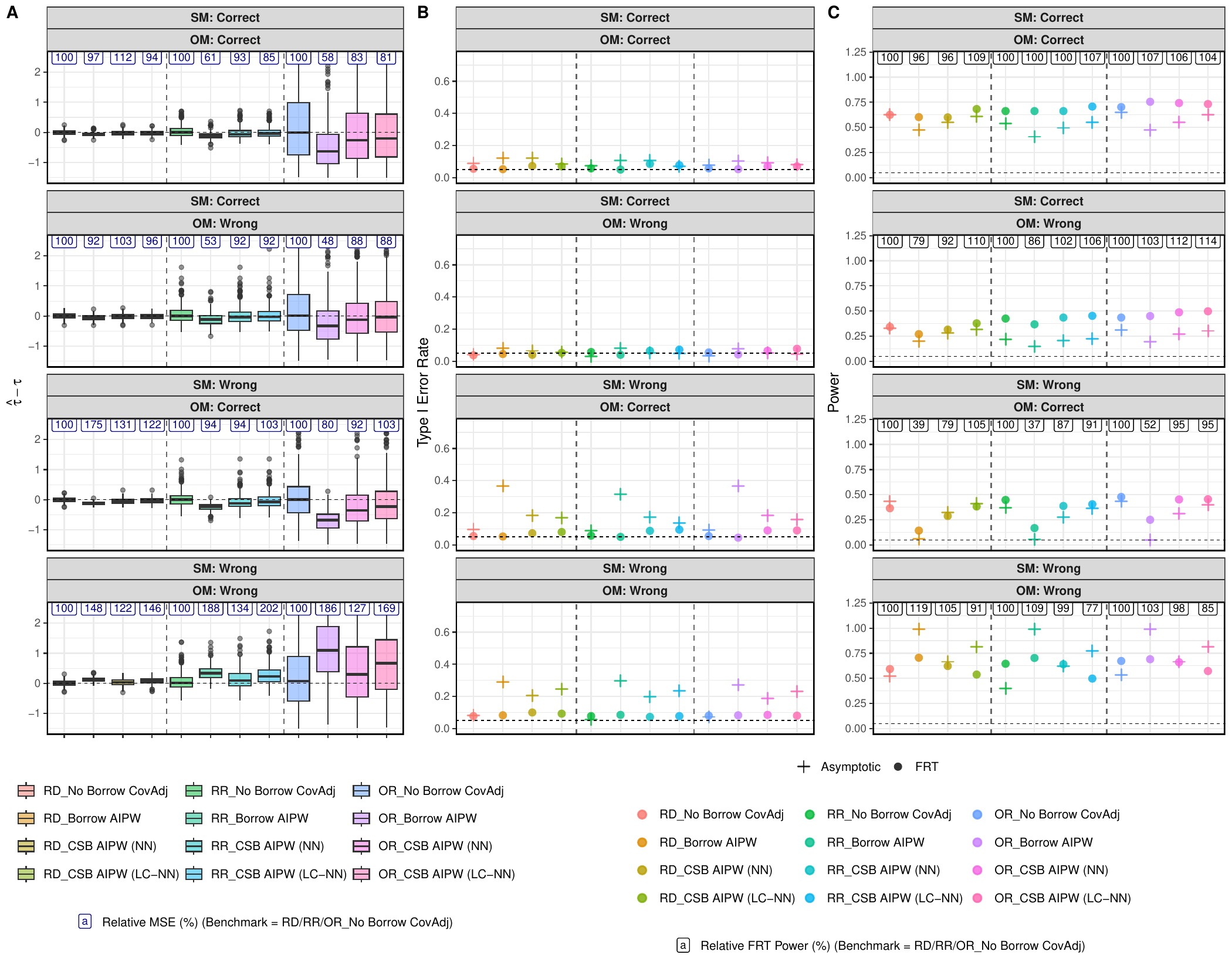}
    \vspace{-10pt}
    \caption{Simulation results for three different estimands given $p=5$ covariates($b=6$)}
    \label{fig:sim_est_inf_6_dimX5}
\end{figure}

\begin{figure}[ht]
    \centering
    \includegraphics[width=0.9\linewidth]{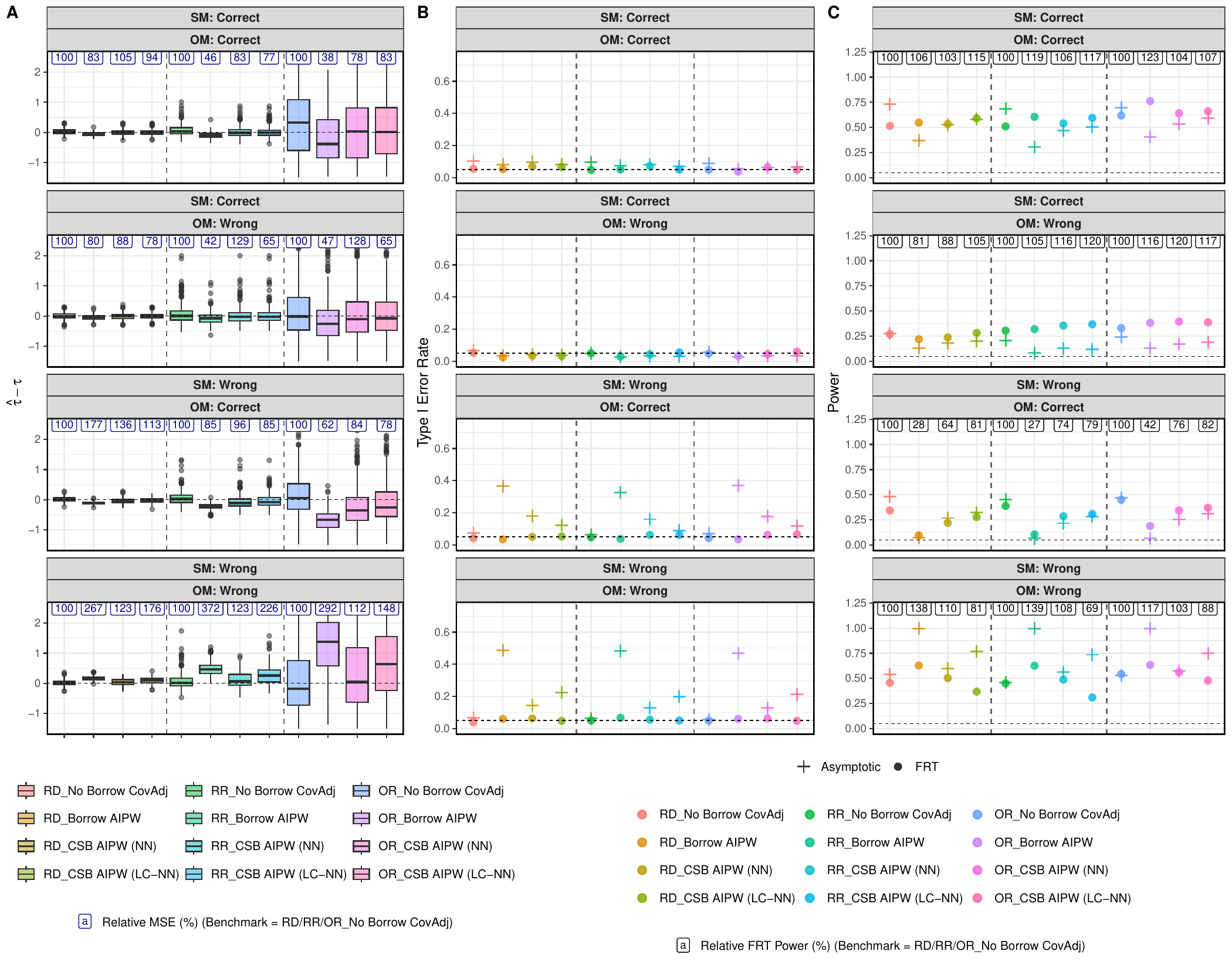}
    \vspace{-10pt}
    \caption{Simulation results for three different estimands given $p=10$ covariates($b=6$)}
    \label{fig:sim_est_inf_6_dimX10}
\end{figure}

The results for $p=5,10$ are shown in Figures \ref{fig:sim_est_inf_6_dimX5} and \ref{fig:sim_est_inf_6_dimX10}, respectively. The overall qualitative conclusions remain consistent with those in the main text. In Panel (A), \texttt{Borrow AIPW} continues to exhibit substantial bias under hidden bias $b=6$, whereas both \texttt{CSB (NN)} and \texttt{CSB (LC-NN)} substantially reduce the bias toward zero. In Panel (B), the type I error rate remains well controlled across methods when using the FRT procedure, indicating that increasing the covariate dimension does not compromise inferential validity. In Panel (C), the \texttt{CSB} methods, especially \texttt{CSB (LC-NN)}, still provide some power improvement over \texttt{No Borrow} in several settings, although the magnitude of the gain is smaller than that observed in lower-dimensional scenarios.

Overall, these results suggest that the proposed \texttt{CSB+FRT} framework remains reasonably robust in moderate-dimensional settings relevant to HCTs. Even at $p=10$, it continues to mitigate hidden bias and preserve type I error control, while still offering efficiency gains in some scenarios. The reduced power gain is consistent with the known deterioration of Euclidean-distance-based nearest-neighbor methods in higher dimensions, and it points to useful future extensions for genuinely high-dimensional settings.

\subsection{Additional simulation under weak null}\label{Supp:sim_weak_null}

We conducted an additional simulation to evaluate the proposed framework under a zero-ATE weak null with treatment effect heterogeneity. In this setting, the average treatment effect satisfies $\mathbb{E}\{Y(1)-Y(0)\}=0$, while individual treatment effects may vary across subjects. This setting is distinct from the sharp null targeted by the FRT, since the null does not require $Y(1)=Y(0)$ for every individual. 

\begin{figure}[ht]
    \centering
    \includegraphics[width=0.9\linewidth]{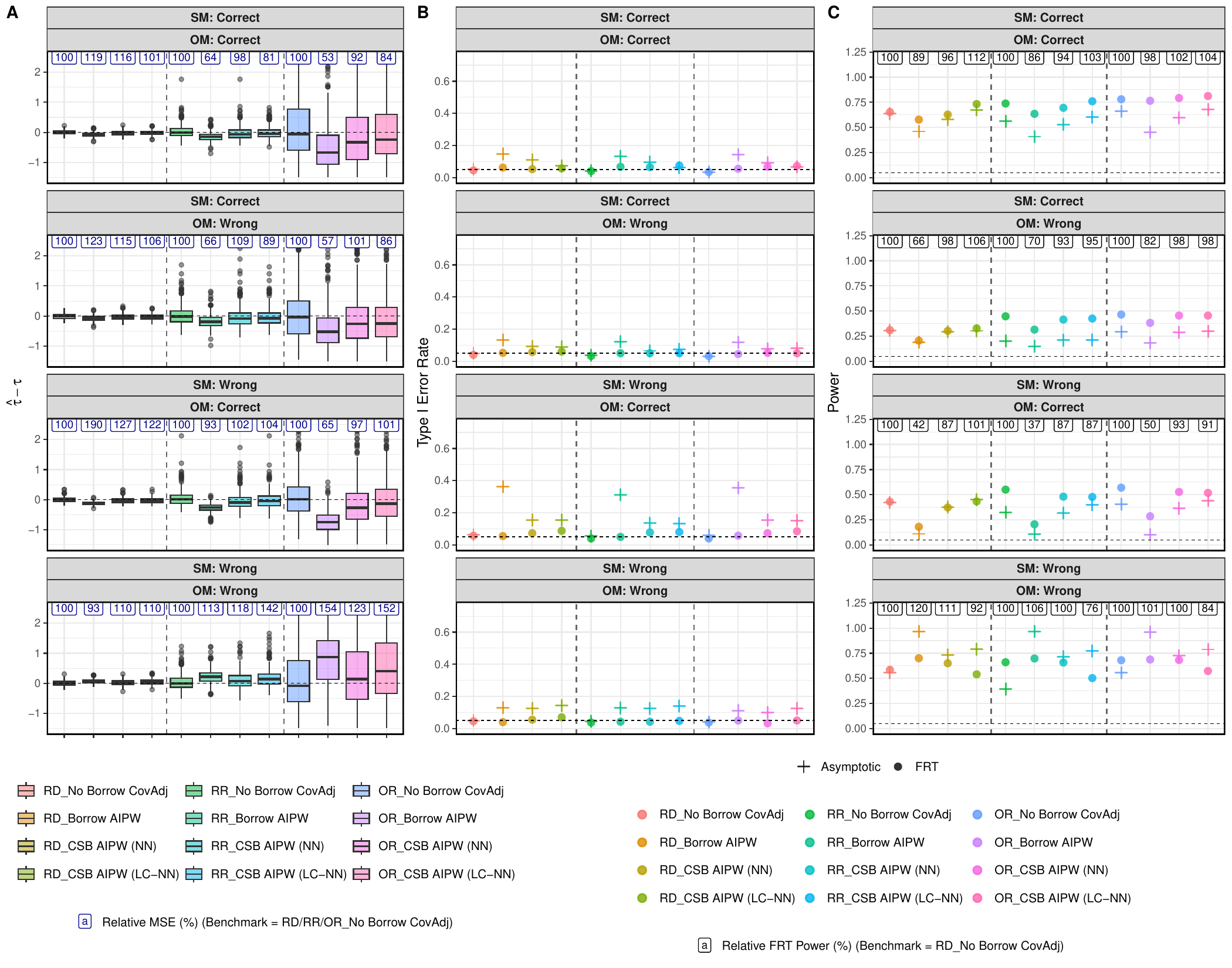}
    \vspace{-10pt}
    \caption{Simulation results for three different estimands under weak null ($b=6$)}
    \label{fig:sim_est_inf_6_wknull_hetero}
\end{figure}

Figure~\ref{fig:sim_est_inf_6_wknull_hetero} presents the results. Full borrowing \texttt{Borrow AIPW} can show substantial bias when all external controls are incorporated, particularly when the outcome model for the external controls is misspecified. In contrast, \texttt{CSB AIPW} reduces this bias by excluding external controls that appear less consistent with the RCT controls. This leads to improved type I error control under asymptotic inference compared with \texttt{Borrow AIPW}. Although FRT is primarily motivated by the sharp null, it also shows reasonably good type I error control in this weak-null setting. In terms of power, selective borrowing generally preserves or improves efficiency relative to no borrowing, especially when the selection and outcome models are correctly specified.

Therefore, when the ATE is the primary estimand, \texttt{CSB AIPW} with asymptotic inference may be considered a useful inferential approach, with validity supported by double robustness conditions together with the selection consistency of CSB.

\section{Additional Case Study Results}
\subsection{Summary table for hybrid controlled dataset}\label{Supp:case_Table1s}
\begin{table}[ht]
\centering
\caption{335 CALGB 9633 + 335 NCDB: Patient Characteristics}
\label{tab:baseline}
\begin{adjustbox}{max width=\textwidth}
\begin{tabular}{llccc}
\toprule
\textbf{}  & \textbf{C9633 controlled} & \textbf{C9633 treated} & \textbf{NCDB controlled} & \textbf{Total} \\
\textbf{}  & \textbf{(N=168)} & \textbf{(N=167)} & \textbf{(N=335)} & \textbf{(N=670)} \\
\midrule
\textbf{Sex}                 &                &                &                &                \\
                    \hspace{1em}Male   & 106 (63.1\%) & 109 (65.3\%) & 225 (67.2\%) & 440 (65.7\%) \\
                    \hspace{1em}Female &  62 (36.9\%) &  58 (34.7\%) & 110 (32.8\%) & 230 (34.3\%) \\
\textbf{Age (years)}           &                &                &                &                \\
                    \hspace{1em}Mean (SD) & 61.2 (9.28)  & 60.4 (10.2)  & 61.0 (9.73)  & 60.9 (9.73)  \\
                    \hspace{1em}Median [Min, Max] & 62.0 [40.0, 81.0] & 61.0 [34.0, 78.0] & 62.0 [34.0, 81.0] & 61.0 [34.0, 81.0] \\
\textbf{Race}               &                &                &                &                \\
                    \hspace{1em}White  & 148 (88.1\%) & 151 (90.4\%) & 311 (92.8\%) & 610 (91.0\%) \\
                    \hspace{1em}Non-white &  20 (11.9\%) &  16 (9.6\%) &  24 (7.2\%) &  60 (9.0\%) \\
\textbf{Histology}           &                &                &                &                \\
                    \hspace{1em}Squamous &  65 (38.7\%) &  66 (39.5\%) & 131 (39.1\%) & 262 (39.1\%) \\
                    \hspace{1em}Other & 103 (61.3\%) & 101 (60.5\%) & 204 (60.9\%) & 408 (60.9\%) \\
\textbf{Tumor Size (Diameter/cm)}           &                &                &                &                \\
                    \hspace{1em}Mean (SD) &  4.56 (2.05) &  4.60 (2.04) &  5.10 (1.62) &  4.84 (1.86) \\
                    \hspace{1em}Median [Min, Max] & 4.00 [1.00, 12.0] & 4.00 [1.00, 12.0] & 4.80 [3.10, 12.0] & 4.50 [1.00, 12.0] \\
\bottomrule
\end{tabular}
\end{adjustbox}
\end{table}
After preprocessing the dataset for the primary analysis in the case study, we present a summary table of baseline characteristics across the three study arms. As shown in Table \ref{tab:baseline}, the baseline covariates are well balanced among the RCT treated, RCT controlled, and external control (EC) groups, indicating that the dataset following preprocessing and matching is suitable for subsequent analyses.

\subsection{Supplementary analysis for varying sample sizes and allocation ratios}\label{sec:caseRes2}
To assess the applicability of the proposed methods across a broader range of practical scenarios, we examine nine settings that vary by RCT sample size, allocation ratio, and EC size. Specifically, we consider two RCT sample sizes: a small sample $(\nrct=75)$ with allocation ratios of 1:1 and 2:1, and a moderate sample $(\nrct=335)$ with a 1:1 allocation ratio. For each setting, the number of external control subjects varies, with $\nec\in \{75, 150, 300\}$ for the small RCT and $\nec\in \{335, 670, 1005\}$ for the moderate RCT. 

For moderate HCT scenarios, we retain all CALGB 9633 participants as the RCT sample and again apply nearest-neighbor matching with varying ratios to select subsets from NCDB as ECs. For the analysis of small HCTs, we randomly sample subsets from CALGB 9633. For a 1:1 allocation, we select $\nrct=75$ CALGB 9633 patients as RCT, preserving the original trial design. To reflect real-world situations, such as ethical concerns, cost, and patient willingness to be randomized \citep{Sibbald1998,Dumville2006,Deaton2018}, we additionally include a 2:1 ratio. Specifically, we sample $n_1=50$ treated and $n_0=25$ controlled from CALGB 9633. For each of these RCT datasets, we apply nearest-neighbor matching using three different matching ratios to construct the corresponding EC datasets. Finally, to study how $p$-values respond to increasing EC size, we vary the matching ratio to incorporate different volumes of external control data.




\begin{table}[ht]
\caption{Results of case study ($n_1=50, n_0=25, \nec=150$)}
\label{tab:resC2}
\centering
\resizebox{\linewidth}{!}{
\begin{tabular}{llllrrrr}
\hline
 & \multicolumn{1}{c}{} & \multicolumn{3}{c}{Asymptotic Inference} & \multicolumn{1}{c}{FRT} &  \\ \cline{3-5}
Method & Point Est.& SE$^a$ & 95\% CI & $p$-value & $p$-value & Num. of EC$^b$ & ESS$^c$ of EC \\
\hline
No Borrow Unadj & 0.180 & 0.112 & (-0.040, 0.400) & 0.109 & 0.096 & 0 & 0\\
No Borrow CovAdj & 0.167 & 0.120 & (-0.067, 0.402) & 0.162 & 0.128 & 0 & 0\\
Conformal Selective Borrow NN & 0.185 & 0.075 & (0.039, 0.332) & 0.013 & 0.057 & 150 & 144\\
Conformal Selective Borrow LC-NN & 0.185 & 0.075 & (0.039, 0.332) & 0.013 & 0.056 & 150 & 144\\
Borrow Na\"ive  & 0.178 & 0.068 & (0.045, 0.311) & 0.009 & 0.071 & 150 & 150 \\
Borrow IPW  & 0.187 & 0.068 & (0.053, 0.321) & 0.007 & 0.053 & 150 & 144 \\
Borrow CW  & 0.181 & 0.070 & (0.043, 0.319) & 0.009 & 0.052 & 150 & 142 \\
Borrow OM  & 0.185 & 0.070 & (0.048, 0.321) & 0.007 & 0.051 & 150 & 150 \\
Borrow AIPW  & 0.185 & 0.075 & (0.039, 0.332) & 0.013 & 0.052 & 150 & 144 \\
Borrow ACW  & 0.180 & 0.075 & (0.034, 0.327) & 0.016 & 0.064 & 150 & 142 \\
\hline
\end{tabular}
}

\raggedright
\footnotesize{$^a$ SEs of No Borrow Unadj, Borrow CW, Borrow OM, and Borrow IPW are obtained from Bootstrap, while that of other methods are derived from EIF; $^b$ Number of EC subjects borrowed; $^c$ Effective Sample Size}

\end{table}

For example, results for a small HCT with unequal allocation and 1:2 matching ($\{n_1=50, n_0=25, \nec=150\}$) are provided in Table \ref{tab:resC2}, with additional scenarios provided in Section \ref{Supp:case_summaryTables}. Similar pattern as primary analysis can be seen in Table \ref{tab:resC2}. All six Borrow methods yield sharply lower asymptotic $p$-values at around 0.01 compared to larger than 0.1 for No Borrow methods. Under FRT, the $p$-values for CSB and Borrow methods also decrease but not by much. As expected, CSB methods yield point estimates between No Borrow and Borrow, reflecting their selective use of EC data, as they neither entirely keep nor discard the EC set. When no hidden bias is detected, CSB may retain all EC subjects, producing inference results similar to Borrow methods that adjust for covariate imbalance.



\begin{figure}[ht]
\includegraphics[width=\textwidth]{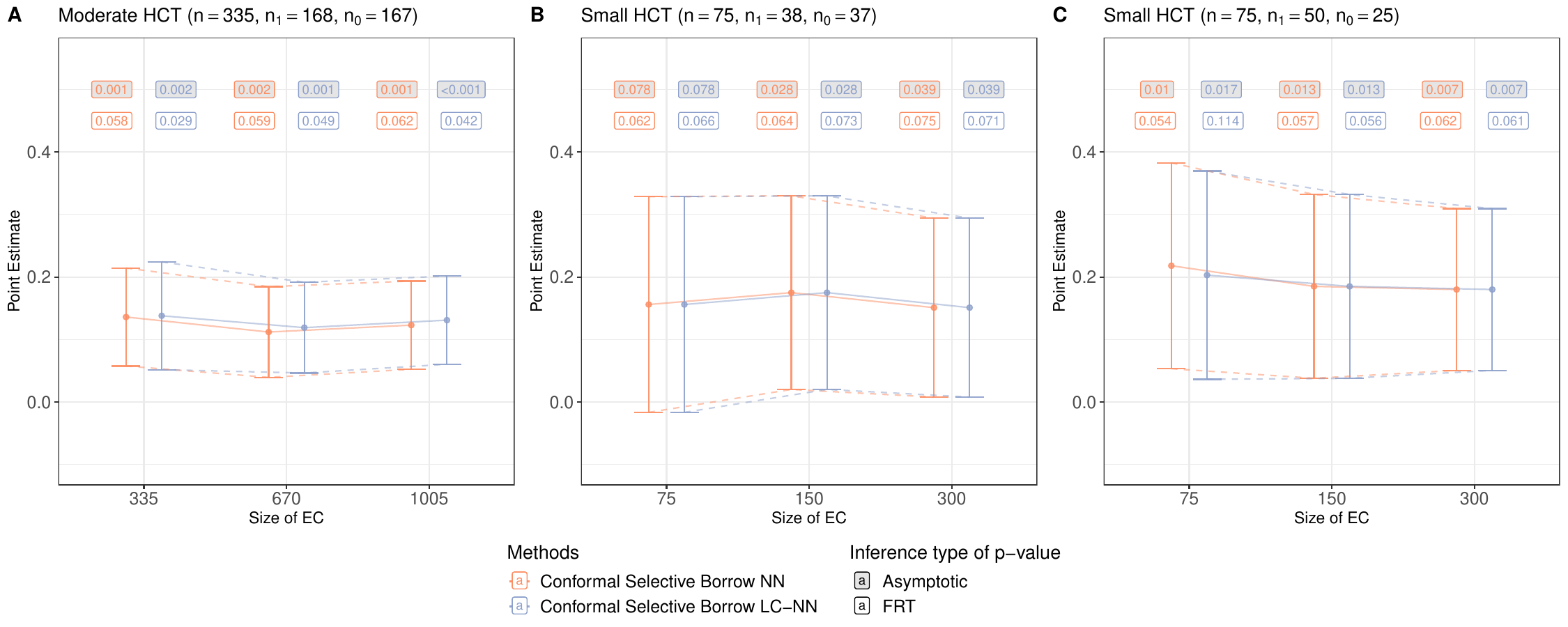}
\caption{Change of estimates as size of EC increases}
\label{fig:rd_estimates}
\end{figure}

Finally, we explore how the results change as size of EC increases. For simplicity, each group of methods has one typical approach to represent in Figure \ref{fig:rd_estimates}, the comparisons within each group are provided in Section \ref{Supp:case_sizeECchange}. In general, the plots indicate that the point estimates are stable as the size of EC increases when doing \texttt{Borrow AIPW}, \texttt{CSB LC-NN}, and \texttt{No Borrow CovAdj} for all the scenarios. The variances of \texttt{No Borrow CovAdj} are larger than the  Borrow and CSB methods. As borrowing more information from EC, the variances declines for both \texttt{Borrow AIPW} and \texttt{CSB LC-NN}. In Figure \ref{fig:rd_estimates} (A), \texttt{CSB LC-NN} has a point estimate between \texttt{Borrow AIPW} and \texttt{No Borrow CovAdj}, with variance slightly larger than \texttt{Borrow AIPW} but noticeably smaller than \texttt{No Borrow CovAdj}. Similar pattern can be seen in Figure \ref{fig:rd_estimates} (C), while the variance of \texttt{CSB LC-NN} seems equal to \texttt{Borrow AIPW}. In Figure \ref{fig:rd_estimates} (B), \texttt{CSB LC-NN} keeps all the subjects from EC and thus has an overlap pattern with \texttt{Borrow AIPW}. 

Within the same setting, all the Borrow methods leads to smaller asymptotic inference $p$-values and FRT $p$-values compared to No Borrow, which implies an efficiency gain after enriching the RCT data with EC. Secondly,FRT $p$-values exhibit a stable pattern as the size of EC increases. The FRT $p$-values does not keep decreasing as borrowing more outside information, which can be explained by that the randomization of FRT approach is within the RCT data \citep{Ding2024}. This stability of FRT is consistent with the expectation that EC will not be allowed to dominate target population RCT, and protects inference results from the bias introduced by EC data. Thirdly, the asymptotic $p$-values are more sensitive to borrowing ECs, which fall sharply even when borrowing the same number of external controlled patients as the RCT. For an instance, when constructing a small HCT with unequal allocation ratio, the asymptotic $p$-values decrease from 0.162 to 0.01. Therefore, the validity of asymptotic inference is sensitive to the size of ECs and require extra caution when using asymptotic methods in small HCTs. 

\subsection{Estimation results for primary analysis}\label{Supp:case_summaryTables_prim}
In this section, we provide the detailed estimation results for OR and RR results in the primary analysis for all the other scenarios in Table \ref{tab:resC1_rr} and Table \ref{tab:resC1_or}. 

\begin{table}[ht]
\caption{Results of case study (RR, $n_1=167, n_0=168, \nec=335$)}
\label{tab:resC1_rr}
\centering
\resizebox{\linewidth}{!}{
\renewcommand{\arraystretch}{1}
\begin{tabular}{llllrrrrr}
\hline
 & \multicolumn{1}{c}{} & \multicolumn{3}{c}{Asymptotic Inference} & \multicolumn{1}{c}{FRT} &  &  \\ \cline{3-5}
Method & Point Est. & SE & 95\% CI & $p$-value & $p$-value & Num. of EC & ESS of EC & FRT Runtime (s)\\
\hline
No Borrow DiM & 1.110 & 0.073 & (0.975, 1.260) & 0.116 & 0.092 & 0 & 0 & 0.001 \\
No Borrow CovAdj & 1.120 & 0.075 & (0.979, 1.270) & 0.100 & 0.079 & 0 & 0 & 22.026 \\
Conformal Selective Borrow NN & 1.210 & 0.066 & (1.080, 1.340) & 0.001 & 0.065 & 302 & 294 & 57.478 \\
Conformal Selective Borrow LC-NN & 1.200 & 0.073 & (1.070, 1.350) & 0.002 & 0.038 & 144 & 138 & 64.492 \\
Borrow Na\"ive  & 1.240 & 0.068 & (1.120, 1.380) & $<$0.001 & 0.055 & 335 & 335 & 36.916 \\
Borrow IPW  & 1.230 & 0.067 & (1.100, 1.370) & $<$0.001 & 0.060 & 335 & 315 & 18.059 \\
Borrow CW  & 1.230 & 0.066 & (1.100, 1.360) & $<$0.001 & 0.055 & 335 & 313 & 23.003 \\
Borrow OM  & 1.240 & 0.066 & (1.110, 1.370) & $<$0.001 & 0.043 & 335 & 335 & 29.474 \\
Borrow AIPW  & 1.240 & 0.068 & (1.110, 1.380) & $<$0.001 & 0.048 & 335 & 315 & 37.312 \\
Borrow ACW  & 1.230 & 0.067 & (1.100, 1.370) & $<$0.001 & 0.046 & 335 & 313 & 44.060 \\
\hline
\end{tabular}
}
\end{table}

\begin{table}[ht]
\caption{Results of case study (OR, $n_1=167, n_0=168, \nec=335$)}
\label{tab:resC1_or}
\centering
\resizebox{\linewidth}{!}{
\renewcommand{\arraystretch}{1}
\begin{tabular}{llllrrrrr}
\hline
 & \multicolumn{1}{c}{} & \multicolumn{3}{c}{Asymptotic Inference} & \multicolumn{1}{c}{FRT} &  &  \\ \cline{3-5}
Method & Point Est. & SE & 95\% CI & $p$-value & $p$-value & Num. of EC & ESS of EC & FRT Runtime (s)\\
\hline
No Borrow DiM & 1.480 & 0.395 & (0.877, 2.500) & 0.142 & 0.055 & 0 & 0 & 0.001 \\
No Borrow CovAdj & 1.520 & 0.389 & (0.923, 2.510) & 0.100 & 0.051 & 0 & 0 & 22.026 \\
Conformal Selective Borrow NN & 1.930 & 0.409 & (1.270, 2.920) & 0.002 & 0.057 & 302 & 294 & 57.478 \\
Conformal Selective Borrow LC-NN & 1.900 & 0.429 & (1.220, 2.960) & 0.004 & 0.032 & 144 & 138 & 64.492 \\
Borrow Na\"ive  & 2.060 & 0.419 & (1.380, 3.070) & $<$0.001 & 0.057 & 335 & 335 & 36.916 \\
Borrow IPW  & 2.000 & 0.446 & (1.290, 3.100) & 0.002 & 0.058 & 335 & 315 & 18.059 \\
Borrow CW  & 1.990 & 0.436 & (1.300, 3.060) & 0.002 & 0.055 & 335 & 313 & 23.003 \\
Borrow OM  & 2.050 & 0.443 & (1.340, 3.130) & 0.001 & 0.044 & 335 & 335 & 29.474 \\
Borrow AIPW  & 2.050 & 0.431 & (1.360, 3.100) & 0.001 & 0.048 & 335 & 315 & 37.312 \\
Borrow ACW  & 2.020 & 0.425 & (1.340, 3.060) & 0.001 & 0.047 & 335 & 313 & 44.060 \\
\hline
\end{tabular}
}
\end{table}

\subsection{Estimation results for supplementary analysis}\label{Supp:case_summaryTables}
In this section, we provide estimation results for all the other varying sample sizes and allocation ratios scenarios for Supplementary Analysis in Table \ref{tab:resC3} - \ref{tab:resC9}. 

\begin{table}[ht]
\caption{Results of case study ($n_1=38, n_0=37, \nec=75$)}
\label{tab:resC3}
\centering
\resizebox{\linewidth}{!}{
\renewcommand{\arraystretch}{1}
\begin{tabular}{llllrrrr}
\hline
 & \multicolumn{1}{c}{} & \multicolumn{3}{c}{Asymptotic Inference} & \multicolumn{1}{c}{FRT} &  \\ \cline{3-5}
Method & Point Est.& SE & 95\% CI & $p$-value & $p$-value & Num. of EC & ESS of EC \\
\hline
No Borrow Unadj & 0.152 & 0.105 & (-0.053, 0.358) & 0.147 & 0.206 & 0 & 0\\
No Borrow CovAdj & 0.132 & 0.118 & (-0.099, 0.362) & 0.263 & 0.240 & 0 & 0\\
Conformal Selective Borrow NN & 0.156 & 0.088 & (-0.017, 0.328) & 0.078 & 0.062 & 75 & 73\\
Conformal Selective Borrow LC-NN & 0.156 & 0.088 & (-0.017, 0.328) & 0.078 & 0.066 & 75 & 73\\
Borrow Na\"ive & 0.154 & 0.082 & (-0.007, 0.314) & 0.061 & 0.084 & 75 & 75 \\
Borrow IPW & 0.142 & 0.083 & (-0.021, 0.306) & 0.082 & 0.096 & 75 & 73\\
Borrow CW  & 0.137 & 0.084 & (-0.028, 0.303) & 0.096 & 0.068 & 75 & 73\\
Borrow OM & 0.156 & 0.084 & (-0.009, 0.320) & 0.056 & 0.065 & 75 & 75\\
Borrow AIPW  & 0.156 & 0.088 & (-0.017, 0.329) & 0.078 & 0.068 & 75 & 73\\
Borrow ACW & 0.147 & 0.088 & (-0.026, 0.319) & 0.096 & 0.074 & 75 & 73\\
\hline
\end{tabular}
}
\end{table}

\begin{table}[ht]
\caption{Results of case study ($n_1=38, n_0=37, \nec=150$)}
\label{tab:resC4}
\centering
\resizebox{\linewidth}{!}{
\renewcommand{\arraystretch}{1}
\begin{tabular}{llllrrrr}
\hline
 & \multicolumn{1}{c}{} & \multicolumn{3}{c}{Asymptotic Inference} & \multicolumn{1}{c}{FRT} &  \\ \cline{3-5}
Method & Point Est.& SE & 95\% CI & $p$-value & $p$-value & Num. of EC & ESS of EC \\
\hline
No Borrow Unadj & 0.152 & 0.105 & (-0.053, 0.358) & 0.147 & 0.206 & 0 & 0 \\
No Borrow CovAdj & 0.132 & 0.118 & (-0.099, 0.362) & 0.263 & 0.226 & 0 & 0 \\
Conformal Selective Borrow NN & 0.175 & 0.079 & (0.019, 0.330) & 0.028 & 0.064 & 150 & 148\\
Conformal Selective Borrow LC-NN & 0.175 & 0.079 & (0.019, 0.330) & 0.028 & 0.073 & 150 & 148\\
Borrow Na\"ive  & 0.180 & 0.074 & (0.034, 0.326) & 0.016 & 0.065 & 150 & 150 \\
Borrow IPW  & 0.163 & 0.076 & (0.015, 0.312) & 0.031 & 0.097 & 150 & 148 \\
Borrow CW  & 0.162 & 0.076 & (0.013, 0.310) & 0.030 & 0.068 & 150 & 148 \\
Borrow OM  & 0.175 & 0.076 & (0.025, 0.324) & 0.022 & 0.071 & 150 & 150 \\
Borrow AIPW  & 0.175 & 0.079 & (0.019, 0.330) & 0.028 & 0.069 & 150 & 148 \\
Borrow ACW  & 0.171 & 0.079 & (0.016, 0.327) & 0.031 & 0.069 & 150 & 148 \\
\hline
\end{tabular}
}
\end{table}

\begin{table}[ht]
\caption{Results of case study ($n_1=38, n_0=37, \nec=300$)}
\label{tab:resC5}
\centering
\resizebox{\linewidth}{!}{
\renewcommand{\arraystretch}{1}
\begin{tabular}{llllrrrr}
\hline
 & \multicolumn{1}{c}{} & \multicolumn{3}{c}{Asymptotic Inference} & \multicolumn{1}{c}{FRT} &  \\ \cline{3-5}
Method & Point Est.& SE & 95\% CI & $p$-value & $p$-value & Num. of EC & ESS of EC \\
\hline
No Borrow Unadj & 0.152 & 0.105 & (-0.053, 0.358) & 0.147 & 0.206 & 0 & 0 \\
No Borrow CovAdj & 0.132 & 0.118 & (-0.099, 0.362) & 0.263 & 0.236 & 0 & 0 \\
Conformal Selective Borrow NN & 0.151 & 0.073 & (0.008, 0.294) & 0.039 & 0.075 & 300 & 295\\
Conformal Selective Borrow LC-NN & 0.151 & 0.073 & (0.008, 0.294) & 0.039 & 0.071 & 300 & 295\\
Borrow Na\"ive & 0.156 & 0.069 & (0.021, 0.291) & 0.024 & 0.070 & 300 & 300 \\
Borrow IPW  & 0.141 & 0.072 & (0.000, 0.282) & 0.049 & 0.098 & 300 & 295 \\
Borrow CW  & 0.141 & 0.071 & (0.001, 0.280) & 0.048 & 0.065 & 300 & 294 \\
Borrow OM  & 0.151 & 0.073 & (0.009, 0.294) & 0.037 & 0.071 & 300 & 300 \\
Borrow AIPW  & 0.151 & 0.073 & (0.008, 0.294) & 0.039 & 0.069 & 300 & 295 \\
Borrow ACW  & 0.149 & 0.073 & (0.006, 0.293) & 0.041 & 0.076 & 300 & 294 \\
\hline
\end{tabular}
}
\end{table}

\begin{table}[ht]
\caption{Results of case study ($n_1=50, n_0=25, \nec=75$)}
\label{tab:resC6}
\centering
\resizebox{\linewidth}{!}{
\renewcommand{\arraystretch}{1}
\begin{tabular}{llllrrrr}
\hline
 & \multicolumn{1}{c}{} & \multicolumn{3}{c}{Asymptotic Inference} & \multicolumn{1}{c}{FRT} &  \\ \cline{3-5}
Method & Point Est.& SE & 95\% CI & $p$-value & $p$-value & Num. of EC & ESS of EC \\
\hline
No Borrow Unadj & 0.180 & 0.112 & (-0.040, 0.400) & 0.109 & 0.096 & 0  & 0\\
No Borrow CovAdj & 0.167 & 0.120 & (-0.067, 0.402) & 0.162 & 0.124 & 0 & 0\\
Conformal Selective Borrow NN & 0.218 & 0.084 & (0.052, 0.383) & 0.010 & 0.054 & 75 & 73\\
Conformal Selective Borrow LC-NN & 0.203 & 0.085 & (0.036, 0.370) & 0.017 & 0.114 & 70 & 69\\
Borrow Na\"ive & 0.216 & 0.077 & (0.064, 0.368) & 0.005 & 0.073 & 75 & 75 \\
Borrow IPW  & 0.218 & 0.077 & (0.068, 0.369) & 0.005 & 0.055 & 75 & 73 \\
Borrow CW  & 0.216 & 0.080 & (0.060, 0.372) & 0.007 &0.050 & 75 & 72 \\
Borrow OM  & 0.218 & 0.076 & (0.068, 0.368) & 0.004 & 0.059 & 75 & 75 \\
Borrow AIPW  & 0.218 & 0.084 & (0.052, 0.383) & 0.010 &0.055 & 75 & 73 \\
Borrow ACW  & 0.213 & 0.085 & (0.047, 0.379) & 0.012 &0.067 & 75 & 72 \\
\hline
\end{tabular}
}
\end{table}

\begin{table}[ht]
\caption{Results of case study ($n_1=50, n_0=25, \nec=300$)}
\label{tab:resC7}
\centering
\resizebox{\linewidth}{!}{
\renewcommand{\arraystretch}{1}
\begin{tabular}{llllrrrr}
\hline
 & \multicolumn{1}{c}{} & \multicolumn{3}{c}{Asymptotic Inference} & \multicolumn{1}{c}{FRT} &  \\ \cline{3-5}
Method & Point Est.& SE & 95\% CI & $p$-value & $p$-value & Num. of EC & ESS of EC \\
\hline
No Borrow Unadj & 0.180 & 0.112 & (-0.040, 0.400) & 0.109 & 0.096 & 0 & 0 \\
No Borrow CovAdj & 0.167 & 0.120 & (-0.067, 0.402) & 0.162 & 0.135 & 0 & 0 \\
Conformal Selective Borrow NN & 0.180 & 0.066 & (0.050, 0.310) & 0.007 & 0.062 & 300 & 294\\
Conformal Selective Borrow LC-NN & 0.180 & 0.066 & (0.050, 0.310) & 0.007 & 0.061 & 300 & 294\\
Borrow Na\"ive & 0.179 & 0.061 & (0.059, 0.299) & 0.003 & 0.079 & 300 & 300 \\
Borrow IPW  & 0.182 & 0.062 & (0.060, 0.303) & 0.003 & 0.043 & 300 & 294 \\
Borrow CW  & 0.178 & 0.063 & (0.055, 0.301) & 0.004 & 0.051 & 300 & 292 \\
Borrow OM  & 0.180 & 0.062 & (0.059, 0.302) & 0.003 & 0.057 & 300 & 300 \\
Borrow AIPW  & 0.180 & 0.066 & (0.050, 0.310) & 0.007 & 0.055 & 300 & 294 \\
Borrow ACW  & 0.177 & 0.066 & (0.047, 0.307) & 0.008 & 0.064 & 300 & 292 \\
\hline
\end{tabular}
}
\end{table}

\begin{table}[ht]
\caption{Results of case study ($n_1=167, n_0=168, \nec=670$)}
\label{tab:resC8}
\centering
\resizebox{\linewidth}{!}{
\renewcommand{\arraystretch}{1}
\begin{tabular}{llllrrrrr}
\hline
 & \multicolumn{1}{c}{} & \multicolumn{3}{c}{Asymptotic Inference} & \multicolumn{1}{c}{FRT} &  &  \\ \cline{3-5}
Method & Point Est. & SE & 95\% CI & $p$-value & $p$-value & Num. of EC & ESS of EC & FRT Runtime (s)\\
\hline
No Borrow Unadj & 0.076 & 0.048 & (-0.018, 0.169) & 0.110 & 0.120 & 0 & 0 & 0.001 \\
No Borrow CovAdj & 0.081 & 0.049 & (-0.015, 0.178) & 0.097 & 0.096 & 0 & 0 & 22.271\\
Conformal Selective Borrow NN & 0.112 & 0.037 & (0.040, 0.184) & 0.002 & 0.059 & 641 & 620 & 64.679\\
Conformal Selective Borrow LC-NN & 0.119 & 0.037 & (0.047, 0.191) & 0.001 & 0.049 & 670 & 634 & 70.172\\
Borrow Na\"ive  & 0.121 & 0.036 & (0.050, 0.191) & 0.001 & 0.055 & 670 & 670 & 60.640\\
Borrow IPW  & 0.116 & 0.037 & (0.043, 0.188) & 0.002 & 0.062 & 670& 634 & 14.872 \\
Borrow CW  & 0.114 & 0.036 & (0.042, 0.185) & 0.002 & 0.060 & 670& 630 & 23.589\\
Borrow OM & 0.119 & 0.036 & (0.049, 0.189) & 0.001 & 0.041 & 670& 670 & 25.123\\
Borrow AIPW  & 0.119 & 0.037 & (0.047, 0.191) & 0.001 & 0.048 & 670& 634 & 34.450\\
Borrow ACW  & 0.117 & 0.037 & (0.045, 0.190) & 0.001 & 0.045 & 670& 630 & 43.075\\
\hline
\end{tabular}
}
\end{table}

\begin{table}[ht]
\caption{Results of case study ($n_1=167, n_0=168, \nec=1005$)}
\label{tab:resC9}
\centering
\resizebox{\linewidth}{!}{
\renewcommand{\arraystretch}{1}
\begin{tabular}{llllrrrrr}
\hline
 & \multicolumn{1}{c}{} & \multicolumn{3}{c}{Asymptotic Inference} & \multicolumn{1}{c}{FRT} &  &  \\ \cline{3-5}
Method & Point Est. & SE & 95\% CI & $p$-value & $p$-value & Num. of EC & ESS of EC & FRT Runtime (s)\\
\hline
No Borrow Unadj & 0.076 & 0.048 & (-0.018, 0.169) & 0.110 & 0.120 & 0 & 0 & 0.001 \\
No Borrow CovAdj & 0.081 & 0.049 & (-0.015, 0.178) & 0.097 & 0.096 & 0 & 0 & 23.642\\
Conformal Selective Borrow NN & 0.123 & 0.036 & (0.053, 0.193) & $<$0.001 & 0.062 & 965 & 942 & 67.065\\
Conformal Selective Borrow LC-NN & 0.131 & 0.036 & (0.061, 0.201) & $<$0.001 & 0.042 & 986 & 954 & 76.068\\
Borrow Na\"ive & 0.132 & 0.035 & (0.063, 0.200) &  $<$0.001 & 0.057 & 1005 & 1005 & 47.362\\
Borrow IPW  & 0.128 & 0.035 & (0.058, 0.197) & $<$0.001 & 0.067 & 1005 & 965 & 18.651\\
Borrow CW  & 0.127 & 0.036 & (0.056, 0.197) & $<$0.001 & 0.057 & 1005 & 961 & 34.456\\
Borrow OM  & 0.131 & 0.035 & (0.063, 0.199) & $<$0.001 & 0.047 & 1005 & 1005 & 31.633\\
Borrow AIPW & 0.131 & 0.036 & (0.061, 0.201) & $<$0.001 & 0.048 & 1005 & 965 & 40.194\\
Borrow ACW & 0.131 & 0.036 & (0.061, 0.201) & $<$0.001 & 0.048 & 1005 & 961 & 74.764\\
\hline
\end{tabular}
}
\end{table}

\subsection{Plots for the estimation results as size of EC increases}\label{Supp:case_sizeECchange}
\begin{figure}[ht]
    \centering
    \includegraphics[width=\linewidth]{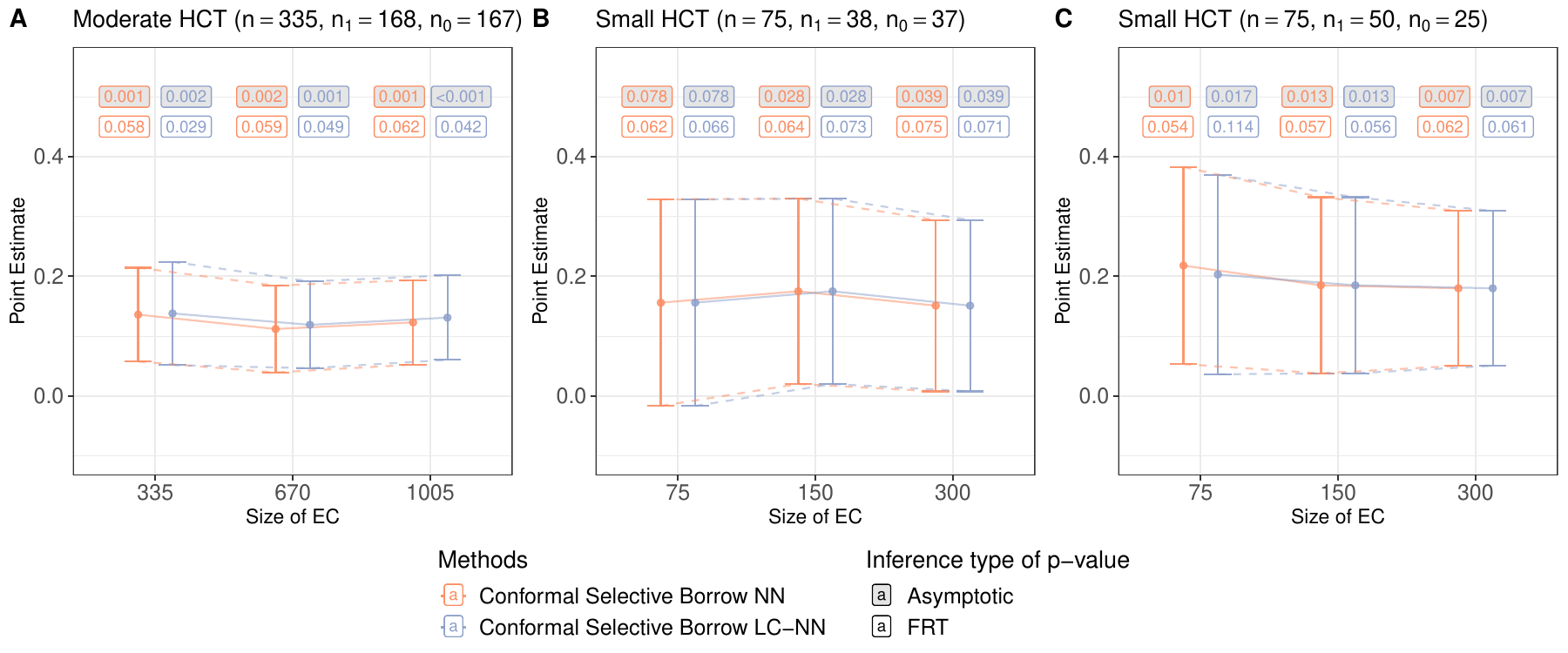}
    \caption{Change of estimates as size of EC increases (Conformal Selective Borrow focused)}
    \label{fig:rd_estimates_csb}
\end{figure}

\begin{figure}[ht]
    \centering
    \includegraphics[width=0.9\linewidth]{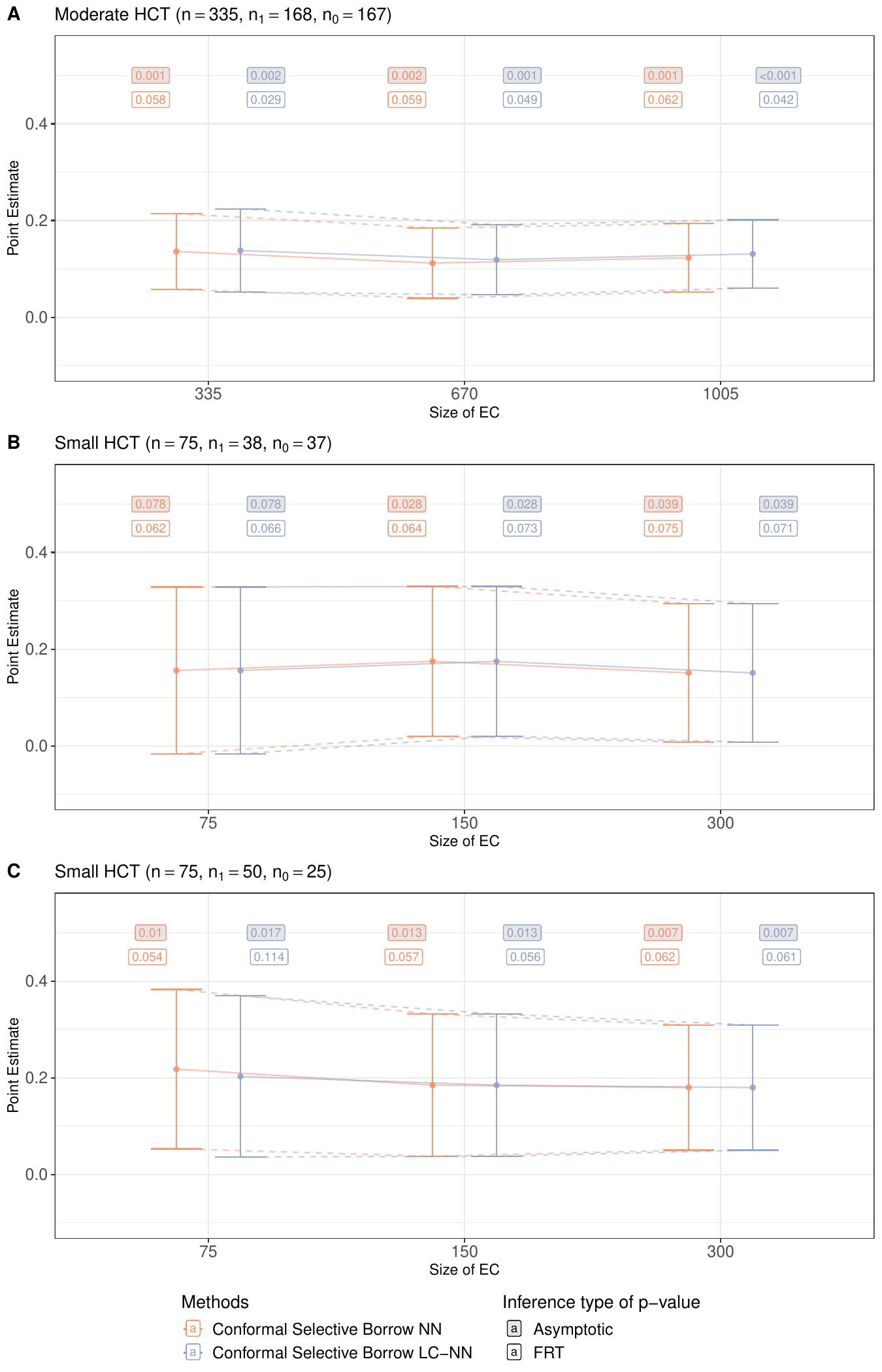}
    \caption{Change of estimates as size of EC increases (Full Borrow focused)}
    \label{fig:rd_estimates_fullBorrow}
\end{figure}

This section provides the figures (Figure \ref{fig:rd_estimates_csb} - \ref{fig:rd_estimates_fullBorrow}) demonstrating how the size of EC impacts the change of point estimates and $p$-values. Both Borrow methods and CSB Borrow methods are considered. 

\subsection{Additional diagnostic assessment of outcome comparability}\label{Supp:case_diag_outcome}
In this section, we evaluate the outcome comparability after applying CSB methods by performing a diagnostic test. Specifically, we compared the RCT controls ($S=1$, $A=0$) with the ECs ($S=0$, $A=0$) using covariate-adjusted logistic regression models for the binary outcome \citep{ventz2022design}. We considered a reduced model including only the baseline covariates, age, sex, tumor size, histology, and race, and a full model that additionally allowed the outcome model to differ by data source through the source indicator $S$ and its interactions with the baseline covariates:
\begin{align*}
    \text{Reduced model: } Y \sim X,\qquad \text{Full model: } Y \sim X + S + X \times S.
\end{align*}

We then used a likelihood ratio test to assess whether allowing source-specific outcome models significantly improved model fit. A smaller $p$-value indicates stronger evidence that the outcome model differs between the RCT controls and the EC controls after accounting for observed covariates. 
The results show statistically significant evidence of violation for the full EC set ($p=0.017$), while the CSB-selected subset yields a larger $p$-value ($p=0.158$), suggesting improved comparability between the selected EC subset and the RCT controls. 
Therefore, while this diagnostic test cannot verify Assumption~\ref{as2} definitively, it provides empirical support that Assumption~\ref{as2} is more plausible for the CSB-selected comparable subgroup than for the full borrowing sample.

Lastly, we emphasize that this should be viewed as a diagnostic assessment of Assumption~\ref{as2} rather than a formal decision rule, since such testing may introduce additional post-selection inference issues \citep{yang2023elastic}. Moreover, the likelihood ratio test relies on correct specification of the working model; for {nonparametric alternatives}, see \cite{luedtke2019omnibus}.

\end{document}